\documentclass{article}
\usepackage{graphicx}
\usepackage[a4paper]{geometry}
\usepackage{ulem}
\usepackage{color}
\usepackage{acronym}
\usepackage{xcolor}
\usepackage{url}
\usepackage{amsmath,amssymb}
\usepackage{longtable}
\usepackage{caption}
\usepackage{subcaption}
\usepackage{slashed}
\usepackage{jheppub}
\usepackage{cleveref}
\usepackage{physics}
\usepackage{dsfont}
\usepackage{float}
\usepackage{comment}

\crefname{section}{sec.}{secs.}
\crefname{table}{tab.}{tabs.}
\crefname{figure}{fig.}{figs.}
\crefname{equation}{eq.}{eqs.}
\crefname{appendix}{appendix}{appendices}

\newcommand{\gt}{\Tilde{g}}
\newcommand{\gV}{g_{V\pi\pi}}
\newcommand{\hc}{\mathrm{h.c.}} 
\newcommand{\SO}{\mathrm{SO}}
\newcommand{\SU}{\mathrm{SU}}
\newcommand{\U}{\mathrm{U}}
\newcommand{\Sp}{\mathrm{Sp}}

\def\lsim{\mathrel{\rlap{\lower4pt\hbox{\hskip1pt$\sim$}}
    \raise1pt\hbox{$<$}}}         
\def\gsim{\mathrel{\rlap{\lower4pt\hbox{\hskip1pt$\sim$}}
    \raise1pt\hbox{$>$}}}

\newcommand{\ii}{\mathrm{i}}

\acrodef{CH}{Composite Higgs}
\acrodef{QCD}{quantum chromodynamics}
\acrodef{pNGB}{pseudo Nambu--Goldstone Boson}
\acrodefplural{pNGB}[pNGBs]{pseudo Nambu--Goldstone Bosons}
\acrodef{SM}{Standard Model}
\acrodef{EW}{electroweak}
\acrodef{irrep}{irreducible representation}
\acrodefplural{irrep}[irreps]{irreducible representations}
\acrodef{VEV}{vacuum expectation value}
\acrodef{UFO}{Universal FeynRules Output}
\acrodef{PC}{partial compositeness}

\title{Electroweak spin-1 resonances in Composite Higgs models}
\author[a]{R.~Caliri,}
\affiliation[a]{Institut f\"{u}r Theoretische Physik und Astrophysik, Uni W\"{u}rzburg, Emil-Hilb-Weg 22, D-97074 W\"{u}rzburg, Germany}
\emailAdd{rosy.caliri@uni-wuerzburg.de}
\author[a]{J.~Hadlik,}
\emailAdd{jan.hadlik@uni-wuerzburg.de}
\author[a]{M.~Kunkel,}
\emailAdd{manuel.kunkel@uni-wuerzburg.de}
\author[a]{W.~Porod,}
\emailAdd{werner.porod@uni-wuerzburg.de}
\author[b]{Ch.~Verollet}
\affiliation[b]{Institut de Physique des 2 Infinis de Lyon (IP2I), 69100 Villeurbanne Cedex,
France}
\emailAdd{c.verollet@ip2i.in2p3.fr}
\date{\today}

\begin{document}

\abstract{Composite Higgs models predict the existence
of various bound states. Among these are spin-1 resonances.
We investigate  models containing 
SU(2)$_L\times$SU(2)$_R$ as part of the unbroken subgroup
in the new strong sector. These models predict that there
are two neutral and one charged  spin-1 resonances mixing
sizably with the SM vector bosons. 
 As a consequence, these can be singly produced
in Drell-Yan processes at the LHC. We explore
their rich LHC phenomenology and show that there are
still viable scenarios consistent with existing LHC data where the masses of these states can be as low
as about 1.5 TeV. 
}

\maketitle

\section{Introduction}

\ac{CH} models offer a possible explanation for the nature of the Higgs
boson discovered at CERN and a dynamical origin for the breaking of the electroweak symmetry 
in the \ac{SM} \cite{Englert:1964et,Higgs:1964pj,Guralnik:1964eu}. 
These models solve the problem of the hierarchy between the \ac{EW} 
scale and the Planck scale by interpreting the Higgs boson as a composite particle originating from a new strongly interacting sector.
Similarly to \ac{QCD}, 
the breaking scale is dynamically generated via confinement and condensation of a new
strong interaction. This idea was first implemented in
the context of technicolor models \cite{Weinberg:1975gm,Dimopoulos:1979es,Dimopoulos:1979za}
containing no Higgs at all. This was then further developed
to models with the  Higgs boson 
being a \ac{pNGB} \cite{Kaplan:1983fs,Kaplan:1983sm}. 
This got then extended to \ac{PC} \cite{Kaplan:1991dc}
including linear interactions between top-quarks and 
so-called top-partners to explain the heaviness of the top-quark.
Composite model building has gotten a further push
thanks to the idea of holography 
\cite{Contino:2003ve,Agashe:2004rs,Hosotani:2005nz} which 
has been freely adapted from 
duality conjectures \cite{Maldacena:1997re}.
We refer to \cite{Contino:2010rs,Cacciapaglia:2020kgq} for reviews on model-building aspects.

We will focus here on models based on an underlying gauge-fermion description in which
properties and quantum numbers of the resonances can be systematically classified.
We denote the new fermions as hyperfermions to distinguish them from the SM fermions.  
Consistent models with a single species of hyperfermions can only be based on $\SU(3)$ 
\cite{Vecchi:2015fma} or $G_2$ \cite{Ferretti:2013kya} with fermions in the fundamental
representation. Models with two separate species in different \acp{irrep} of the gauge group \cite{Barnard:2013zea,Ferretti:2013kya} offer a significantly larger possibility to realize top-partners
of which some have non-standard phenomenology \cite{Cacciapaglia:2021uqh}. 
 Recently, theoretical and 
phenomenological considerations have led to the definition of 12 minimal models.
They are fully specified \cite{Ferretti:2016upr,Belyaev:2016ftv} in terms of 
the confining gauge group and the irreps and multiplicities of the two species of hyperfermions. Both species condense upon confinement as confirmed by lattice results for 
$\SU(4)$ and $\Sp(4)$ gauge symmetries \cite{Ayyar:2017uqh,Bennett:2023wjw}
generating two sets of \acp{pNGB} \cite{Ferretti:2016upr}. 
These models contain
top partners which emerge as so-called ``chimera'' baryons formed of the two species, 
where two different patterns can be realized: $\psi \psi \chi$ or $\psi \chi \chi$, depending on the specific model. 
Here $\psi$ carries only electroweak charges while $\chi$ carries QCD color and hypercharge.

The phenomenology of various resonances predicted
by these 12 models has been studied in the literature
covering different sectors. So far, studies have focused on the  pNGBs  charged under electroweak quantum numbers 
\cite{Ferretti:2016upr,Agugliaro:2018vsu,Cacciapaglia:2022bax,Flacke:2023eil}, 
the singlets stemming from the global $\U(1)$'s \cite{Ferretti:2016upr,Belyaev:2016ftv,Cacciapaglia:2019bqz,BuarqueFranzosi:2021kky}, 
QCD colored pNGBs \cite{Cacciapaglia:2015eqa,Belyaev:2016ftv,Cacciapaglia:2020vyf}, 
top partners with non-standard decays \cite{Bizot:2018tds,Xie:2019gya,Cacciapaglia:2019zmj} 
or color assignment \cite{Cacciapaglia:2021uqh}, spin-1 resonances carrying electroweak charges \cite{BuarqueFranzosi:2016ooy} or QCD charges \cite{Cacciapaglia:2024wdn}. 
For completeness, we also note that the spectra and couplings of such resonances can 
be computed on the Lattice, and some results are available for models based on $\Sp(4)$ \cite{Bennett:2017kga,Bennett:2019cxd,Bennett:2019jzz,Bennett:2020hqd,Bennett:2020qtj,Bennett:2022yfa,Kulkarni:2022bvh,Bennett:2023gbe,Bennett:2023mhh,Bennett:2023qwx,Bennett:2024wda}
and based on $\SU(4)$ \cite{Ayyar:2017qdf,Ayyar:2018glg,Ayyar:2018ppa,Ayyar:2018zuk,Ayyar:2019exp,Golterman:2020pyx,Hasenfratz:2023sqa}. 
Computations based on holography are also available \cite{Erdmenger:2020lvq,Erdmenger:2020flu,Elander:2020nyd,Elander:2021bmt,Elander:2023aow,Erdmenger:2023hkl,Erdmenger:2024dxf}.

In the present work, we will focus on the 
phenomenology of spin-1 resonances which emerge
as bound states of 
the $\psi$ species and which carry 
electroweak charges. Their properties depend on the
corresponding coset for which we consider here
$\SU(4)/\Sp(4)$, $\SU(5)/\SO(5)$ and $\SU(4)\times \SU(4)/\SU(4)$. These cosets are symmetric implying 
that the spin-1 resonances fall into two categories: 
(i) states decaying into two 
\acp{pNGB} which will be called vector states $\mathcal 
V^\mu$ and (ii) states decaying into three \acp{pNGB} which
will be called axial-vectors 
$\mathcal A^\mu$. We will show that some of these 
states mix with the electroweak vector bosons. It turns out 
that all three cosets considered here contain one charged 
state mixing with the $W$-boson and two neutral states 
mixing with $Z$-boson.  
This is a consequence of the fact that
$\Sp(4)$, $\SO(5)$ and $\SU(4)$ contain $\SU(2)_L \times \SU(2)_R$ as a subgroup.  These states correspond essentially to the ones discussed in \cite{Belyaev:2008yj,Becciolini:2014eba}. Their s-channel 
production is constrained by existing 
LHC data. However, these states do not only decay 
into SM fermions but also via various other channels:   
$ V\, H$ ($V=W,Z$), two electroweak vector bosons
as well as two  \acp{pNGB}.
In this paper, we therefore evaluate to which
extent existing data constrain these models. We will
show that the additional decay possibilities imply
that masses as low as 1.5~TeV are still allowed.

The paper is organized as follows: In \cref{sec:mod} we first 
summarize the relevant features of the models considered here, including relevant parts of the effective Lagrangian. In \cref{sec:pheno} we will discuss phenomenological aspects of the spin-1 resonances mixing with the SM electroweak vector bosons as these are the ones which can
be singly produced at the LHC. This motivates the
study of four limiting scenarios which will
 be used in \cref{sec:lhc} to 
present bounds in mass-coupling planes from existing LHC data.
In \cref{sec:outlook} we draw our conclusions
and present an outlook. This is complemented 
by various appendices on model details in \cref{{app:modeldetails}},
additional information on the LHC constraints in \cref{app:pheno},
as well as 
formulae for the partial widths of
three body decays of a \ac{pNGB} 
into three vector bosons in \cref{sec:eta30decay} which have not been given in
the literature so far.

\section{Model aspects}
\label{sec:mod}

In the models proposed in ref.~\cite{Ferretti:2016upr} the \ac{EW} sector is 
contained in one of three cosets: $\SU(5)/\SO(5)$, 
$\SU(4)/\Sp(4)$, and $\SU(4)^2/\SU(4)$ depending
on whether the hyperfermions are in a real, pseudo-
real or complex \ac{irrep} of $G_\mathrm{HC}$, 
respectively. Below we will denote these cosets
generically as  $G/H$. 
In the following, we collect the main ingredients for
the construction of the effective Lagrangian of
these models in a generic way 
following the lines of
refs.~\cite{BuarqueFranzosi:2016ooy,BuarqueFranzosi:2023xux}  to which we refer for further details.
The model specific details can be found in
\cref{app:models}.

\subsection{Vacuum alignment}

We work in a basis where the \ac{pNGB} fields are defined around a true vacuum which includes the source of electroweak symmetry breaking.
One can show that the vacuum alignment can be described in terms of a single parameter, $\theta$, and the corresponding true vacuum $\Sigma_\theta$ can be expressed as
\begin{align}
\Sigma_\theta =  \Omega(\theta) \Sigma_0 \Omega^T(\theta) \,,
\end{align}
where $\Sigma_0$ is the vacuum which leaves the subgroup $H$ invariant and
\begin{align}
\Omega(\theta) = \exp\left( \sqrt{2} \ii \theta X^H  \right) \,,
\end{align}
with $X^H$ being the broken generator of the Higgs \ac{pNGB} 
fulfilling like any broken generator 
\begin{align} 
\label{eq:defX}
 X^H \cdot \Sigma_0 - \Sigma_0 \cdot {X^H}^T = 0\,.
\end{align}
For $\theta = 0$, i.e. $\Omega (0) = \mathds{1}$, the electroweak symmetry is unbroken and we denote
the $\SU(2)_L$ by $T^i_L = T^i$ ($i=1,2,3$) and the ones of $\SU(2)_R$ by $T^i_R = T^{i+3}$ ($i=1,2,3$). Here the $T^j$ are the generators
of the unbroken subgroup $H$ fulfilling
\begin{align} 
\label{eq:defT}
 T^j \cdot \Sigma_0 + \Sigma_0 \cdot {T^j}^T = 0
 \quad (j=1,\dots,\text{dim}(H) )\,.
\end{align}
The  $\U(1)_Y$ generator is given by $T^3_R$ in our convention. 
In the phase where $\theta$ is non-zero, the unbroken generators $\tilde T^a$ satisfy
\begin{align} 
\label{eq:deftT}
\tilde T^a \cdot \Sigma_\theta + \Sigma_\theta \cdot {{\tilde T}^a}{}^T = 0\,,
\end{align}
and are no longer aligned with the gauged generators $T^i$ ($i=1,2,3,6)$ of $G$.
Analogously, the broken generators now fulfill
\begin{align} 
\label{eq:deftX}
\tilde X^I \cdot \Sigma_\theta - \Sigma_\theta \cdot {\tilde X^I}{}^T = 0\,.
\end{align}
The Goldstone matrix is given by
\begin{align}
U = \exp \left( \ii \frac{\sqrt{2}}{f_\pi} \sum_{I=1}^{n} \pi^I \tilde X^I \right) = \exp \left( \ii \frac{\sqrt{2}}{f_\pi} \tilde \Pi \right) \,,
\end{align}
with $n=\dim G - \dim H$.
The decay constant
$f_\pi$ is related to the misalignment angle by
\begin{align}
f_\pi \sin \theta = v_{\text{SM}} = 246~\text{GeV}\,.
\label{eq:def_fpi}  
\end{align}

\subsection{Hidden symmetry approach}

One can construct a chiral-type theory  based on custodial symmetry and gauge invariance 
to describe the new strong sector while remaining as general as possible. 
Following along the lines of Refs.~\cite{BuarqueFranzosi:2016ooy,Cacciapaglia:2024wdn}, we employ the hidden gauge symmetry approach \cite{Bando:1987br} which 
introduces a local copy of the global  symmetry to 
obtain a description of the spin-1 resonances.
In the limit in which these states decouple
one obtains the Lagrangian of the non--linear $\sigma$-model, describing the Goldstone bosons 
associated to the breaking of $G \to H$. In order to achieve this we initially extend the global
group $G$ to a product of two copies: $G_0\times G_1$. Here $G_0$ corresponds to the usual global 
symmetry leading to the Higgs as a composite \ac{pNGB}, and the electroweak gauge bosons are introduced via its partial gauging. As
indicated below, the physical \acp{pNGB} are a linear combination of the \acp{pNGB} of the two sectors.
The new group $G_1$ allows one to introduce a new set of massive ``gauge'' bosons transforming as a 
complete adjoint irrep of $G_1$.
These correspond to the spin-1 resonances studied in this work.
The states corresponding to the unbroken and 
broken generators are called vectors 
$\mathcal{V}_\mu$ and axial vectors $\mathcal{A}_\mu$,
 respectively.

The factors $G_i$ ($i=0,1$) are  spontaneously broken to $H_i$ via the introduction of 
two Goldstone matrices $U_i$ containing the same number of \acp{pNGB} each
\begin{align}
U_0 = \exp\left( \frac{\ii \sqrt 2}{f_0} \sum_{I=1}^n \pi_0^I  \tilde X^I  \right),   \qquad
U_1 = \exp\left( \frac{\ii \sqrt 2}{f_1} \sum_{I=1}^n \pi_1^I  \tilde X^I \right)\,,
\end{align}
which transform non-linearly as 
\begin{align}
U_i\to U^\prime_i=g_i U_i h(g_i,\pi_i)^\dagger\,.
\end{align}
Here $g_i$ is an element of the corresponding factor $G_i$ and $h$ an element of 
the respective subgroup $H_i$.
As discussed below, a linear combination of these \acp{pNGB} will give rise to the longitudinal
components of the axial vector bosons whereas the second set corresponds to observable states
apart from the ones providing the masses to W- and Z-bosons.  
The final low energy Lagrangian is then characterized in terms of the breaking of the extended symmetry $H_0\times H_1$ down to a single $H$ by a sigma field $K$, containing $m=\dim H$ Goldstone bosons $k^a$. 
They give rise to the longitudinal components of the vector resonances.

We define a Maurer-Cartan form for each sector,
\begin{align}
    \Omega_{i,\mu} = \ii U_i^\dagger  D_\mu U_i
\end{align}
where the covariant derivatives are given by
\begin{align}
    D_\mu U_0 &= \left( \partial_\mu - \ii\hat g\, \tilde W^i_\mu T^i_L - \ii\hat g'\, B_\mu T_R^3 \right) U_0\,, \label{eq:DmuU0}\\
    D_\mu U_1 &= \left( \partial_\mu -\ii \tilde g\, \mathcal V^a_\mu \tilde T^a - \ii \tilde g\, \mathcal A^I_\mu \tilde X^I \right) U_1\,. \label{eq:DmuU1}
\end{align}
For a more compact notation we sometimes write $\boldsymbol{\tilde W}_\mu = \tilde W^i_\mu T^i_L$ etc.
The couplings in \cref{eq:DmuU0} carry hats to indicate that these are not the usual \ac{EW} gauge couplings as we will see below.
Note that we use misaligned generators in \cref{eq:DmuU1} but non-rotated generators in \cref{eq:DmuU0} since these are the ones corresponding to $\SU(2)_L \times \U(1)_Y$.
This ensures on the one hand
 correct quantum numbers of the underlying hyperfermions. 
On the other hand, the
spin-1 resonances are excitations around the true vacuum.
From the Maurer-Cartan forms we define the one-forms
\begin{align}
    d_{i,\mu} &= \Tr(\Omega_{i,\mu} \tilde X^I) \tilde X^I \\
    e_{i,\mu} &=  \Tr(\Omega_{i,\mu} \tilde T^a) \tilde T^a
\end{align}
for use in the CCWZ construction \cite{Coleman:1969sm,Callan:1969sn}.
We further define a Goldstone matrix for the $k^a$ fields
\begin{align}
    K = \exp( \frac{\ii}{f_K} \sum_{a=1}^{m} k^a \tilde T^a ) \,,
\end{align}
with covariant derivative
\begin{align}
    D_\mu K = \partial_\mu K - \ii e_{0,\mu} K + \ii K e_{1,\mu}.
\end{align}
Finally, in our conventions the field strength tensor of a generic gauge field $V_\mu$ reads
\begin{align}
    \mathbf V_{\mu\nu} = \partial_\mu \mathbf V_\nu - \partial_\nu \mathbf V_\mu - \ii  g [\mathbf V_\mu, \mathbf V_\nu],
\end{align}
with the appropriate gauge coupling $g$.
We recall that the full $G_1$ is gauged, so the corresponding gauge field is $\boldsymbol {\mathcal F}_\mu = \boldsymbol {\mathcal V}_\mu + \boldsymbol {\mathcal A}_\mu$. 

We now have all the ingredients in place to write down the Lagrangian, which is given at leading order by \cite{BuarqueFranzosi:2016ooy}
\begin{align}
    \mathcal{L} &= -\frac{1}{4} \Tr \boldsymbol {\mathcal F}_{\mu \nu} \boldsymbol {\mathcal F}^{\mu \nu} - \frac{1}{4} \Tr \mathbf W_{\mu \nu} \mathbf W^{\mu \nu} - \frac{1}{4} \Tr \mathbf B_{\mu \nu} \mathbf B^{\mu \nu} \nonumber\\
    &+ \frac{f_0^2}{4} \Tr d_{0\mu} d_0^{\mu} + \frac{f_1^2}{4} \Tr d_{1\mu} d_1^{\mu} + \frac{r f_1^2}{2} \Tr d_{0\mu} K d_1^\mu K^\dag + \frac{f_K^2}{4} \Tr D^\mu K  (D_\mu K)^\dag \,.
  \label{eq:generic_lagrangian}  
\end{align}
Here we have normalized the generators as $\Tr T^A T^B = \delta^{AB}$.

\subsection{Physical states}

The $r$-term in the Lagrangian induces a mixing of the \acp{pNGB}. 
We refer to \cite{BuarqueFranzosi:2016ooy} for details and only recall here that a linear combination 
denoted as $\pi_U$ gives mass to the $\mathcal A_\mu$, while the orthogonal combination $\pi_P$ are the physical \acp{pNGB}
\begin{align}
    \pi_0 = \frac{f_0}{f_\pi} \pi_P \,,\qquad
    \pi_1 = \pi_U - \frac{r f_1}{f_\pi} \pi_P
    \label{eq:physical-pions}
\end{align}
with
\begin{align}
   f_\pi = \sqrt{f^2_0 - r^2 f_1^2} = v_{\text{SM}}/\sin \theta \label{eq:pion-decay-constant}\, ,
\end{align}
see also \cref{eq:def_fpi}.
The resulting $\pi_P$ are summarized in the first column of \cref{tab:cosets} for or three cosets, where $\varphi$ are the longitudinal components of the $W$ and $Z$ bosons and the $H$ is the Higgs boson.

\begin{table}[t]
    \resizebox{\textwidth}{!}{
    \begin{tabular}{|c|ccc|ccc|ccc|}
    \hline
      coset/particles   & \multicolumn{3}{c|}{pNGBs} &  \multicolumn{3}{c|}{$\mathcal A_\mu$} & \multicolumn{3}{c|}{$\mathcal V_\mu$} \\
      &  $\SU(2)^2$
      &  $\SU(2)_D$ & name  &  $\SU(2)^2$
      &  $\SU(2)_D$ & name  &  $\SU(2)^2$
      &  $\SU(2)_D$ & name \\ \hline
    SU(4)/Sp(4)  & (2,2) & 3 & $\varphi$ 
                 & (2,2) & 3 & $a_\mu$ 
                 & (2,2) & 3 & $\hat r_\mu$ \\
                 &  & 1 & $H$
                 &  & 1 & $\hat y_{1\mu}$      
                 &  & 1 & $\hat x_{1\mu}$    \\
    in M8-M9             & (1,1) & 1 & $\eta$ 
                 & (1,1) & 1 & $\hat y_{2\mu}$ 
                 & (3,1)+(1,3) & 3 &  $v_{1\mu}$ \\
                & & & & & & 
          &  & 3 &  $v_{2\mu}$ \\ \hline        
    SU(5)/SO(5)  & (2,2) & 3 & $\varphi$ 
                 & (2,2) & 3 & $a_\mu$ 
                 & (2,2) & 3 & $\hat r_\mu$ \\
                 &  & 1 & $H$
                 &  & 1 & $\hat y_{1\mu}$      
                 &  & 1 & $\hat x_{1\mu}$    \\
    in M1-M7            & (1,1) & 1 & $\eta$ 
                 & (1,1) & 1 & $\hat y_{2\mu}$ 
                & (3,1)+(1,3) & 3 &  $v_{1\mu}$ \\
                 & (3,3) & 5 & $\eta_5 $ 
                & (3,3) & 5 &  $\hat a_{5\mu}$
          &  & 3 &  $v_{2\mu}$ \\       
                & & 3 &  $\eta_3$
                & & 3 &  $\hat a_{3\mu}$
          &  &  &  \\       
                & & 1 &  $\eta_1$
                & & 1 &  $\hat a_{1\mu}$
          &  &  &  \\ \hline        
    SU(4)$^2$/SU(4)  & (2,2) & 3 & $\varphi$ 
                 & (2,2) & 3 & $a_\mu$ 
                 & (2,2) & 3 & $\hat r_\mu$ \\
                 &  & 1 & $H$
                 &  & 1 & $\hat y_{1\mu}$      
                 &  & 1 & $\hat x_{1\mu}$    \\
    in M10-M12              & (2,2) & 3 & $\phi_1$ 
                 & (2,2) & 3 & $\hat a_\mu$ 
                 & (2,2) & 3 & $ r_\mu$ \\
                 &  & 1 & $\phi_2$
                 &  & 1 & $\hat y_{3\mu}$      
                 &  & 1 & $\hat x_{3\mu}$    \\
                 & (1,1) & 1 & $\eta$ 
                 & (1,1) & 1 & $\hat y_{2\mu}$ 
                 & (1,1) & 1 & $\hat x_{2\mu}$ \\
                & (3,1)+(1,3) & 3 &  $\eta_{1}$
                & (3,1)+(1,3) & 3 &  $b_{1\mu}$
          & (3,1)+(1,3) & 3 &  $v_{1\mu}$ \\        
                &  & 3 &  $\eta_{2}$
                &  & 3 &   $b_{2\mu}$ 
          &  & 3 &  $v_{2\mu}$ \\ \hline        
    \end{tabular}
    }
    \caption{List of pNGBs, axial vector and vector states for the three cosets. For each particle
    we give first the $\SU(2)^2 \equiv \SU(2)_L \times \SU(2)_R$ representation, the $\SU(2)_D$ representation and the name used for the latter. 
    Moreover, we list in the first column the models from ref.~\cite{Belyaev:2016ftv} that feature the corresponding coset.}
    \label{tab:cosets}
\end{table}

Expanding the Lagrangian to second order in the spin-1 fields, we also find mass and mixing terms.
In particular, some of the resonances mix with the elementary gauge fields.
These states drive the LHC phenomenology of the models since the mixing allows for single production.
In the second and third columns of \cref{tab:cosets} we collect the axial vector and vector states\footnote{We note that the designation of ``vector'' and ``axial vector'' is strictly speaking only correct for cosets of the form $\SU(N)^2/\SU(N)$. We find it appropriate however, since the $\mathcal V_\mu$/$\mathcal A_\mu$ couple to two/three \acp{pNGB} to lowest order, just as in QCD.}, respectively.
The spin-1 states which do \textit{not} mix with the \ac{EW} vector bosons are indicated by a hat on the corresponding name.
They have the universal masses
\begin{align}
    M_A^2 = \frac{f_1^2 \Tilde{g}^2}{2} \,, \qquad
    M_V^2 = \frac{f_K^2 \Tilde{g}^2}{2} \,.\label{eq:mass-parameter}
\end{align}
For the states mixing with the \ac{SM} vector bosons we
obtain the following mass matrices for the cosets $\SU(4)/\Sp(4)$ and $\SU(5)/\SO(5)$. We note for completeness
that they coincide with the results of ref.~\cite{BuarqueFranzosi:2016ooy} for the $\SU(4)/\Sp(4)$ coset.
In the basis
$(\tilde W^+_\mu,\, a^+_\mu,\, v_{1\mu}^+,\, v_{2\mu}^+)$, the mass matrix in the charged sector reads
\begin{align}
    \mathcal{M}_C^2 &= \mqty(
        \frac{ \hat g^2 M_{V}^2 (1 + \omega s^2_{\theta})}{ \tilde{g}^2} & \hat g \frac{r s_\theta M_A^2}{\sqrt{2}\Tilde{g}} & - \hat g \frac{M_V^2}{\sqrt{2} \Tilde{g}} & - \hat g \frac{M_V^2 c_\theta}{\sqrt{2} \Tilde{g}} \\
        \hat g \frac{r s_\theta M_A^2}{\sqrt{2}\Tilde{g}} & M_A^2 & 0&0 \\
        - \hat g \frac{M_V^2}{\sqrt{2} \Tilde{g}} & 0& M_V^2 & 0\\
        - \hat g \frac{M_V^2 c_\theta}{\sqrt{2} \Tilde{g}} &0 &0 & M_V^2
    ) \,,
 \label{eq:MC}   
\end{align}
where $\omega=(f_0^2 / f_K^2 - 1)/2$. 
Only a linear combination of $v_{1\mu}^+$ and $v_{2\mu}^+$ mixes with $\tilde W^+$. We denote this heavy mass eigenstate by $V^+_{1\mu}$ in the following.
Moreover, the mixing with $a^+_\mu$ vanishes in the limit $\sin\theta\to 0$.
In the neutral sector we take the basis $(B_\mu, \tilde W^3_\mu,\, a^0_\mu,\, v_{1\mu}^0,\, v_{2\mu}^0)$, which yields the mass matrix
\begin{align}
    \mathcal{M}_N^2 &= \mqty(
        \frac{\hat g'^2 M_{V}^2 (1 + \omega s^2_{\theta})}{ \tilde{g}^2} & -\frac{\hat g' \hat g M_{V}^2 \omega  s^2_{\theta}}{\tilde{g}^2} & - \hat g' \frac{r s_\theta M_A^2}{\sqrt{2} \Tilde{g}} & - \hat g' \frac{M_V^2}{\sqrt{2} \Tilde{g}} &  \hat g' \frac{M_V^2 c_\theta}{\sqrt{2} \Tilde{g}} \\
        -\frac{ \hat g'  \hat g M_{V}^2 \omega  s^2_{\theta}}{\tilde{g}^2} & \frac{ \hat g^2 M_{V}^2 (1 + \omega s^2_{\theta})}{ \tilde{g}^2} &  \hat g \frac{r M_A^2 s_\theta}{\sqrt{2} \Tilde{g}} & - \hat g \frac{M_V^2}{\sqrt{2} \Tilde{g}} & - \hat g \frac{M_V^2 c_\theta}{\sqrt{2} \Tilde{g}} \\
        - \hat g' \frac{r s_\theta M_A^2}{\sqrt{2} \Tilde{g}} &  \hat g \frac{r M_A^2 s_\theta}{\sqrt{2} \Tilde{g}} & M_A^2 & 0&0 \\
        - \hat g' \frac{M_V^2}{\sqrt{2} \Tilde{g}} & - \hat g \frac{M_V^2}{\sqrt{2} \Tilde{g}} &0 & M_V^2 & 0\\
         \hat g' \frac{M_V^2 c_\theta}{\sqrt{2} \Tilde{g}} & - \hat g \frac{M_V^2 c_\theta}{\sqrt{2} \Tilde{g}} & 0&0 & M_V^2 ) \,.
\end{align}
In this sector the two states $v_{1\mu}^0$ and 
$v_{2\mu}^0$ mix with the photon and the $Z$-boson
whereas the mixing with $a_{\mu}^0$ is suppressed
and vanishes in the limit $\sin\theta\to 0$.
We denote the corresponding heavy mass eigenstates which mix sizably with the electroweak vector bosons by $V^0_{1\mu}$ and $V^0_{2\mu}$.  
In the $\SU(4)\times \SU(4)/\SU(4)$ coset the situation is quite similar, and we give the details in \cref{app:su4su4-mass}.
We note that in all cosets the states $a_\mu$, $v_{1\mu}$ and  $v_{2\mu}$ mix with the SM vector bosons. 
The reason for this is that in all cosets $G/H$ one has $\SU(2)_L\times \SU(2)_R$ as a subgroup of $H$.

Both mass matrices are diagonalized by orthogonal
rotation matrices which we denote by $\mathcal{C}$ and $\mathcal{N}$ 
for the charged and neutral sectors, respectively: 
\begin{align}
    \mqty( \tilde W^+_\mu \\ a^+_\mu \\ v_{1\mu}^+ \\ v_{2\mu}^+ ) = \mathcal{C} \mqty( W^+_\mu \\ A^+_\mu \\ V_{1\mu}^+ \\ V_{2\mu}^+ ) = \mathcal{C} R_\mu^+ \,, \qquad \mqty( B_\mu \\ \tilde W^3_\mu \\ a^0_\mu \\ v_{1\mu}^0 \\ v_{2\mu}^0 ) = \mathcal{N} \mqty( A_\mu \\ Z_\mu \\ A^0_\mu \\ V_{1\mu}^0 \\ V_{2\mu}^0 ) = \mathcal{N} R_\mu^0 \,, \label{eq:mass-eigenstates-basis}
\end{align}
denoting the mass eigenstate vectors by $R_\mu^+$
and $R_\mu^0$. The eigenvector of the massless photon can be obtained analytically:
\begin{align}
    A_\mu &= \frac{e}{\hat g} \tilde W_\mu^3 + \frac{e}{ \hat g'} B_\mu + \frac{2 e}{\gt} v_{1\mu}^0 \label{eq:photon}
\end{align}
with
\begin{align}
    \frac{1}{e^2} &= \frac{1}{ \hat g^2} + \frac{1}{ \hat g'^2} + \frac{2}{\Tilde{g}^2} \,.
\label{eq:electric-charge}    
\end{align}

We note for completeness that this coset 
contains an additional $\U(1)$, which 
implies a further \ac{SM} singlet vector state. 
However, it does not mix with the other spin-1 resonances and, thus, is irrelevant for our considerations here.
Moreover, in models  containing additional  hyperfermions 
charged with respect to $\SU(3)_C$, e.g.\ in
the ones of ref.~\cite{Ferretti:2016upr,Belyaev:2016ftv},
there is an additional U(1) vector $\tilde V$ stemming
from the embedding of the SU(3) sector which
mixes with $\U(1)_Y$ \cite{Cacciapaglia:2024wdn}. 
 An inclusion of this state
would introduce quite some model dependence
which will be part of a future investigation
\cite{2nd_paper}. There are essentially two possibilities:
(i)  The additional state $\tilde V$ is heavier
then the $V_1$ states. This will in particular
be the case if the underlying hyperfermions $\chi$
belong to a 2-index representation of $G_{HC}$
and the electroweak spin-1 resonances are formed
from hyperfermions $\psi$ belong to the fundamental
representation of $G_{HC}$ as indicated by studies on 
the lattice \cite{Ayyar:2018zuk,Bennett:2019cxd,Bennett:2022yfa,Bennett:2023qwx} 
as well as using gauge/gravity duality
\cite{Erdmenger:2020lvq,Erdmenger:2020flu}. The mass difference between the spin-1 states
will be further enhanced if the $\chi$ have a larger mass than the $\psi$ as the masses
of the vector states increases with the mass
of the underlying hyperfermions \cite{Kulkarni:2022bvh,Erdmenger:2023hkl,Erdmenger:2024dxf}. In such scenarios, the main effect of
the additional state will be a slight decrease
of the smallest mass and thus a slight increase of the mass difference between $V_{1\mu}^0$ and $V^0_{2\mu}$.
Moreover, the entries of the mixing matrix
$\mathcal{N}$ will change slightly but the impact
on the branching ratios for the final states discussed below is rather small. Consequently
the main features of our findings below will
still be correct.
(ii) The additional state $\tilde V$ has about the same mass or is even lighter
than the electroweak spin-1 resonances. In such
scenarios we expect stronger exclusion limits from
LHC data compared to those presented in \cref{sec:pheno}. The investigation of such
a scenario is left for a future study as it is more
model dependent.

\subsection{Relevant interactions}

We now collect the interactions that facilitate either the production or the decay of the heavy spin-1 resonances. Here we focus on the states $V_1^0, V_2^0, V_1^+$ that mix with \ac{SM} vector bosons even for $\sin\theta \to 0$.
The reason for this focus is that the mixing generates couplings between these spin-1 resonances and SM fermions, allowing for single production:
\begin{align}
    \mathcal{L}_\mathrm{CC} &= \frac{\hat g}{\sqrt{2}} \sum_{i,f,f'} \mathcal{C}_{1i} \Bar{\psi}_f \slashed{R}^+_i P_\mathrm{L} (V_{\tiny \mathrm{CKM}})_{ff'}\psi_{f'} + \hc \,, \label{eq:LagCC}\\
    \mathcal{L}_\mathrm{NC} &=  \sum_{i,f} \Bar{\psi}_f \slashed{R}^0_i \left( g_{\mathrm{L}i}^f P_\mathrm{L} + g_{\mathrm{R}i}^f P_\mathrm{R} \right) \psi_f  \,,\label{eq:LagNC}
\end{align}
with
\begin{align}
g_{\mathrm{L}i}^f = \hat g T^3_f \mathcal{N}_{2i} + \hat g' Y_{fL} \mathcal{N}_{1i}  \quad \text{ and } \quad
g_{\mathrm{R}i}^f = \hat g' Y_{fR} \mathcal{N}_{1i}.
\end{align}
Here $T^3_f$ is the weak isospin of the fermion $f$ and $Y_{fL,fR}$ the corresponding hypercharges. 
Note that \cref{eq:photon} implies
\begin{align}
    \hat g' \mathcal{N}_{11} = \hat g \mathcal{N}_{21} = e = g' c_W = g s_W
\end{align}
with $g'$ and $g$ being the usual \ac{SM} couplings, 
$c_W=\cos \theta_W$, $s_W=\sin \theta_W$ and
$\theta_W$ the Weinberg angle. 

In models with \ac{PC} the third generation quarks get an additional contribution from the mixing between the elementary fields and the top partners, which we parameterize as
\begin{align}\label{eq:simplifiedmodelVtt}
    \mathcal{L}_\mathrm{PC} &= \Bar{t} \left(\slashed{V}^0_{1}+\slashed{V}^0_{2} \right)  \left( g_{t,L} P_L + g_{t,R} P_R \right) t + \Bar{b} \left(\slashed{V}^0_{1}+\slashed{V}^0_{2} \right)  \left( g_{b,L} P_L  \right) b +  g_{tb,L}
    \Bar{t} \slashed{V}^+_{1}  P_L b \,.
\end{align}
Due to the small mixing\footnote{In practice it is sufficient that the mixing of the left-chiral quarks with their corresponding partners is a factor $\sqrt{3}$ smaller the one of the right-chiral top quark with its partner. This implies a relative factor 3 for the corresponding couplings.}
of the bottom quark with its partner the $g_{b,L}$ will be small.
Here we have assumed for simplicity that
the couplings of $V^0_1$ and $V^0_2$ are the
same. In practice they differ slightly due
to the difference in the corresponding entries
of $\mathcal{N}$. We have checked that the
corresponding
entries are very similar justifying our ansatz.
The couplings depend on the model dependent mixing between the elementary fields and the top partners which we take as free parameters encoded in the corresponding couplings.

The hidden symmetry Lagrangian also induces couplings 
of one spin-1 resonance to two electroweak vector bosons:
They originate from the terms
\begin{align}
    \mathcal{L} \supset -\mathrm{i} &\Big( \hat g \tilde W^{+\nu} \tilde W_\mu^- \partial^\mu \tilde W_\nu^3 + \frac{\gt}{\sqrt{2}} \left( (a^{+\nu} v^-_{1\mu} + v^{+\nu}_1 a^-_\mu) \partial^\mu a_\nu^0 + (v^{+\nu}_1 v^-_{2\mu} + v^{+\nu}_2 v^-_{1\mu}) \partial^\mu v_{2\nu}^0 \right) \nonumber \\
    &+ \frac{\gt}{\sqrt{2}} \left(a^{+\nu} a^-_\mu + v^{+\nu}_1 v^-_{1\mu} + v_2^{+\nu} v^-_{2\mu} \right) \partial^\mu v_{1\nu}^0  \Big) + \text{permutations } 
    \label{eq:int_3v}
\end{align}
once the mixing in \cref{eq:mass-eigenstates-basis} is taken into account. Note that all terms in \cref{eq:int_3v} give contributions of similar size:
the first one has a small gauge coupling but requires only
one small mixing between a heavy state and and electroweak vector bosons whereas the other contributions have a large gauge coupling but require two small mixing entries.
We further find couplings to two \acp{pNGB} of the form
\begin{align}
    \mathcal{L}_{V\pi\pi} &=
    \frac{\ii}{2} g_{V \pi \pi} \cdot \Tr\left(\boldsymbol{\mathcal{V}}_\mu \comm{\tilde \Pi_P}{\partial_\mu \tilde \Pi_P}\right)
    - \frac{\ii(\gV + 2\gt)}{2\gt} \Tr \left(\left(\hat g T(\boldsymbol{\tilde W}_\mu) + \hat g' T(\boldsymbol B_\mu) \right) \comm{\tilde \Pi_P}{\partial_\mu \tilde \Pi_P} \right)  \label{eq:gVpipi-lagrangian} \,,
\end{align}
where the vectors are taken before mixing but we already rotate the \acp{pNGB} to the physical $\Pi_P$.
We defined the vector-pNGB-pNGB coupling constant
\begin{align}
    g_{V \pi \pi} &= \frac{\Tilde{g} f_K^2 (r^2-1)}{f_\pi^2}  \label{eq:gVpipi}
\end{align}
and use the shortcut
\begin{align}
    T(\boldsymbol{\tilde W}_\mu) = \tilde W_{\mu}^i \mathrm{Tr} \Big( T_L^i \tilde T^a \Big) \tilde T^a \,, \quad T(\boldsymbol B_\mu) = B_{\mu} \mathrm{Tr} \Big( T_R^3 \tilde T^a \Big) \tilde T^a
    \label{eq:Tprojection}
\end{align}
which is the projection of the \ac{SM} gauge bosons onto the unbroken subgroup of the misaligned vacuum.
Note that the second term in \cref{eq:gVpipi-lagrangian} implies that even for $g_{V\pi\pi}=0$
there is a non-vanishing coupling of the spin-1
resonances to two \acp{pNGB}. 
Neither $H$ nor $\varphi$ participate in the couplings in \cref{eq:gVpipi-lagrangian} as a consequence of our choice of vacuum and the fact that we work in unitary gauge.

The spin-1 resonances do couple to the Higgs and one \ac{SM} vector boson, however.
We can generically write the couplings as
\begin{align}
    \mathcal{L} &\supset c_{H R^+ R^-} \cdot H R_{i\mu}^+ R^{-\mu}_j + \frac{1}{2} c_{H R^0 R^0} \cdot H R_{i\mu}^0 R^{0\mu}_j \,.
\end{align}
with the details given in \cref{app:higgsappendix}.

\subsection{Independent parameters}

For the following, we will swap the original parameters $f_0$,  $f_1$, $f_K$ and $r$ of the
effective action in \cref{eq:generic_lagrangian}
for the vector mass parameter $M_V$ (see \cref{eq:mass-parameter}), 
the ratio of axial to vector mass $\xi=M_A/M_V$, 
the coupling scale of vectors to two \acp{pNGB} $g_{V\pi\pi}$ (see \cref{eq:gVpipi}) 
and the decay constant $f_\pi$ (see \cref{eq:pion-decay-constant}). 
They are related to the original set by
\begin{gather}
    f_K = \frac{\sqrt{2} M_V}{\gt}  \,, \qquad
    f_1 = \frac{\sqrt{2}}{\gt} M_V \xi  \,,\qquad
    r = \sqrt{1 + \frac{f_\pi^2 g_{V\pi\pi} \gt}{2 M_V^2}} \,,\label{eq:r-parameter} \qquad \\
    f_0 = \sqrt{f_\pi^2 + r^2 f_1^2} = \sqrt{f_\pi^2 + \frac{2 M_V^2 \xi^2}{\gt^2} + \frac{\xi^2 f_\pi^2 g_{V\pi\pi}}{\gt}} \,.
\end{gather}
In addition we have the strong coupling $\gt$ of
the new sector as a free parameter.
In the following we will fix the pion decay constant $f_\pi$ to $1\,\mathrm{TeV}$.
Varying $f_\pi$ while keeping $\gV$ only mildly affects the decay channels of interest.
Both Lattice studies \cite{Ayyar:2018zuk,Bennett:2019jzz,Bennett:2019cxd,Bennett:2023qwx} 
and holographic models using
gauge/gravity duality \cite{Erdmenger:2020lvq,Erdmenger:2020flu,Elander:2020nyd,Elander:2021bmt,Erdmenger:2023hkl,Erdmenger:2024dxf} yield $\xi > 1$ and, thus, we set $\xi=1.4$ in
the following.
This also
implies that the cross sections for the 
axial vectors are smaller than the ones for
the vectors due to the kinematics.
Additionally, we use the \ac{SM} values of the electric charge $e$ and mass of the $Z$ boson $M_Z$ as input parameters to get an output expression for the coupling constants $\hat g$, $\hat g'$ derived from the conditions
\begin{align}
    \frac{1}{e^2} = \frac{1}{\hat g^2} + \frac{1}{\hat g'^2} + \frac{2}{\gt^2} \,, \qquad  \det\left(M_N^2 - M_Z^2 \mathds{1}_5\right) = 0 \,.
\end{align}

\section{Phenomenological aspects}
\label{sec:pheno}
\begin{figure}[t]
    \centering
    \includegraphics[width=0.5\linewidth]{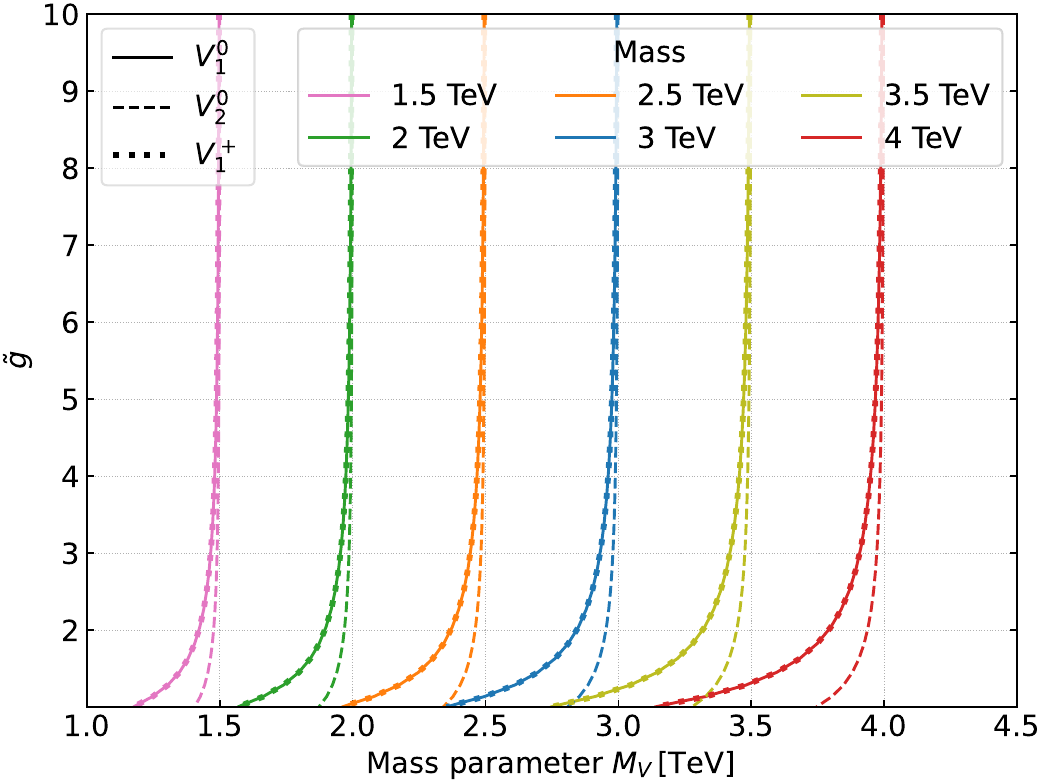}
    \caption{Contour lines for the masses of $V^+_{1\mu}$, $V^0_{1\mu}$ and $V^0_{2\mu}$ in the $M_V$-$\gt$ plane,
    where $M_V$ is given by \cref{eq:mass-parameter}. The results look nearly identical for each coset $\SU(4)/\Sp(4)$, $\SU(5)/\SO(5)$ and $\SU(4)\times \SU(4)/\SU(4)$.}
    \label{fig:mass-contour-plot}
\end{figure}

We focus here on those states which mix with
the SM electroweak bosons even in the limit
$\sin\theta \to 0$. These states can be singly produced at the LHC as we will see below. We denote
them as $V^+_1$, $V^0_1$ and $V^0_2$.
The first two states stem essentially from $(3,1)$ of $\SU(2)_L\times \SU(2)_R$
whereas $V^0_2$ is mainly the neutral state of
$(1,3)$ 
mixing primarily with the hypercharge boson.
This can also be seen from \cref{fig:mass-contour-plot} 
where we show corresponding contour lines for the 
masses of these states in the $M_V$-$\gt$ plane
setting $f_\pi=1$\,TeV.
Note that for $\gt \gsim 4$ all states are nearly mass degenerate.

In view of LHC phenomenology we group the various
decay channels as follows
\begin{alignat}{3}
    &\mathcal V^0 \to q\bar q, \, l^+ l^-, \,\nu\bar \nu, \qquad\qquad &&\mathcal V^0 \to t\bar t, \qquad\qquad &&\mathcal V^0 \to \pi\pi,\, HZ,\, W^+ W^-, \\
    &\mathcal V^+ \to q\bar q', \, l^+ \nu, \qquad &&\mathcal V^+ \to t\bar b, \qquad &&\mathcal V^+\to \pi\pi , \, W^+ Z,\, W^+ H.
\end{alignat}
The phenomenology of the spin-1 resonances obviously depends on various unknown parameters. We therefore
consider four different scenarios which are
combinations of couplings to pNGBs and the top quark.
For the latter we consider
\begin{alignat}{4}
    &\mathbf{SM}\, \boldsymbol t: \qquad &&g_{t,L/R} = g_{Ztt,L/R}^\mathrm{SM}, \qquad \qquad  \qquad &&g_{b,L} = g_{Zbb,L}^\mathrm{SM}, \qquad \qquad  &&g_{tb,L} = g_{Wtb}^\mathrm{SM}\,,\\
    &\textbf{PC}\, \boldsymbol t: &&g_{t,L}=\frac{1}{\sqrt{10}}\,,\, g_{t,R}=\frac{3}{\sqrt{10}}\,,&&g_{b,L}=\frac{1}{\sqrt{10}}\,,  && g_{tb,L}=\frac{1}{\sqrt{5}}\,,
\end{alignat}
which implies that the $t\bar t$ channel dominates over $b\bar b$ for the neutral states.
For the \ac{pNGB} couplings we consider
\begin{alignat}{2}
    &\mathbf {weak}\, \boldsymbol \pi: \qquad &&g_{V\pi\pi} = 0\,, \\
    &\mathbf {strong}\, \boldsymbol \pi: \qquad &&g_{V\pi\pi} = 4 \,.
\end{alignat}
We expect that a realistic scenario will be in between these extreme cases.

\begin{figure}[t]
    \centering
    \begin{subfigure}{.32\textwidth}
        \includegraphics[width=\linewidth]{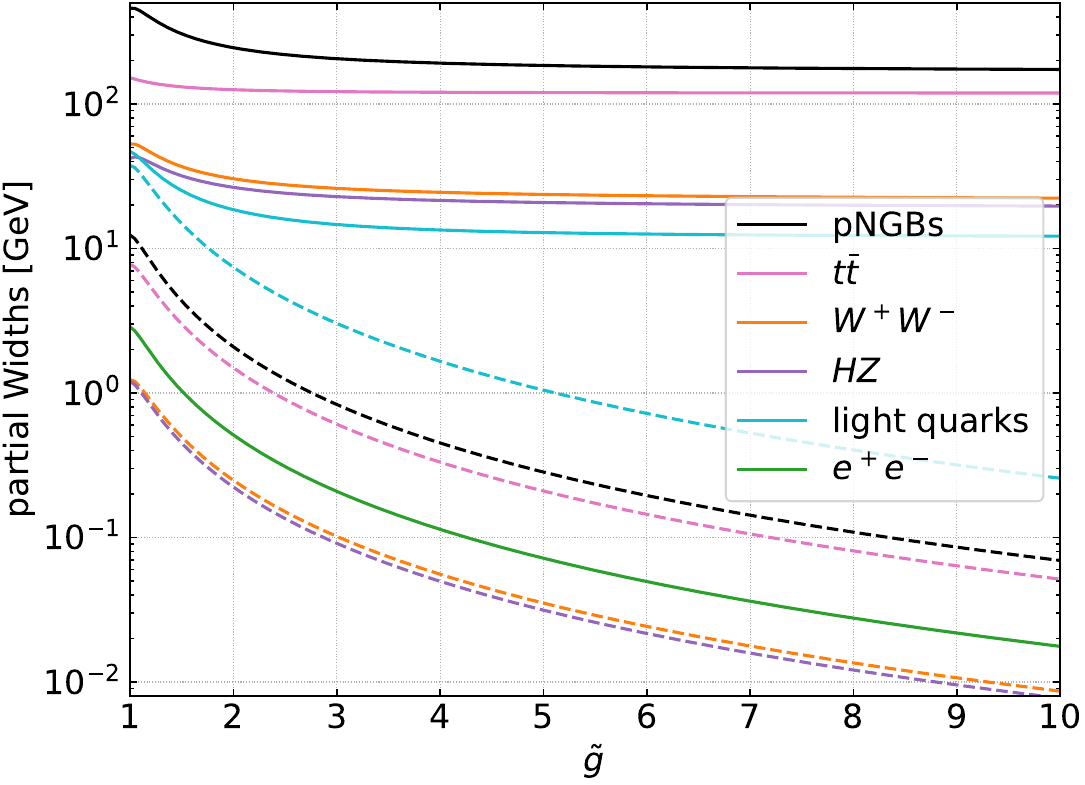}
        \caption{partial widths of $V_{1\mu}^0$}
    \end{subfigure}
    \begin{subfigure}{.32\textwidth}
        \includegraphics[width=\linewidth]{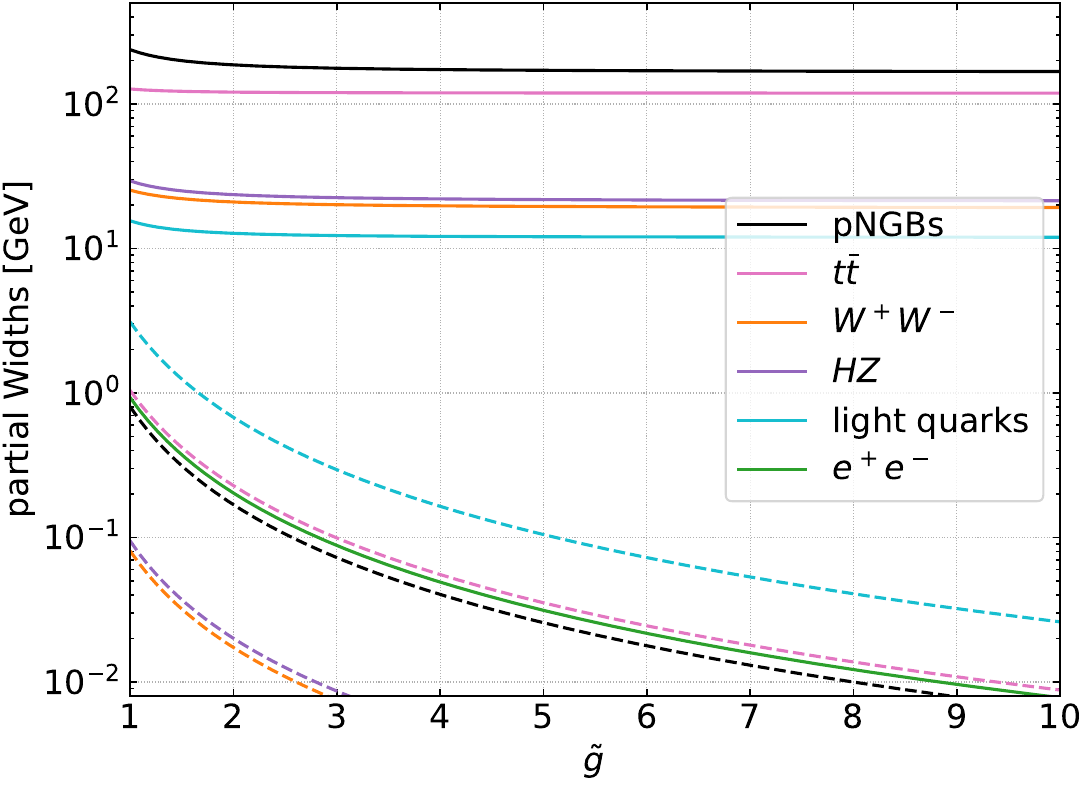}
        \caption{partial widths of $V_{2\mu}^0$}
    \end{subfigure}
    \begin{subfigure}{.32\textwidth}
        \includegraphics[width=\linewidth]{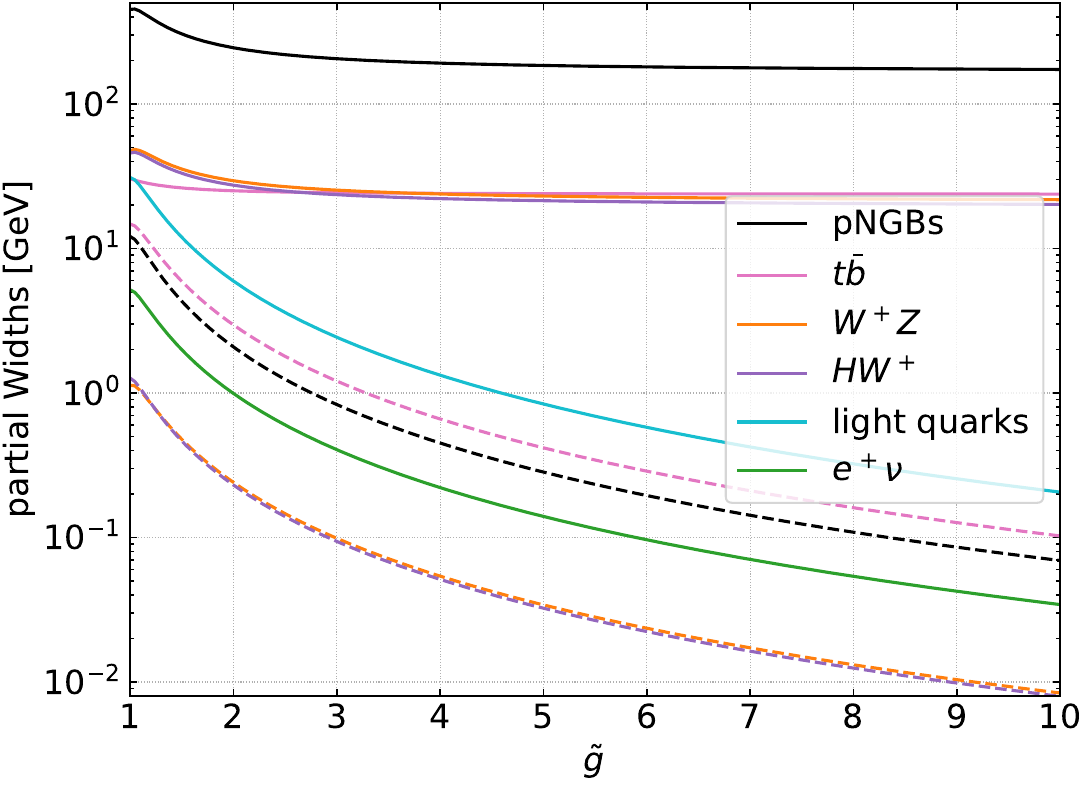}
        \caption{partial widths of $V_{1\mu}^+$}
    \end{subfigure}
    \caption{Partial decay widths of selected spin-1 resonances for the $\SU(5) / \SO(5)$ coset.  
  The solid lines of the pNGB, $W^+ W^-$, $HZ$, $W^+Z$ and $H W^+$  channels correspond to a scenario with $\gV=4$, while the corresponding dashed lines correspond to $\gV=0$. For the top quark channels, the solid lines correspond to $g_t = 1$ and the dashed lines to SM-like couplings. We have set   $M_V=3000$~GeV and $M_\pi=700$~GeV.
     These also represent the partial widths for the $\SU(4)/\Sp(4)$ coset for which the black lines (additional pNGB channels) are absent.
}
  \label{fig:partial-widths}
\end{figure}

\begin{figure}[t]
    \centering
    \begin{subfigure}{.32\textwidth}
        \includegraphics[width=\linewidth]{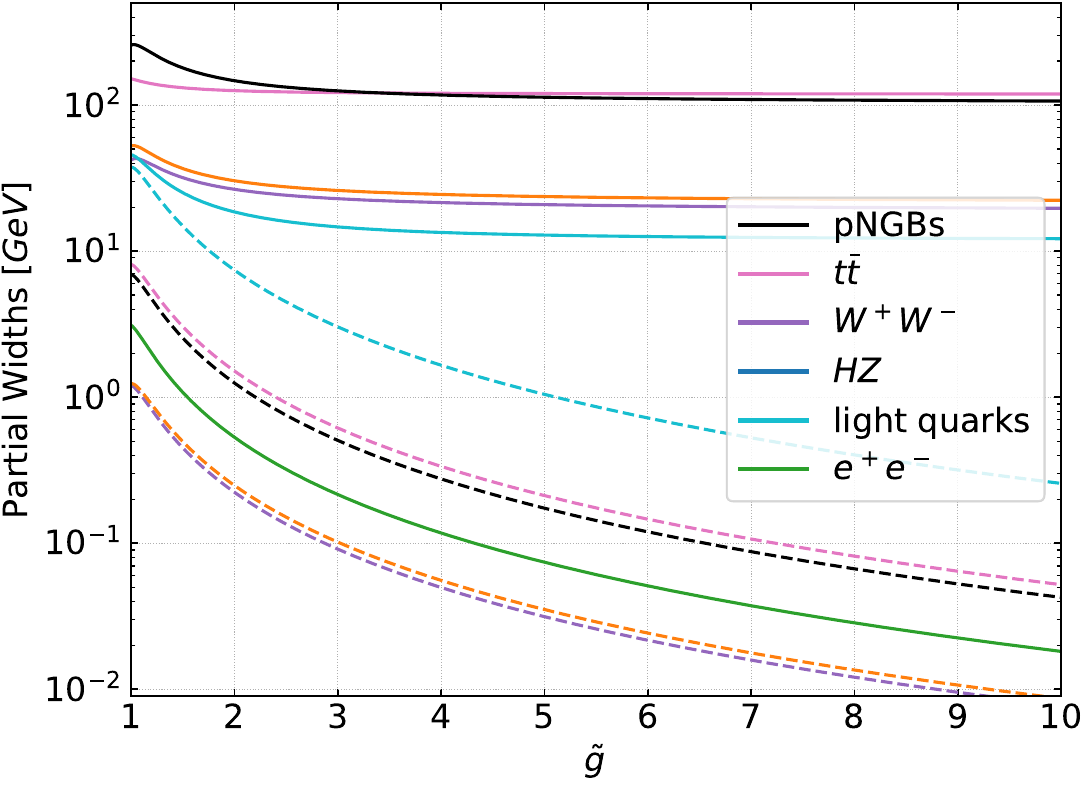}
        \caption{partial widths of $V_{1\mu}^0$}
    \end{subfigure}
    \begin{subfigure}{.32\textwidth}
        \includegraphics[width=\linewidth]{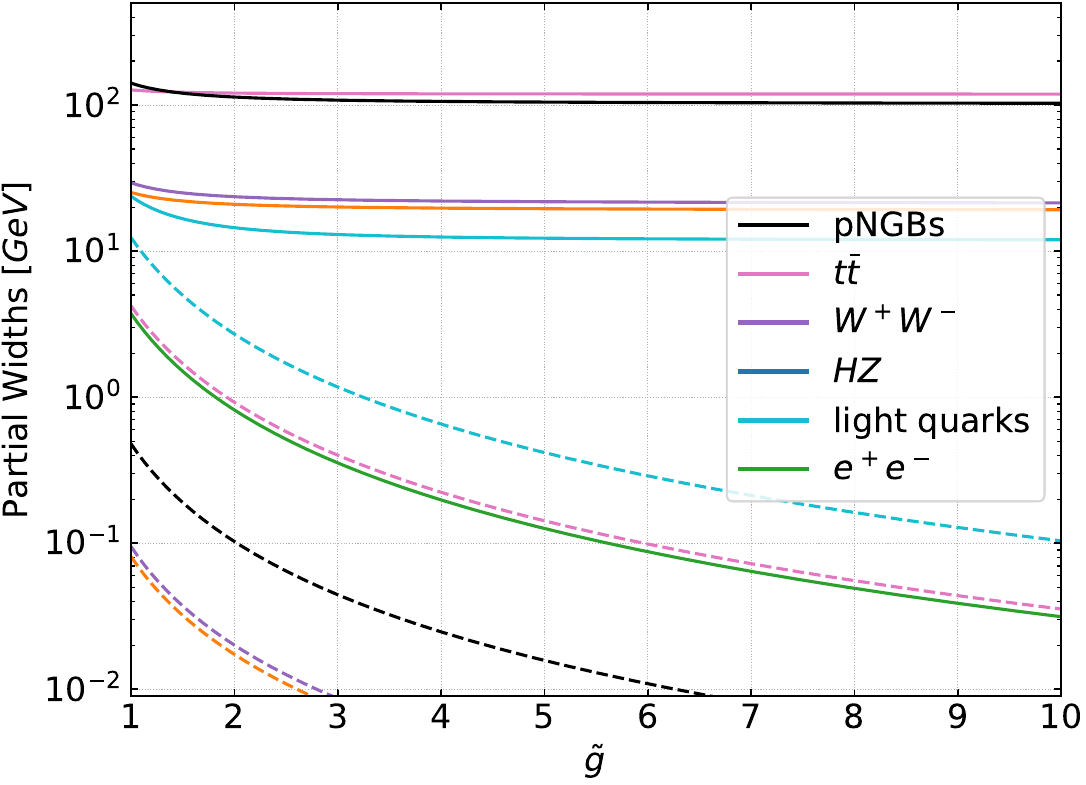}
        \caption{partial widths of $V_{2\mu}^0$}
    \end{subfigure}
    \begin{subfigure}{.32\textwidth}
        \includegraphics[width=\linewidth]{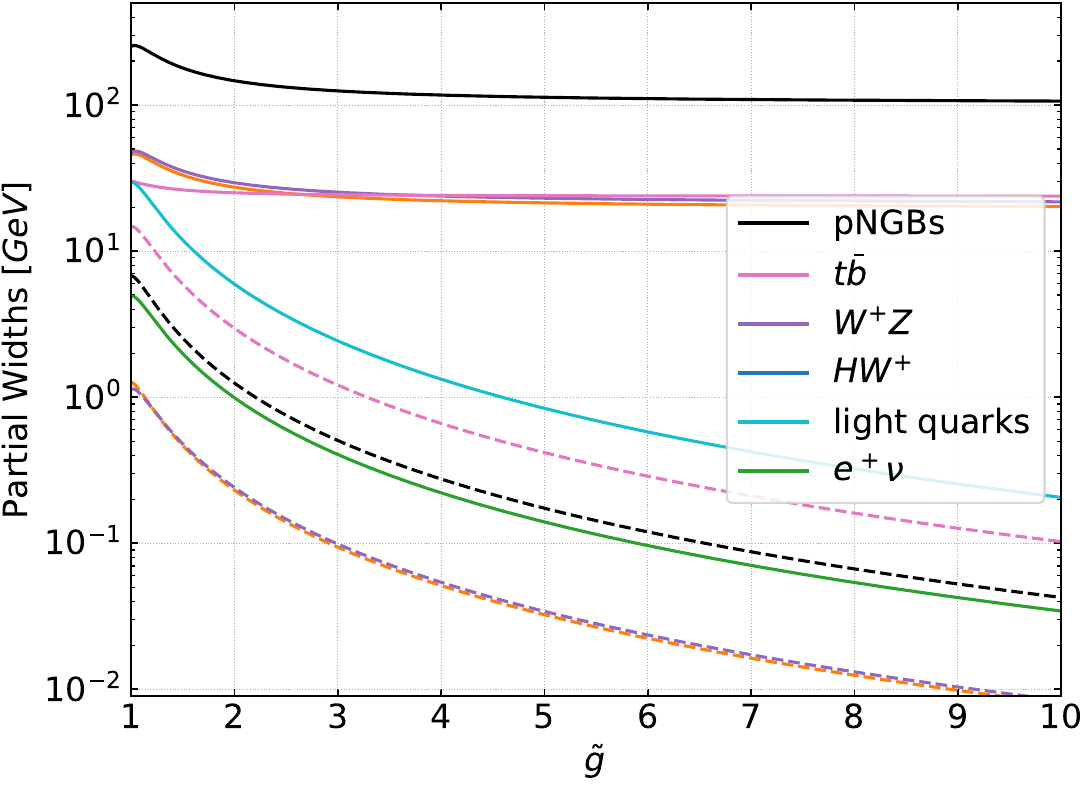}
        \caption{partial widths of $V_{1\mu}^+$}
    \end{subfigure}
    \caption{Partial decay widths of selected spin-1 resonances for the  $\SU(4) \times \SU(4)/\SU(4)$ coset.   
  The solid lines of the pNGB, $W^+ W^-$, $HZ$, $W^+Z$ and $H W^+$  channels correspond to a scenario with $\gV=4$, while the dashed lines correspond to $\gV=0$. For the top quark channels, the solid lines correspond to $g_t = 1$ and the dashed lines to SM-like couplings. We have set  $M_V=3000$~GeV and $M_\pi=450$~GeV.
   }\label{fig:partial-widths-su4su4}
\end{figure}

In \cref{fig:partial-widths} we show the partial decay widths for the different scenarios for the $\SU(5)/\SO(5)$ and the $\SU(4)/\Sp(4)$ cosets. For the latter, the black lines representing the decays into the additional \acp{pNGB} are absent as there is
no coupling of the gauge singlet $\eta$ to any combination of the electroweak vector bosons and any of the considered spin-1
resonances\footnote{However, there are couplings to combinations of the electroweak vector bosons and certain spin-1 resonances that
do not mix with the electroweak vector bosons \cite{BuarqueFranzosi:2016ooy}.}. We have fixed the \ac{pNGB} mass to 700~GeV such that
we are above existing LHC bounds \cite{Cacciapaglia:2022bax} and the vector mass parameter to $M_V = 3000$~GeV.
For the decays into two bosons we show the widths for the cases $g_{V\pi\pi} = 4$ and 0 as solid and dashed lines, respectively. 
Analogously, for the top quark channel we distinguish the PC $t$ and SM $t$ scenarios by solid and dashed lines.
The most important features  can be
summarized as follows:
\begin{itemize}
 \item $\textbf{PC}\, \boldsymbol t,\, \mathbf{strong}\, \boldsymbol \pi$: In scenarios where $g_{V\pi\pi}$ and the additional top couplings are large, the \mbox{spin-1} resonances will dominantly
    decay into the additional \acp{pNGB} and $t\bar{t}$ followed by $HV$ and $WV$ ($V=Z,W$)  in case of $\SU(5)/\SO(5)$. In case of
    of $\SU(4)/\Sp(4)$ the dominant channel will be $t\bar{t}$ followed by $HV$ and $WV$. Note that the enhancement of the
    $HV$ channel is caused by the longitudinal components of the vector bosons.   
 \item $\textbf{PC}\, \boldsymbol t,\, \mathbf{weak}\, \boldsymbol \pi$: In scenarios with $g_{V\pi\pi} \lsim \mathcal O(0.1)$ and large additional top Yukawa couplings, the $t\bar{t}$ channel will 
    dominate in case of both cosets.     
    \item $\textbf{SM}\, \boldsymbol t,\, \mathbf{strong}\, \boldsymbol \pi$: In case of large  $g_{V\pi\pi} $ and SM-like couplings to top quarks, the decays into the additional \acp{pNGB}
    dominate followed by the $HV$ and $WV$ channels in case of  $\SU(5)/\SO(5)$ whereas the latter channels 
    dominate in case of $\SU(4)/\Sp(4)$. 
    \item $\textbf{SM}\, \boldsymbol t,\, \mathbf{weak}\, \boldsymbol \pi$: In case that the additional couplings are small, the decay patterns are similar as for $W$ and $Z$ bosons but for
    the additional decays into top quarks. Moreover, the decays into the additional \acp{pNGB} are rather important in
    case of $\SU(5)/\SO(5)$.
\end{itemize}
We see in \cref{fig:partial-widths-su4su4} that for the $\SU(4) \times \SU(4)/\SU(4)$ coset the same generic features hold as
for $\SU(5)/\SO(5)$ coset
while details differ. We have checked using the recast tools described below that for this coset there are no  mass bounds on the additional \acp{pNGB} 
from existing LHC data. We have fixed the masses to 450~GeV to be above the $t\bar{t}$ threshold.
We find in particular that the dominance of the additional \ac{pNGB}, $HV$ and $WV$ ($V=Z,W$) channels is 
somewhat less pronounced. In case of the additional \acp{pNGB} this is due to the different multiplet structures, whereas 
in case of the other channels this is mainly due to the slightly different mixing patterns. 

The obvious importance of decays into
\acp{pNGB} implies that we have to
take cascade decays via intermediate
\acp{pNGB} into account. We summarize here the possible decay modes
and refer to the literature for further
details on the \ac{pNGB} decays \cite{Agugliaro:2018vsu,Cacciapaglia:2022bax,Ferretti:2016upr}.

As one limiting case -- dubbed the fermiophilic scenario -- we consider the case that the \acp{pNGB} dominantly decay into third generation quarks in all cosets.
These are mainly induced
from the mixing of the top-partners with the top and
bottom quarks. 
One finds for the
decays of a neutral state
\begin{align}
S^0 \to t \bar{t}\,,\quad b \bar{b}     
\end{align}
where the $b \bar{b}$ channel is suppressed by
the ratio $(m_b/m_t)^2$ and only becomes important
if $S^0 \to t \bar{t}$ is kinematically suppressed or even forbidden. 
$S^0$ denotes any of the neutral \acp{pNGB} in \cref{tab:cosets} except\footnote{Our assumptions about the vacuum imply that $H$ does not mix with any of the other neutral \acp{pNGB}.} the
$H$. Similarly, $S^+$ denotes any
of the singly charged states given in this table
which decays as
\begin{align}
    S^+ \to t \bar{b}.
\end{align}
The coset SU(5)/SO(5) features a doubly charged
scalar which decays as
\begin{align}
    \eta^{++}_5 \to W^+ t \bar{b}
\end{align}
via a $S^+$ \cite{Cacciapaglia:2022bax}.

In case that these \ac{pNGB} couplings to quarks are absent -- fermiophobic scenario --
then decays into two SM vector bosons induced by the anomalous WZW terms become relevant. If there are mass splittings between the \acp{pNGB} then cascade decays into a vector boson and another \ac{pNGB} are also important \cite{Cacciapaglia:2022bax}.
We take here the SU(5)/SO(5) coset as an example
where all \acp{pNGB} but $\eta_3^0$ have anomaly induced
couplings. 
As long as the triplet is the lightest state, which we will assume in the following, the CP-even $\eta_3^0$ can only undergo 3-body decays via an off-shell \ac{pNGB}:
\begin{align}
    \eta_3^0 &\to W^\mp {\eta^\pm_{3,5}}^* \to W^+ W^- \gamma,\,W^+ W^-Z \\
    \eta_3^0 &\to Z {\eta^0_{1,5}}^* \to Z Z Z ,\, Z Z \gamma ,\, Z \gamma \gamma
    \,.
\end{align}
Their analytic expressions of the corresponding partial widths are given in \cref{sec:eta30decay}.
We note for completeness that in case of the 
SU(4)/Sp(4) coset the $\eta$ also has
anomaly induced couplings but this particle is not
relevant for our investigations here. 
In the SU(4)$\times$SU(4)/SU(4) coset this state is also present with the same anomaly couplings, whereas the other \acp{pNGB} do not couple to the anomaly \cite{Banerjee:2022izw}.

\section{Constraints from LHC data}\label{sec:lhc}
\begin{figure}[t]
    \centering
    \begin{subfigure}{0.33\linewidth}
        \includegraphics[width=\linewidth]{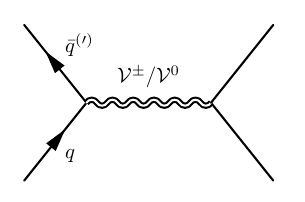}
        \vspace{5ex}
    \end{subfigure} \quad
    \begin{subfigure}{0.50\linewidth}
        \includegraphics[width=\linewidth]{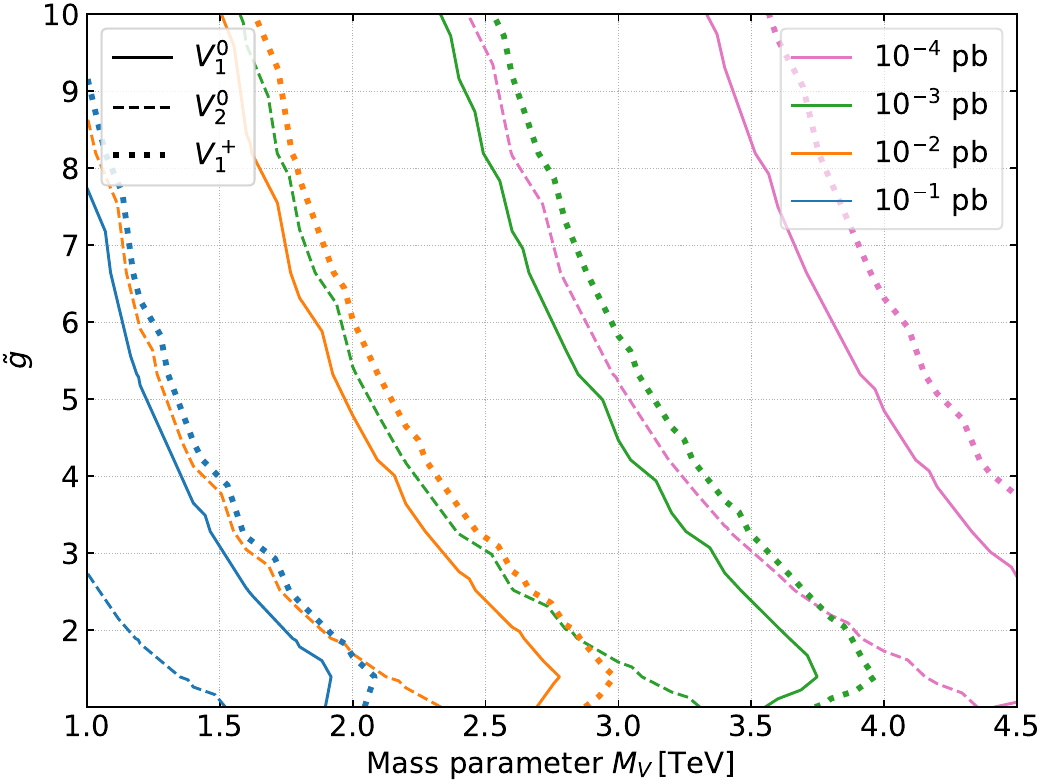}
    \end{subfigure}
    \caption{Drell-Yan production of heavy vectors. The left panel shows typical Feynman diagrams.
    The right panel shows the production cross sections at $\sqrt s = 13$~TeV of the heavy vector states in the $\SU(5)/\SO(5)$ coset in the $M_V$-$\tilde g$-plane assuming a small $g_{V\pi\pi}$ coupling and (nearly) SM-like couplings to the top-quarks.
   }
    \label{fig:feynman-dy}
\end{figure}

The states $V^0_{1,2}$ and $V^+_1$ have sizable 
couplings to quarks of the first two generations
as indicated in \cref{eq:LagCC,eq:LagNC}.
Thus, they can be singly produced at the LHC as shown in \cref{fig:feynman-dy} for $\SU(5)/\SO(5)$.
The cross section can reach $\mathcal O(0.1)$~pb for masses of about 1~TeV. 
Note that the production cross section of $V_2^0$ is about one order of magnitude smaller than that of $V_1$.
In case of $\SU(4)/\Sp(4)$ the results are the same, whereas they differ slightly for $\SU(4)\times \SU(4)/\SU(4)$ due to the somewhat different mixing patterns.

Combining the single production of the vector states with the decay channels outlined in the previous sections leads to multiple signatures that have been searched for at the LHC.
Specifically, searches for heavy gauge bosons are relevant for us, such as
\begin{itemize}
    \item an ATLAS search for $Z' \to \ell^+ \ell^-$ using 139~fb$^{-1}$ \cite{ATLAS:2019erb},
    \item an ATLAS search for $Z' \to t\bar t$ using 139~fb$^{-1}$ \cite{ATLAS:2020lks},
    \item an ATLAS search for $W' \to \ell^+ \nu$ using 139~fb$^{-1}$ \cite{ATLAS:2019lsy},
    \item an ATLAS search for $W' \to t\bar b$ using 139~fb$^{-1}$ \cite{ATLAS:2023ibb}.
\end{itemize}
In the following we use these searches to constrain the parameter space of our models.
To this end we implemented all relevant vertices in the \texttt{FeynRules} \cite{Alloul:2013bka} format to obtain a \ac{UFO} library \cite{Degrande:2011ua}.
We then load the \ac{UFO} into \texttt{MadGraph5\_aMC@NLO} \cite{Alwall:2014hca} v3.5.3 and generate events of the respective process at $\sqrt s = 13$~TeV.
We use dynamical renormalization and factorization scales and the \texttt{NNPDF~2.3} set of parton distribution functions \cite{Ball:2012cx} implemented in \texttt{LHAPDF} \cite{Buckley:2014ana}.
This way we calculate the cross sections of a given process for a grid of parameter points and compare them to the upper limits obtained from 
the above searches to derive exclusion limits in the $M_V$-$\tilde g$-plane. 

The decays of the spin-1 resonances into two bosons are not covered by any experimental search.
For these we instead derive bounds from recast searches.
First, we shower and hadronize the events with \texttt{Pythia8} \cite{Sjostrand:2014zea} to produce a \texttt{HepMC} file \cite{Dobbs:2001ck}.
We then pass the hadronized events to \texttt{MadAnalysis5} \cite{Conte:2012fm,Conte:2014zja,Dumont:2014tja,Conte:2018vmg} v1.10.9beta and \texttt{CheckMATE} \cite{Drees:2013wra,Dercks:2016npn} commit number \texttt{1cb3f7}.
Both tools cluster the jets with the anti-$k_T$ algorithm \cite{Cacciari:2008gp} implemented in the \texttt{FastJet} library \cite{Cacciari:2011ma} and simulate the detector response with \texttt{Delphes 3} \cite{deFavereau:2013fsa}.
The events are then run through the kinematic cuts of the recast searches, and from the number of remaining events an exclusion value is calculated with the CL$_s$ method \cite{Read:2002hq} for each signal region.
For every search we collect the observed exclusion for the signal region that had the strongest expected bound, as per the default prescription.
We further run the events against the \ac{SM} measurements implemented in \texttt{Rivet} \cite{Bierlich:2019rhm} v3.1.8 and evaluate the results with \texttt{Contur} \cite{Butterworth:2019wnt,Buckley:2021neu} v2.4.4, which also reports an exclusion value.
As the final result we report the strongest exclusion from any individual search.
In particular we do not perform any statistical combination beyond what is implemented in the tools.
We then draw the contour of the exclusion at 95\% CL as the bound in the $M_V$-$\tilde g$-plane.
Note that the regions with small $\tilde g \lesssim 2$ are not entirely reliable for the scenarios with strong \ac{pNGB} coupling because the width of the vector resonances $V_{1\mu}^{0,\pm}$ clearly exceeds $10\%$ of its mass going up to about $M_V/4$. 

\begin{figure}[p]
    \centering
    \begin{subfigure}{0.47\linewidth}
        \includegraphics[width=\linewidth]{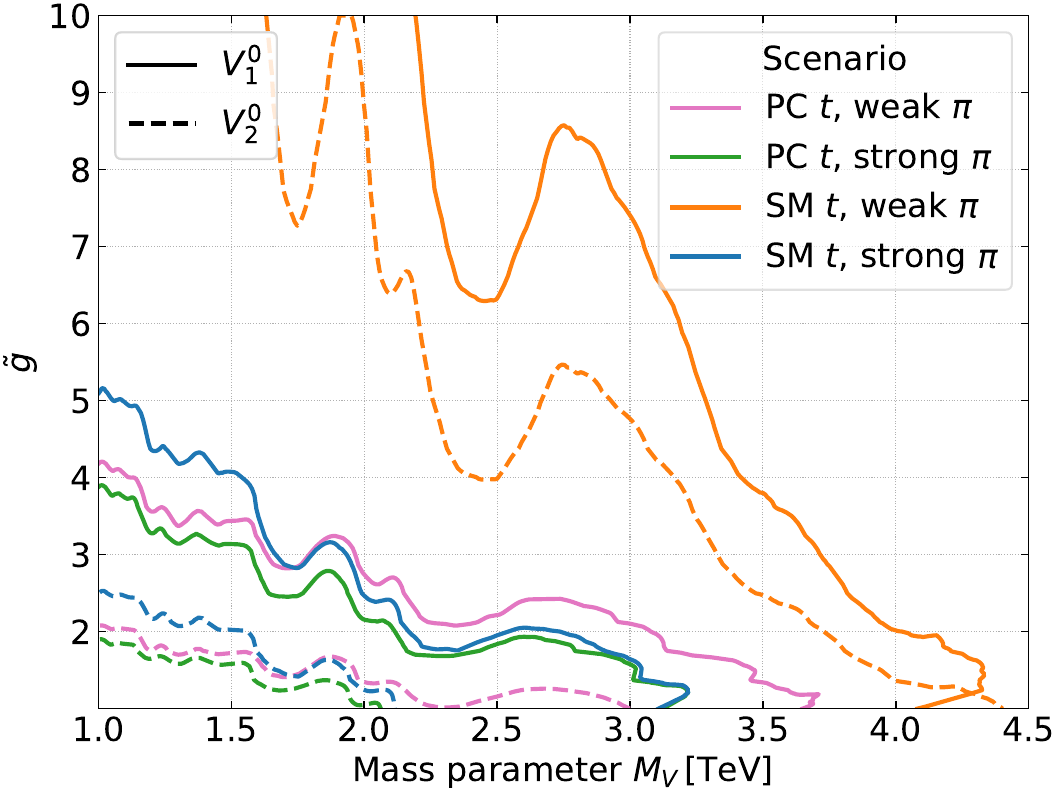}
        \caption{Bounds from $\mathcal V^0\to \ell^+ \ell^-$}
        \label{fig:bounds_su5_ll}
    \end{subfigure}\quad
    \begin{subfigure}{0.47\linewidth}
        \includegraphics[width=\linewidth]{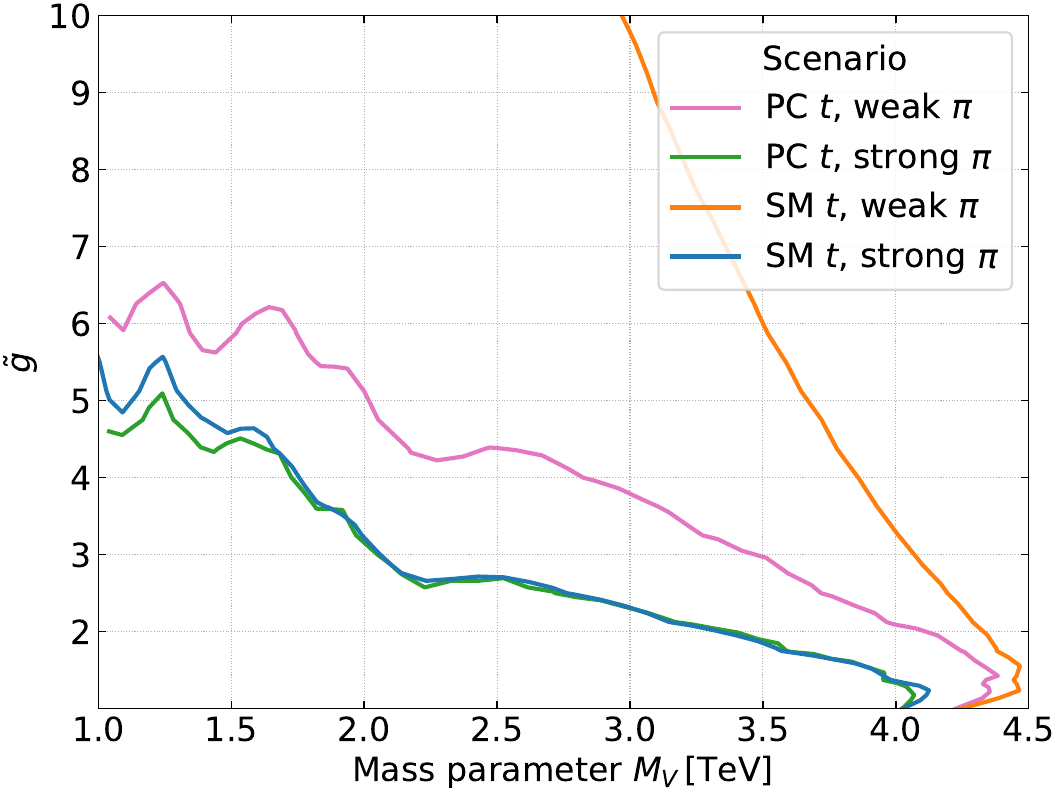}
        \caption{Bounds from $\mathcal V^\pm \to \ell^\pm \nu$}
        \label{fig:bounds_su5_lv}
    \end{subfigure}\vspace{1ex}

    \begin{subfigure}{0.47\linewidth}
        \includegraphics[width=\linewidth]{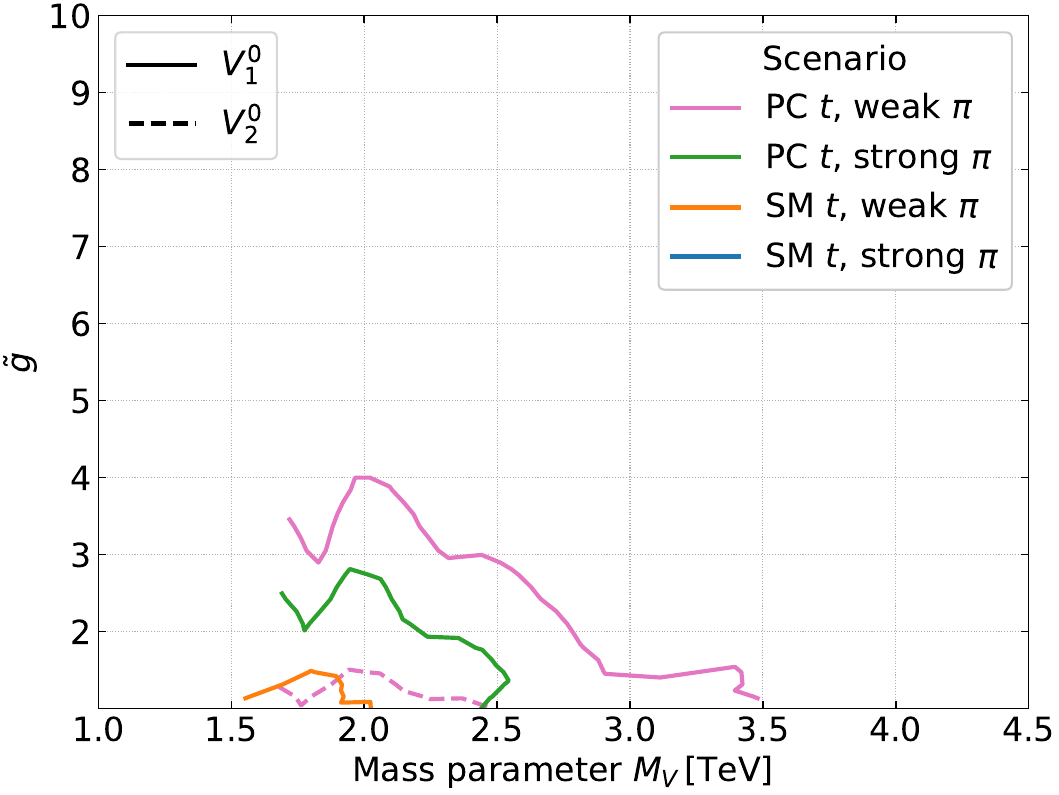}
        \caption{Bounds from $\mathcal V^0 \to t\bar t$}
        \label{fig:bounds_su5_tt}
    \end{subfigure}\quad
    \begin{subfigure}{0.47\linewidth}
        \includegraphics[width=\linewidth]{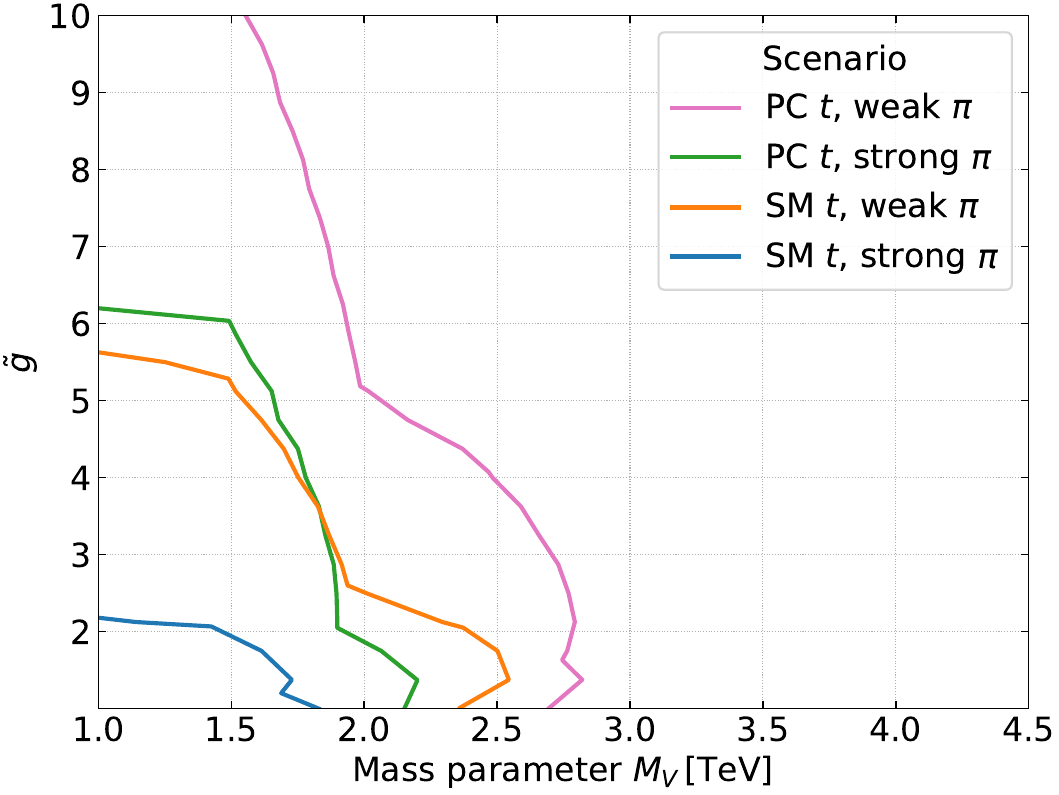}
        \caption{Bounds from $\mathcal V^\pm \to tb$}
        \label{fig:bounds_su5_tb}
    \end{subfigure}

    \begin{subfigure}{0.47\linewidth}
        \includegraphics[width=\linewidth]{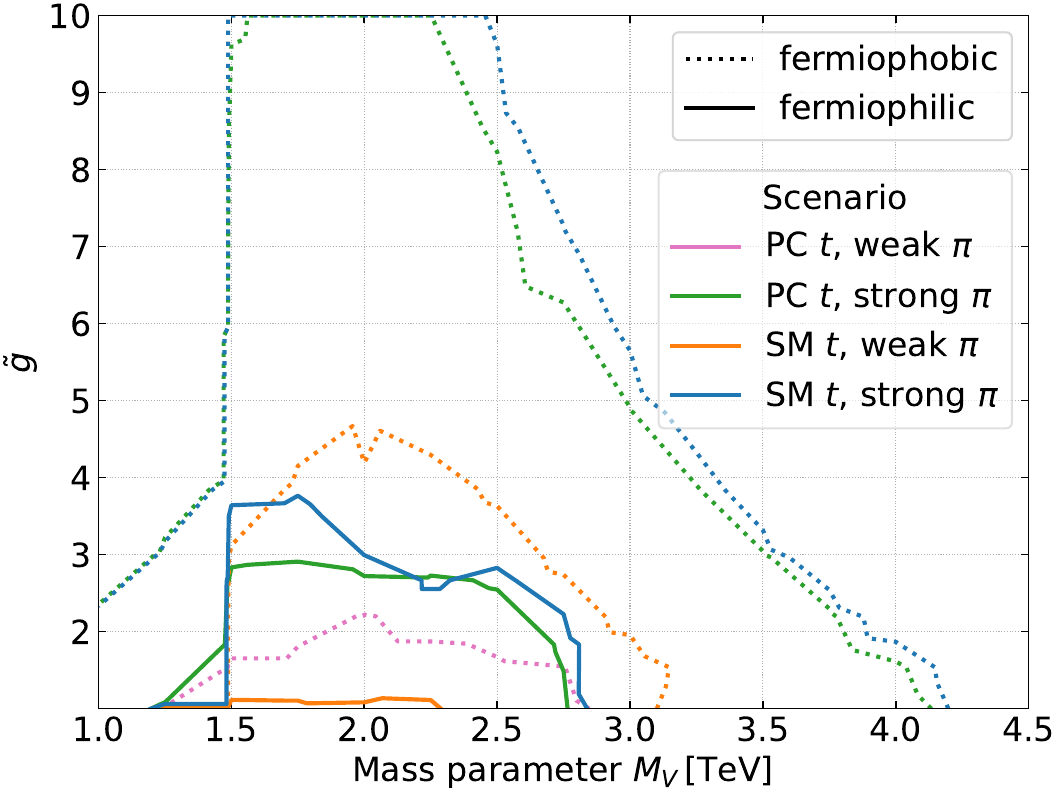}
        \caption{Bounds from $\mathcal V\to \pi\pi$}
        \label{fig:bounds_su5_pipi}
    \end{subfigure}\quad
    \begin{subfigure}{0.47\linewidth}
        \includegraphics[width=\linewidth]{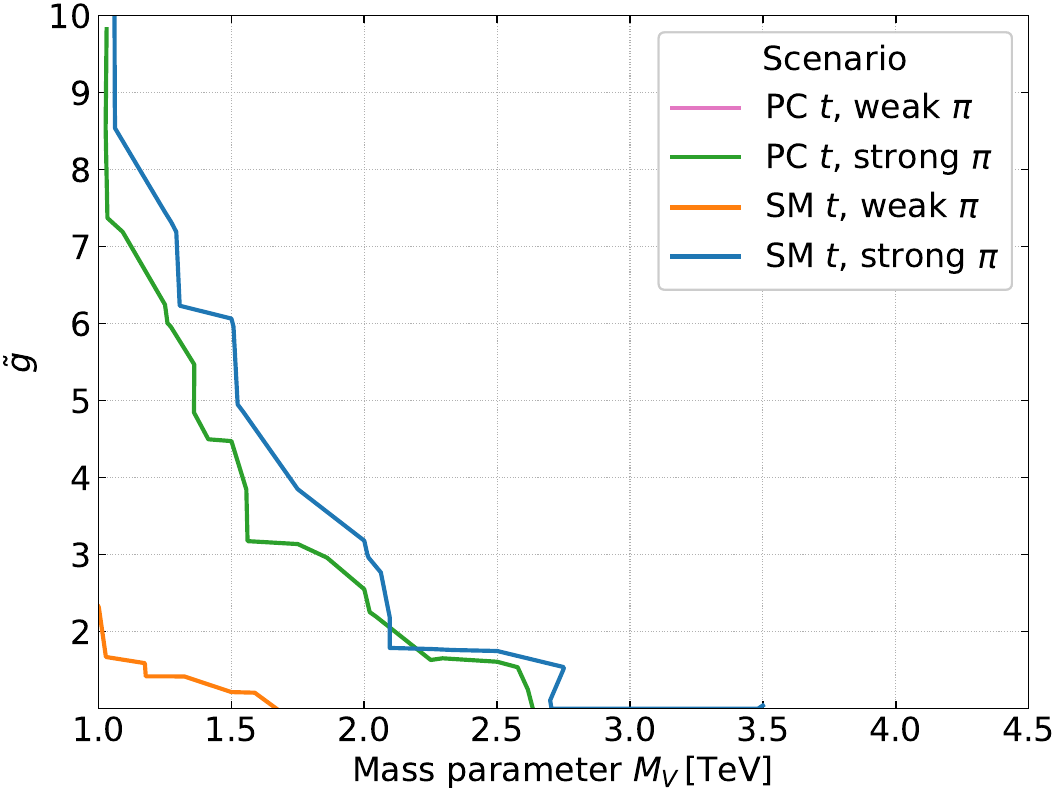}
        \caption{Bounds from $\mathcal V \to H Z, H W^\pm, W^+ W^-, W^\pm Z$}
        \label{fig:bounds_su5_higgs}
    \end{subfigure}
    \caption{Bounds on the single production of heavy vectors in the $\SU(5)/\SO(5)$ coset for a \ac{pNGB} mass of 700~GeV. 
    In the scenarios, ``SM $t$'' means the couplings of the $\mathcal V^0/ \mathcal V^\pm$ to $tt/tb$ are equal to the quark couplings to $Z/W^\pm$, whereas for ``PC $t$'' these couplings are set to 1. 
    For the \acp{pNGB}, ``weak'' and ``strong $\pi$'' refers to couplings $g_{V\pi\pi}=0$ and $g_{V\pi\pi}=4$, respectively.
    In (a)-(d) the upper limits on the cross sections are taken from direct searches \cite{ATLAS:2019erb,ATLAS:2020lks,ATLAS:2019lsy,ATLAS:2023ibb}.
    In (e) we distinguish further between fermiophobic and fermiophilic decay of the pNGBs. The bounds are derived from recasts of \cite{ATLAS:2018nud} and \cite{ATLAS:2021twp,ATLAS:2021fbt,CMS:2019zmd,CMS:2017abv}, respectively.
    The bounds in (f) are derived from recasts of \cite{CMS:2017moi,CMS:2019xjf,Mrowietz:2020ztq,CMS:2019xud,Conte:2021xtt,CMS:2019ius,ATLAS:2020wzf,ATLAS:2022nrp,ATLAS:2021jgw,CMS:2022ubq,ATLAS:2019zci,ATLAS:2019rob}.
    The regions with small $\tilde g \lesssim 2$ are not entirely reliable for scenarios with strong $\pi$ since the resonances are no longer narrow.}
    \label{fig:allbounds_su5so5}
\end{figure}
The resulting bounds are shown in \cref{fig:allbounds_su5so5} for the coset
$\SU(5)/\SO(5)$ which has the richest 
\ac{pNGB} sector, and in appendix \ref{app:pheno-bounds} for the other two cosets.
We stress that the x-axis is not the physical mass of the vectors but the mass parameter $M_V$ defined in \cref{eq:mass-parameter} and refer to \cref{fig:mass-contour-plot} for the corresponding physical masses.
In \cref{fig:bounds_su5_ll,fig:bounds_su5_lv} we get strong bounds when both the couplings to top quarks and \acp{pNGB} are small, leaving a large branching ratio into leptons.
In the other three scenarios the bounds are similar to each other.
The bounds from
$\mathcal V^\pm \to t b$ are significantly stronger
than the bounds from the decays of the neutral resonances into $t\bar{t}$ as can be seen from
\cref{fig:bounds_su5_tt,fig:bounds_su5_tb}.
This can be partly attributed to the increased cross section of the charged channel. 
Note that the cross section limits of \cite{ATLAS:2020lks}, used for \cref{fig:bounds_su5_tt}, are only given for vector boson masses above $1.75\,$TeV.

In \cref{fig:bounds_su5_pipi} we show the decays into \acp{pNGB} in the fermiophilic (solid lines) and fermiophobic (dotted lines) scenarios.
The latter  are strongly constrained by the recast of ref.~\cite{ATLAS:2018nud}, a search for photonic signatures of supersymmetry, which is implemented in \texttt{CheckMATE}. 
The bounds are derived for a common \ac{pNGB} mass $M_\pi = 700$~GeV to evade constraints from Drell-Yan production of \acp{pNGB}~\cite{Cacciapaglia:2022bax} where $\pi$ denotes all physical \acp{pNGB} but the Higgs boson.
Note that this also explains the sudden drop of the exclusion lines at $M_V \approx 2 M_\pi$ due to kinematic suppression of the pNGB channels.
 The bounds on \ac{pNGB} decays into quarks are considerably weaker.
The searches contributing to these bounds are refs.~\cite{ATLAS:2021twp,ATLAS:2021fbt} included in \texttt{CheckMATE} and refs.~\cite{CMS:2019zmd,CMS:2017abv} implemented in \texttt{MadAnalysis5} \cite{cmssus16033,Mrowietz:2020ztq}.
Figure~\ref{fig:bounds_su5_higgs} shows bounds derived from the decay into two gauge bosons or one gauge and one Higgs boson. These are calculated from the recast searches in \texttt{MadAnalysis5} \cite{CMS:2017moi,CMS:2019xjf,Mrowietz:2020ztq,CMS:2019xud,Conte:2021xtt,CMS:2019ius}, \texttt{CheckMATE} \cite{ATLAS:2020wzf} and \texttt{Rivet}/\texttt{Contur} \cite{ATLAS:2022nrp,ATLAS:2021jgw,CMS:2022ubq,ATLAS:2019zci,ATLAS:2019rob}.
For small masses, these channels are the dominant decays in the \textbf{strong}\,$\boldsymbol{\pi}$ scenarios, but get strongly suppressed above the threshold for pNGB pair production.

\begin{figure}[t]
    \centering
    \includegraphics[width=0.6\linewidth]{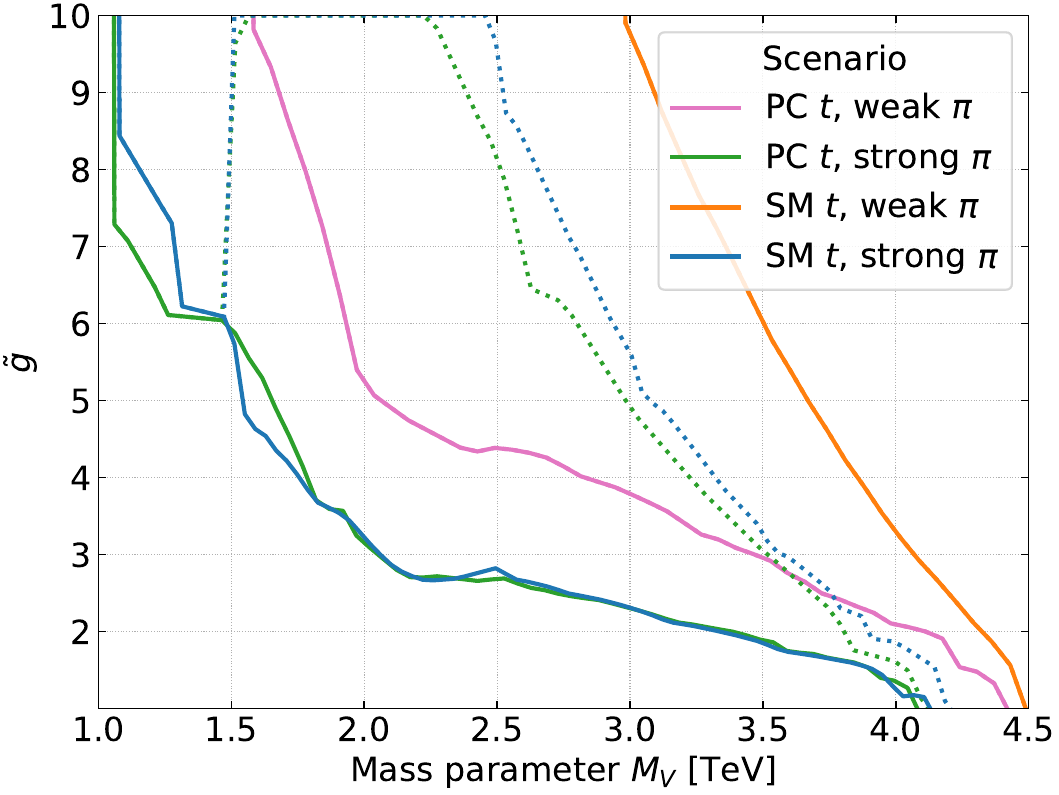}
    \caption{Bounds on the single production of heavy vectors in the $\SU(5)/\SO(5)$ coset. For each scenario we show the envelope of the bounds from the individual channels shown in \cref{fig:allbounds_su5so5}, i.e.\ the strongest bound at every point. The solid lines correspond to the fermiophilic, the dotted lines to the fermiophobic model, both with $M_\pi = 700\,\mathrm{GeV}$.}
    \label{fig:envelopes_su5so5}
\end{figure}

From \cref{fig:allbounds_su5so5} we can conveniently read off which processes yield which constraints.
In the end, however, we are interested in what regions of the parameter space are still viable.
To illustrate this, we show the combined bounds for all four coupling scenarios in \cref{fig:envelopes_su5so5}.
Each line represents the envelope of all channels, thus showing the strongest bound at each parameter point.
In the two scenarios with strong coupling to the \acp{pNGB} we show the fermiophilic scenario as a solid line and indicate where the fermiophobic case differs with a dotted line.
The scenario with weak couplings to top and \acp{pNGB} (orange) is strongly constrained yielding  $M_V > 3$~TeV -- 4.5~TeV depending on $\gt$.
If only the \ac{PC} couplings are turned on (pink), the bounds are considerably weaker, with $M_V$ down to 2~TeV remaining viable with only moderate $\tilde g$.
The shape of the bounds is further changed if we also include a large $g_{V\pi\pi}$ (green). 
The fermiophobic scenario is more strongly constrained than the fermiophilic one, with the latter leaving $\tilde g>4$ allowed for $M_V > 2$~TeV.
Finally we have the case with SM couplings to the top and a strong coupling to the \acp{pNGB} (blue), which has a similar shape to the previous case.
All in all, the scenario with large  $\gV$ and \ac{SM} like couplings to the top-quarks 
leaves the largest portion of parameter space open, especially in the fermiophilic case.

\begin{figure}[t]
    \centering
    \begin{subfigure}{0.47\linewidth}
        \includegraphics[width=\linewidth]{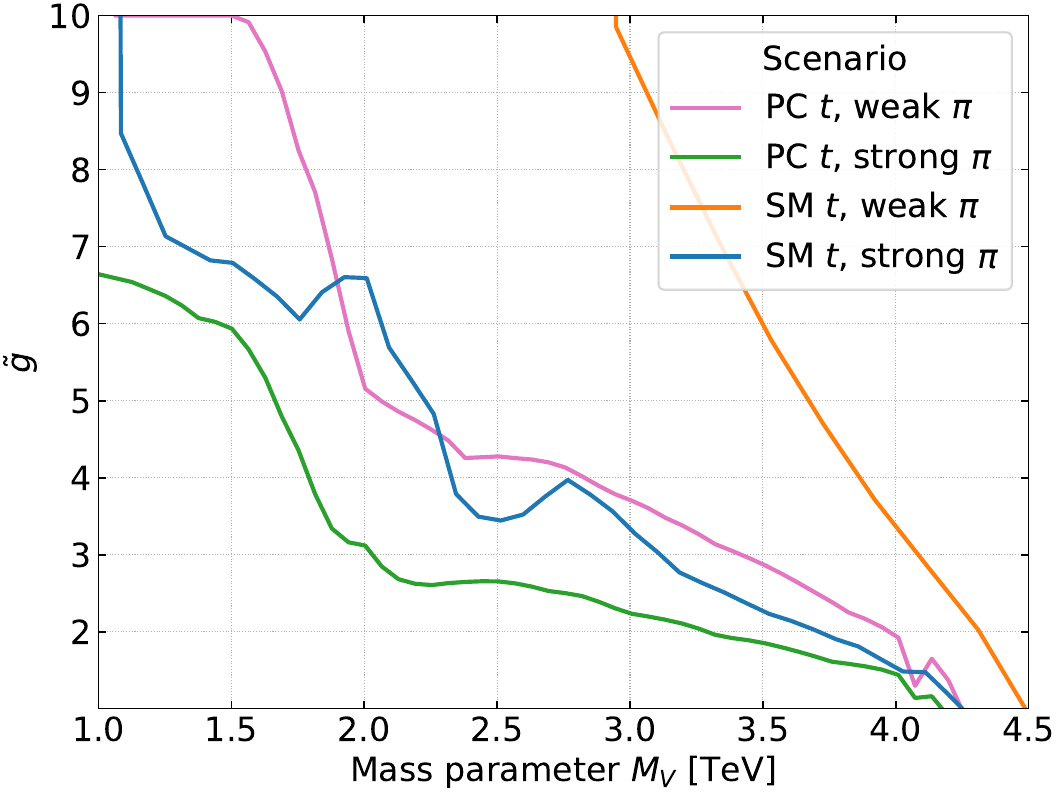}
        \caption{$\SU(4)/\Sp(4)$}\label{fig:envelopes_su4sp4}
    \end{subfigure}
    \begin{subfigure}{0.47\linewidth}
        \includegraphics[width=\linewidth]{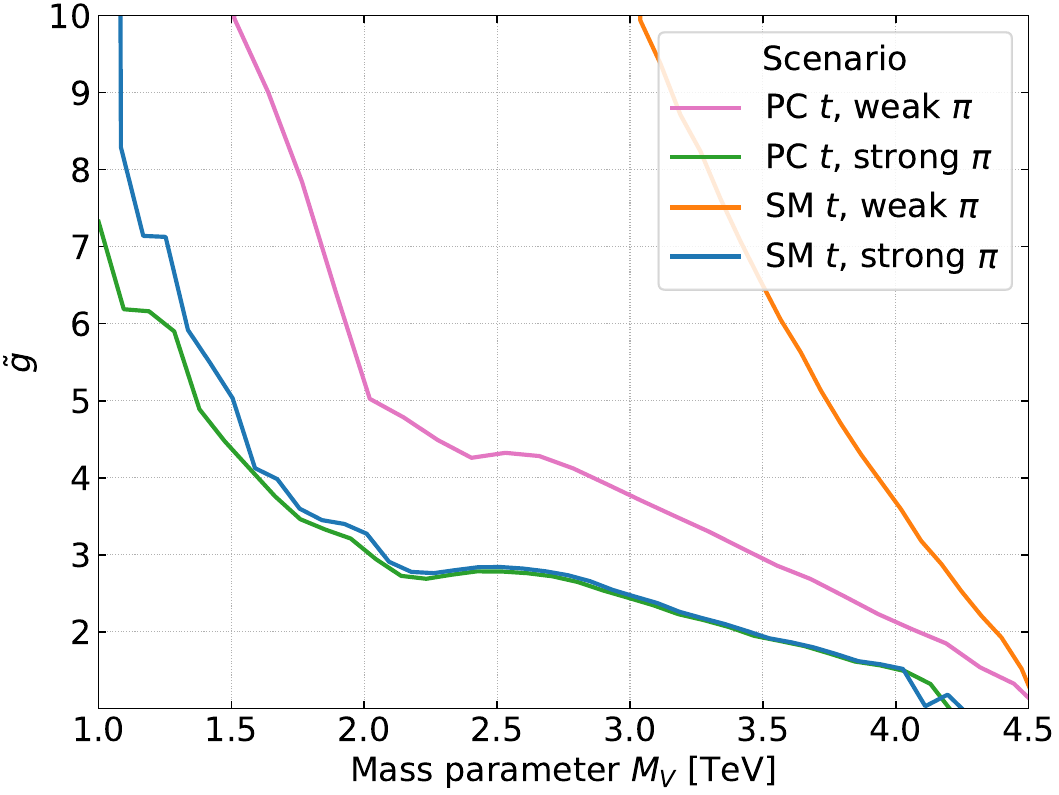}
        \caption{$\SU(4)^2/\SU(4)$}\label{fig:envelopes_su4xsu4}
    \end{subfigure}
    
    \caption{Bounds on the single production of heavy vectors in the $\SU(4)/\Sp(4)$ (left) and $\SU(4)^2/\SU(4)$ coset (right). In the latter coset we fixed the \ac{pNGB} masses to $M_\pi = 450$~GeV. For each scenario we show the envelope of the bounds from the individual channels analogously to \cref{fig:envelopes_su5so5}.}
    \label{fig:envelopes_su4}
\end{figure}

The results so far have been for the case of the
$\SU(5)/\SO(5)$ coset. The other two
cosets differ mainly in the \ac{pNGB} sector.
The overall results are nevertheless very similar
to the previous coset as can be seen in \cref{fig:envelopes_su4}. 
The bounds on the individual channels are given in \cref{app:pheno-bounds}.
The comparison of all three
cosets demonstrates that in particular 
the scenario with no enhancement for the couplings to top quarks and \acp{pNGB} are strongly constrained. The reason is that in this case direct decays of the spin-1 resonances to leptons are large enough to give strong constraints. 
 In practice however we always expect  an enhancement of the top and \ac{pNGB} couplings.
In scenarios where decays into \acp{pNGB} dominate, masses as low as about 1.5 TeV are still viable if $\gt \gsim 4$.

\section{Conclusions and outlook}
\label{sec:outlook}

We have investigated the phenomenology of spin-1 resonances in
Composite Higgs Models related to the electroweak sector, with particular attention
to bounds from existing LHC data.  
Here, we have focused on models which allow for
fermionic UV completions \cite{Ferretti:2016upr,Belyaev:2016ftv} as
they provide detailed information on the quantum numbers and
properties of the bound states. The corresponding cosets are
 $\SU(4)/\Sp(4)$, $\SU(5)/\SO(5)$ and $\SU(4) \times \SU(4)/\SU(4)$.
The considered cosets are symmetric and therefore contain two sets
of spin-1 resonances: vector states that couple to two \acp{pNGB} and
axial-vector states that couple to three 
\acp{pNGB}.

We have paid particular attention to those states which can mix
with the electroweak vector bosons of the SM. This mixing implies that
these states can be singly produced at the LHC. We have found
that independent of the coset there is always one charged spin-1 resonance mixing
sizably with
the W-boson and two neutral spin-1 resonances mixing  sizably with the Z-boson. This
is  a consequence of the fact that in all cases by construction the unbroken subgroup
contains the custodial
group $\SU(2)_L \times \SU(2)_R$. 

We have derived bounds 
in the mass-coupling plane for all cosets.
In case of decays into
two SM fermions we use
direct searches for heavy resonances in the s-channel at the 
LHC. In case of decays
into two bosons, either 
\acp{pNGB} and/or SM vector bosons, we have used recast searches.
 We have considered four different
scenarios to study the effect of unknown model dependent couplings. 
In scenarios with sizable couplings of the spin-1 resonances to \acp{pNGB}, masses as low as 1.5~TeV are still allowed by current LHC data.
In such scenarios, also the states which
only mix weakly or not at all will have masses
of about 1.5 TeV. Potentially one can further
obtain
bounds on all states from processes like
\begin{align}
  gg &\to b \bar{b} V^0, \,    t \bar{t} V^0 \\
    gg &\to b \bar{t} V^+,\,  t \bar{b} V^-
\end{align}
which we will investigate in a follow-up work.

Last but not least we point out that these
models contain an additional spin-1 resonance
$\tilde V$, not considered here, stemming from the inclusion of the QCD
sector and which mixes with the $\U(1)_Y$ boson.
We expect that the impact of this state is weak if it is heavier than the other spin-1 resonances.
 However, there are scenarios 
in which this state could be lighter which deserve further investigations.

\section*{Acknowledgements}
Ch.V. thanks the Institute of Theoretical Physics and Astrophysics of the University of W\"urzburg for its hospitality during an extended stay. We are grateful to the Mainz Institute for Theoretical Physics (MITP) of the DFG
Cluster of Excellence PRISMA+ (Project ID 39083149) for its hospitality and support.
This work has been supported by DFG, project 
no.~PO-1337/12-1.
M.K. is supported by the ``Studienstiftung des deutschen Volkes''. 
\clearpage
\appendix

\section{Model details}

\label{app:modeldetails}
In this appendix we collect the model details which have been omitted in the main text.
\subsection{Conventions}\label{app:models}

This section gives our notation for the models.
Before looking at the individual cosets, we introduce some general definitions.
Considering a generic coset $G/H$, we separate the generators $T^A$ of $G$ into unbroken ($T^a$) and broken ones ($X^I$), $T^A = \{ T^a, X^I \}$.
The $T^a$ and $X^I$ are determined from
\begin{align}\label{eq:broken_unbroken_generators}
    T^a \Sigma_0 + \Sigma_0 (T^a)^T = 0\,, \qquad X^I \Sigma_0 - \Sigma_0 (X^I)^T = 0\,,
\end{align}
where $\Sigma_0$ is the \ac{EW}-preserving vacuum.
We prefer to define our fields around the misaligned ``true'' vacuum $\tilde \Sigma_0$ of the theory, however, which is obtained by rotating with the misalignment matrix,
\begin{align}
    \tilde \Sigma_0 = \Omega(\theta) \, \Sigma_0 \, \Omega(\theta)^T\,, \qquad
    \Omega(\theta)  = \exp(\sqrt 2  i \theta X^h)\,,
\end{align}
where $X^h$ is the broken generator corresponding to the physical Higgs boson and $\theta$ is the vacuum misalignment angle.
We define misaligned generators by
\begin{align}
    \tilde T^a = \Omega(\theta) \, T^a \,\Omega(\theta)^\dagger\,, \qquad \tilde X^I = \Omega(\theta) \, X^I\, \Omega(\theta)^\dagger\,.
\end{align}
The Goldstone matrix employed in the CCWZ construction is given by
\begin{align}
    U = \exp( \frac{\sqrt 2 i}{f_\pi}\, \tilde \Pi )\,, \qquad \tilde \Pi = \Omega(\theta) \, \Pi^I X^I\, \Omega(\theta)^\dagger \,,
\end{align}
with the \ac{pNGB} decay constant $f_\pi = v/\sin\theta$.

\paragraph{Real case: $\SU(5)/\SO(5)$.}
We begin with the case of 5 \ac{EW} hyperfermions in a real \ac{irrep} of $G_\mathrm{HC}$, as is the case for models M1-M7 in ref.~\cite{Belyaev:2016ftv}, leading to $\SU(5)/\SO(5)$ breaking.
This coset has been explored in detail in ref.~\cite{Agugliaro:2018vsu}, and we follow the presentation therein.
We embed the $\SU(2)_L \times \SU(2)_R$ subgroup into $\SU(5)$ by\footnote{Note that for the $\SU(5)/\SO(5)$ and $\SU(4)\times \SU(4) // \SU(4)$ coset we work with generators normalised as $\Tr T^a T^b=\delta^{ab}$ and $\Tr X^I X^J = \delta^{IJ}$ while we normalize them to $\frac 12$ for $\SU(4)/\Sp(4)$}
\begin{align} \label{eq:SU2_gen_SU5_SO5}
    T^i_L = \frac 12 \mqty( \mathds{1} \otimes \sigma_i & 0\\0 & 0 ), \qquad T^i_R = \frac 12 \mqty( \sigma_i \otimes \mathds{1} & 0\\0 & 0 ).
\end{align}
The \ac{EW}-preserving vacuum reads
\begin{align}
    \Sigma_0 = \mqty(0 &i\sigma_2& 0\\ -i\sigma_2 &0& 0\\ 0&0&1),
\end{align}
and we rotate to the misaligned vacuum by means of
\begin{align}
    \Omega(\theta) =  \mqty( 1&0&0&0&0 \\ 0&c^2_{\theta/2} & s^2_{\theta/2} & 0 & is_\theta/\sqrt{2} \\ 0&s^2_{\theta/2} & c^2_{\theta/2} & 0 & -is_\theta/\sqrt{2} \\ 0&0&0&1&0 \\ 0&is_\theta/\sqrt{2} & -is_\theta/\sqrt{2} & 0 & c_\theta).
\end{align}
The \acp{pNGB} in the $\SU(5)/\SO(5)$ coset have been discussed in detail in the literature \cite{Agugliaro:2018vsu,Cacciapaglia:2022bax}, so we won't discuss them here. 
The spin-1 resonances that we defined in \cref{tab:cosets} are embedded in $\SU(5)$ by
\begin{align}
    \mathcal{V}_\mu &= \frac{1}{\sqrt{2}} \Omega(\theta) \cdot 
    \mqty(
        v_{1\mu}^0 & \frac{v_{1\mu}^+ + v_{2\mu}^+}{\sqrt{2}} & \frac{v_{1\mu}^+ - v_{2\mu}^+}{\sqrt{2}} & 0 & - \hat r_\mu^+ \\
        \frac{v_{1\mu}^- + v_{2\mu}^-}{\sqrt{2}} & - v_{2\mu}^0 & 0 & \frac{v_{1\mu}^+ - v_{2\mu}^+}{\sqrt{2}} & \frac{\hat r_\mu^0 - \ii \hat x_{1\mu}}{\sqrt{2}} \\
        \frac{v_{1\mu}^- - v_{2\mu}^-}{\sqrt{2}} & 0 & v_{2\mu}^0 & \frac{v_{1\mu}^+ + v_{2\mu}^+}{\sqrt{2}} & \frac{\hat r_\mu^0 + \ii \hat x_{1\mu}}{\sqrt{2}} \\
        0 & \frac{v_{1\mu}^- - v_{2\mu}^-}{\sqrt{2}} & \frac{v_{1\mu}^- + v_{2\mu}^-}{\sqrt{2}} & -v_{1\mu}^0 & \hat r_\mu^- \\
        - \hat r_\mu^- & \frac{\hat r_\mu^0 + \ii \hat x_{1\mu}}{\sqrt{2}} & \frac{\hat r_\mu^0 - \ii \hat x_{1\mu}}{\sqrt{2}} & \hat r_\mu^+ & 0
    ) \cdot \Omega(\theta)^\dag \,,\label{eq:vector-matrix}
\end{align}
and
\begin{align}
    \mathcal{A}_\mu &= \frac{1}{\sqrt{2}} \Omega(\theta) \cdot 
    \resizebox{0.75\textwidth}{!}{$\mqty(
        \frac{\sqrt{3} \hat y_{2\mu} + \sqrt{5} \hat{a}_{1\mu}^0 - \sqrt{10} \hat{a}_{5\mu}^0}{\sqrt{30}} & \frac{\ii \hat{a}_{3\mu}^+ + \hat{a}_{5\mu}^+}{\sqrt{2}} & \frac{-\ii \hat{a}_{3\mu}^+ + \hat{a}_{5\mu}^+}{\sqrt{2}} & \sqrt{2}\hat{a}_{5\mu}^{++} & \ii a_\mu^+ \\
        \frac{-\ii \hat{a}_{3\mu}^- + \hat{a}_{5\mu}^-}{\sqrt{2}} & \frac{\sqrt{3}y_{2\mu} - \sqrt{5}\hat{a}_{1\mu}^0 + \sqrt{10} \hat{a}_{5\mu}^0}{\sqrt{30}} & \frac{\ii\sqrt{3} \hat{a}_{3\mu}^0 + \hat{a}_{5\mu}^0 + \sqrt{2} \hat{a}_{1\mu}^0}{\sqrt{3}} & \frac{\ii \hat{a}_{3\mu}^+ - \hat{a}_{5\mu}^+}{\sqrt{2}} & \frac{-\ii a_\mu^0 + \hat y_{1\mu}}{\sqrt{2}} \\
        \frac{\ii \hat{a}_{3\mu}^- + \hat{a}_{5\mu}^-}{\sqrt{2}} & \frac{-\ii \sqrt{3}\hat{a}_{3\mu}^0 + \hat{a}_{5\mu}^0 + \sqrt{2} \hat{a}_{1\mu}^0}{\sqrt{3}} & \frac{\sqrt{3}\hat{y}_{2\mu} - \sqrt{5} \hat{a}_{1\mu}^0 + \sqrt{10} \hat{a}_{5\mu}^0}{\sqrt{30}} & \frac{-\ii \hat{a}_{3\mu}^+ - \hat{a}_{5\mu}^+}{\sqrt{2}} & \frac{-\ii a_\mu^0 - \hat y_{1\mu}}{\sqrt{2}} \\
        \sqrt{2} \hat{a}_{5\mu}^{--} & \frac{-\ii\hat{a}_{3\mu}^- - \hat{a}_{5\mu}^-}{\sqrt{2}} & \frac{\ii \hat{a}_{3\mu}^- - \hat{a}_{5\mu}^-}{\sqrt{2}} & \frac{\sqrt{3}y_{2\mu} + \sqrt{5} \hat{a}_{1\mu}^0 - \sqrt{10} \hat{a}_{5\mu}^0}{\sqrt{30}} & -\ii a_\mu^- \\
         -\ii a_\mu^- & \frac{\ii a_\mu^0 + \hat y_{1\mu}}{\sqrt{2}} & \frac{\ii a_\mu^0- \hat y_{1\mu}}{\sqrt{2}} & \ii a_\mu^+ & \frac{-4}{\sqrt{10}} \hat y_{2\mu} )$}
         \cdot \Omega(\theta)^\dag \,.\label{eq:axial-vector-matrix}
\end{align}
Note that for $\mathcal{A}_\mu$ we choose a slightly different parametrization of the bi-doublet compared to the pNGBs, such that we get real mass mixing matrices.

\paragraph{Pseudo-real case: $\SU(4)/\Sp(4)$.}
If the \ac{EW} hyperfermions live in a pseudo-real \ac{irrep} of $G_\mathrm{HC}$, the vacuum
\begin{align}
    \Sigma_0 = \mqty( \ii\sigma_2 & 0\\0 & -\ii\sigma_2 )\,,
\end{align}
spontaneously breaks $\SU(4) \to \Sp(4)$.
The corresponding embedding of the \ac{EW} generators in $\SU(4)$ is 
\begin{align} \label{eq:SU2_gen_SU4_Sp4}
    T^i_L = \frac 12 \mqty( \sigma_i & 0\\0 &0 ), \qquad T^i_R = \frac 12 \mqty( 0& 0\\0 &-\sigma_i^T ).
\end{align}
We rotate to the misaligned vacuum with 
\begin{align}
    \Omega(\theta) = \mqty( \cos \frac \theta 2 & 0&0 & \sin \frac \theta 2 \\0 & \cos \frac \theta 2 & -\sin \frac \theta 2 & 0 \\ 0& \sin \frac \theta 2 & \cos \frac \theta 2 & 0 \\ -\sin \frac\theta 2 & 0 & 0 & \cos \frac \theta 2).
\end{align}
The \acp{pNGB} have been described in \cite{Cacciapaglia:2014uja}.
For this model the spin-1 resonances have already been studied in \cite{BuarqueFranzosi:2016ooy} using a different set of parameters and a smaller LHC dataset. We expand on their results in this work.
They are embedded by
\begin{align}
    \mathcal{V}_\mu &= \frac{1}{2} \Omega(\theta) \cdot \mqty(
        \frac{1}{\sqrt{2}}(v_{1\mu}^0 - v_{2\mu}^0) & v_{1\mu}^+ + v_{2\mu}^+ & \hat r^+_\mu & \frac{1}{\sqrt{2}}(\hat x_{1\mu} + \ii \hat r^0_\mu) \\
        v_{1\mu}^- + v_{2\mu}^- & \frac{1}{\sqrt{2}}(- v_{1\mu}^0 + v_{2\mu}^0) & \frac{1}{\sqrt{2}}(-\hat x_{1\mu} + \ii \hat r^0_\mu) & \hat r^-_\mu \\
         \hat r^-_\mu & \frac{1}{\sqrt{2}}(-\hat x_{1\mu} - \ii \hat r^0_\mu) & \frac{-1}{\sqrt{2}}(v_{1\mu}^0 + v_{2\mu}^0) & - v_{1\mu}^- + v_{2\mu}^- \\
        \frac{1}{\sqrt{2}}(\hat x_{1\mu} - \ii \hat r^0_\mu) & \hat r_\mu^+ & - v_{1\mu}^+ + v_{2\mu}^+ & \frac{1}{\sqrt{2}}(v_{1\mu}^0 + v_{2\mu}^0)
    ) \cdot \Omega^\dag \,,
\end{align}
and
\begin{align}
    \mathcal{A}_\mu &= \frac{1}{2} \Omega(\theta) \cdot \mqty(
        \frac{1}{\sqrt{2}}\hat y_{2\mu} & 0 & a_\mu^+ & \frac{1}{\sqrt{2}} (a_\mu^0 - \ii \hat y_{1\mu}) \\
        0 & \frac{1}{\sqrt{2}}\hat y_{2\mu} & \frac{1}{\sqrt{2}} (a_\mu^0 + \ii \hat y_{1\mu}) & - a_\mu^- \\
        a_\mu^- & \frac{1}{\sqrt{2}} (a_\mu^0 - \ii \hat y_{1\mu}) & \frac{-1}{\sqrt{2}}\hat y_{2\mu} & 0\\
        \frac{1}{\sqrt{2}} (a_\mu^0 + \ii \hat y_{1\mu}) & - a_\mu^+ & 0 & \frac{-1}{\sqrt{2}}\hat y_{2\mu}
    ) \cdot \Omega^\dag \,.
\end{align}

\paragraph{Complex case: $\SU(4)^2/\SU(4)$.}
If the hyperfermions live in a complex \ac{irrep} of $G_\mathrm{HC}$, the global symmetry breaking is $\SU(4) \times \SU(4) \to \SU(4)$.
We begin with a simplified formalism in terms of $4\times4$ matrices by embedding the $\SU(2)_L \times \SU(2)_R$ in the unbroken $\SU(4)$ by
\begin{align}
    T_L^i = \frac 12 \mqty( \sigma_i & 0\\0 & 0 ), \qquad T_R^i = \frac 12 \mqty( 0 &0 \\0 &  \sigma_i  )\,,
\end{align}
which implies \cite{Ferretti:2016upr}
\begin{align}
    \Omega(\theta) = \mqty( \cos \frac \theta 2 & 0 & \sin \frac \theta 2 & 0 \\0  & \cos \frac \theta 2 & 0 & \sin \frac \theta 2 \\ -\sin \frac \theta 2 & 0 & \cos \frac \theta 2 & 0 \\  0& -\sin \frac \theta 2 & 0 & \cos \frac \theta 2)\,.
\end{align}
Since the vacuum is the identity, $\Sigma_0^{(4)} = \mathds 1_4$, the misaligned vacuum reads
\begin{align}
    \tilde \Sigma_0^{(4)} = \Omega(\theta)^2 = \mqty( \cos \theta \, \mathds 1_2  & \sin \theta \, \mathds 1_2 \\ - \sin \theta \, \mathds 1_2 & \cos \theta \, \mathds 1_2)\,.
\end{align}
However, for a full treatment of the spin-1 states we have to work with $8\times8$ matrices.
To this end we introduce the vacuum
\begin{align}
    \Sigma_{0}^{(8)} = \mqty( 0 & \mathds 1_4 \\ \mathds 1_4 & 0 )\,,
\end{align}
with the corresponding non-rotated generators given by
\begin{align}
    T^a =  \mqty( S^a & 0\\0 & -(S^a)^T ), \qquad X^I =  \mqty( S^I & 0\\0 & (S^I)^T )\,,
\end{align}
where $S^a$ are the $\SU(4)$ generators in the fundamental \ac{irrep} \cite{BuarqueFranzosi:2023xux}.
The misaligned vacuum is given by \cite{Preskill:1980mz}:
\begin{align}
    \tilde \Sigma_0^{(8)} = \mqty( 0& \tilde \Sigma_0^{(4)} \\ \tilde \Sigma_0^{(4),T} &0 )\ ,
\end{align}
and we determine the misaligned generators $\tilde T$ and $\tilde X$ by imposing \cref{eq:broken_unbroken_generators}. 
We parameterize the $\SU(4)$ generators as
 \begin{align}
   S^a = \mqty( A & B \\ C & D )\,,
\end{align}
and get the following $8 \times 8$ misaligned generators
 \begin{align}
    &\tilde{T}^a = \nonumber\\
    &\resizebox{\textwidth}{!}{ $\mqty(
 A & B & 0 &  0\\
 C & D & 0 & 0 \\
  0& 0 & \frac{1}{2} (-2 A^T \cos^2(\theta)+ (B^T+
   C^T) \sin (2 \theta)- 2 D^T \sin^2(\theta)) &
   \frac{1}{2} ((D^T-A^T) \sin
   (2 \theta)+2 B^T  \sin^2(\theta)-2 C^T \cos^2(\theta)) \\
 0& 0 & \frac{1}{2} ((D^T-A^T) \sin (2 \theta)-2 B^T  \cos^2(\theta)+2
   C^T\sin^2(\theta))
   &\frac{1}{2} ( -2 A^T \sin^2(\theta)-(B^T+
   C^T) \sin (2 \theta)-2D^T\cos^2(\theta)) \\
)$}\,,
\end{align}
and
 \begin{align}
 &\tilde{X}^I =  \nonumber\\
 &\resizebox{\textwidth}{!}{$ \mqty(
 A & B & 0 & 0\\
 C & D & 0 & 0 \\
  0& 0 & \frac{1}{2} (2 A^T \cos^2(\theta)- (B^T+
   C^T) \sin (2 \theta)+2 D^T \sin^2(\theta)) &
   \frac{1}{2} ((A^T-D^T) \sin
   (2 \theta)-2 B^T  \sin^2(\theta)+2 C^T \cos^2(\theta)) \\
 0& 0 & \frac{1}{2} ((A^T-D^T) \sin (2 \theta)+2 B^T  \cos^2(\theta)-2
   C^T\sin^2(\theta))
   &\frac{1}{2} ( +2 A^T \sin^2(\theta)+(B^T+
   C^T) \sin (2 \theta)+2D^T\cos^2(\theta)) \\
)$}\,.
\end{align}

\subsection{Mass matrices for $\SU(4)\times \SU(4)/\SU(4)$}\label{app:su4su4-mass}
The mass matrices in the $\SU(4)\times \SU(4)/\SU(4)$ coset differ from those of the other two cosets because more particles
mix initially with the SM bosons. In the charged
sector we find in the basis 
$(\tilde W^+_\mu,\, v_{1\mu}^+,\, v_{2\mu}^+, r^+_\mu,\, \,b^+_{1\mu},\, b^+_{2\mu},a^+_\mu)$ the mass matrix
\begin{align}
    \mathcal{M}_C^2 &=\mqty(
         \frac{ \hat g^2 M_{V}^2 (1 + \omega s^2_{\theta})}{ \tilde{g}^2} & -\frac{ \hat g M_{V}^2  (1 + c^2_{\theta})}{2 \tilde{g}} & -\frac{ \hat g M_{V}^2 s^2_{\theta}}{2 \tilde{g}} & \frac{ \hat g M_{V}^2 s_{\theta} c_{\theta}}{\sqrt{2} \tilde{g}} & \frac{ \hat g M_{A}^2 r s^2_{\theta}}{2 \tilde{g}} & -\frac{ \hat g M_{A}^2 r s^2_{\theta}}{2 \tilde{g}} & \frac{ \hat g M_{A}^2 r s_{\theta} c_{\theta}}{\sqrt{2} \tilde{g}} \\
         -\frac{ \hat g M_{V}^2  (1 + c^2_{\theta})}{2 \tilde{g}} & M_{V}^2 & 0 &0  &0  &0 &0\\
         -\frac{ \hat g M_{V}^2 s^2_{\theta}}{2 \tilde{g}} & 0& M_{V}^2 & 0 & 0&  0& 0 \\
         \frac{ \hat g M_{V}^2 s_{\theta} c_{\theta}}{\sqrt{2} \tilde{g}} & 0 & 0 & M_{V}^2 &  0&0  & 0 \\
         \frac{ \hat g M_{A}^2 r s^2_{\theta}}{2 \tilde{g}} &  0& 0 & 0 & M_{A}^2 & 0 & 0 \\
         -\frac{ \hat g M_{A}^2 r s^2_{\theta}}{2 \tilde{g}} & 0 & 0 & 0 & 0 & M_{A}^2 & 0 \\
         \frac{ \hat g M_{A}^2 r s_{\theta} c_{\theta}}{\sqrt{2} \tilde{g}} & 0 &  0& 0 &0 & 0 & M_{A}^2  )\,.
\end{align}
One can easily see that only a linear combination
of  $v_{1\mu}^+$, $v_{2\mu}^+$, $r^+_\mu$ and a linear combination of $a^+_\mu$, $b^+_{1\mu}$, $b^+_{2\mu}$
mix with $W^+$. One obtains the same situation as
for the other two cosets and also the mixing 
with axial-vector states vanishes again in the
limit $\sin\theta\to0$.
Similarly, we can obtain analogous rotations in
the neutral sector to obtain the previous case
as can be seen from the corresponding mass matrix:
\begin{align}
   & \mathcal{M}_N^2 = \nonumber \\
    &\mqty(
 \frac{ \hat g'^2 M_{V}^2 (1 + \omega s^2_{\theta})}{\tilde{g}^2} &
   -\frac{ \hat g'  \hat g M_{V}^2 \omega  s^2_{\theta}}{\tilde{g}^2} &
   -\frac{ \hat g' M_{V}^2 s^2_{\theta}}{2 \tilde{g}} & -\frac{ \hat g'
   M_{V}^2 (1 + c^2_{\theta})}{2 \tilde{g}} & \frac{ \hat g' M_{V}^2 s_{\theta}  c_{\theta}}{\sqrt{2} \tilde{g}} & -\frac{ \hat g' M_{A}^2 r s^2_{\theta}}{2 \tilde{g}} & \frac{ \hat g' M_{A}^2 r s^2_{\theta}}{2
   \tilde{g}} & \frac{ \hat g' M_{A}^2 r s_{\theta} c_{\theta}}{\sqrt{2}
   \tilde{g}} \\
 -\frac{ \hat g'  \hat g M_{V}^2 \omega  s^2_{\theta}}{\tilde{g}^2} &
   \frac{ \hat g^2 M_{V}^2 (1+\omega  s^2_{\theta})}{\tilde{g}^2} &
   -\frac{ \hat g M_{V}^2 (1+c^2_{\theta})}{2 \tilde{g}} & -\frac{ \hat g
   M_{V}^2 s^2_{\theta}}{2 \tilde{g}} & -\frac{ \hat g M_{V}^2 s_{\theta}  c_{\theta}}{\sqrt{2} \tilde{g}} & \frac{ \hat g M_{A}^2 r s^2_{\theta}}{2 \tilde{g}} & -\frac{ \hat g M_{A}^2 r s^2_{\theta}}{2
   \tilde{g}} & -\frac{ \hat g M_{A}^2 r s_{\theta}  c_{\theta}}{\sqrt{2}
   \tilde{g}} \\
 -\frac{ \hat g' M_{V}^2 s^2_{\theta}}{2 \tilde{g}} & -\frac{ \hat g
   M_{V}^2 (1+c^2_{\theta})}{2 \tilde{g}} & M_{V}^2 & 0 & 0 & 0 & 0 & 0 \\
 -\frac{ \hat g' M_{V}^2 (1+c^2_{\theta})}{2\tilde{g}} & -\frac{ \hat g
   M_{V}^2 s^2_{\theta}}{2 \tilde{g}} &0 & M_{V}^2 &  0&  0& 0 & 0 \\
 \frac{ \hat g' M_{V}^2 s_{\theta}  c_{\theta}}{\sqrt{2} \tilde{g}} &
   -\frac{ \hat g M_{V}^2 s_{\theta}  c_{\theta}}{\sqrt{2} \tilde{g}} & 0 & 0 & M_{V}^2 & 0 & 0 & 0 \\
 -\frac{ \hat g' M_{A}^2 r s^2_{\theta}}{2 \tilde{g}} & \frac{ \hat g
   M_{A}^2 r s^2_{\theta}}{2 \tilde{g}} & 0 &0 & 0 & M_{A}^2 & 0 & 0 \\
    \frac{ \hat g' M_{A}^2 r s^2_{\theta}}{2 \tilde{g}} & -\frac{ \hat g M_{A}^2 r s^2_{\theta}}{2 \tilde{g}} & 0& 0& 0&0 & M_{A}^2 & 0 \\
    \frac{ \hat g' M_{A}^2 r s_{\theta}  c_{\theta}}{\sqrt{2} \tilde{g}} & -\frac{ \hat g M_{A}^2 r s_{\theta} c_{\theta}}{\sqrt{2} \tilde{g}} & 0 & 0 & 0 &0 & 0 & M_{A}^2 )\,.
\end{align}
The only difference is that the
eigenvector for the photon changes and we find
\begin{align}
    A_\mu &= \frac{e}{\hat g}  W_\mu^3 + \frac{e}{ \hat g'} B_\mu + \frac{e}{\gt} v_{1\mu}^0+  \frac{e}{\gt}v_{2\mu}^0  \,,
\label{eq:photon2} 
\end{align}
with $e$ given by eq.~\eqref{eq:electric-charge}.

\subsection{Higgs-vector-vector couplings}\label{app:higgsappendix}

The fact that the longitudinal components of the SM vector bosons are formed by the would-be Goldstone bosons implies couplings of the spin-1 resonances 
to the Higgs boson and one SM vector boson. 
 The contributing terms in the Lagrangian are
\begin{align}
    \mathcal{L} &\supset \frac{\ii}{\sqrt{2} f_\pi} \Tr\bigg( -\big( f_0^2 \hat g X(\boldsymbol{\tilde W}_\mu) +f_0^2 \hat g' X(\boldsymbol{B}_\mu) + r f_1^2 \tilde g \boldsymbol{\mathcal{A}}_\mu \big) \comm{\tilde \Pi_P}{\hat g T(\boldsymbol{\tilde W}_\mu) + \hat g' T(\boldsymbol{B}_\mu)} \nonumber\\
    &\qquad + r f_1^2 \big( \tilde g \boldsymbol{\mathcal{A}}_\mu + r \hat g X(\boldsymbol{\tilde W}_\mu) + r \hat g' X(\boldsymbol{B}_\mu) \big) \comm{\tilde \Pi_P}{\tilde g \boldsymbol{\mathcal{V}}_\mu} \bigg) \nonumber\\
    &\quad + \frac{\ii f_K^2}{\sqrt{2} f_\pi} \Tr\bigg( \big( \tilde g \boldsymbol{\mathcal{V}}_\mu - \hat g T(\boldsymbol{\tilde W}_\mu) - \hat g' T(\boldsymbol{B}_\mu) \big) \comm{\Pi_P}{r \tilde g \boldsymbol{\mathcal{A}}_\mu + \hat g X(\boldsymbol{\tilde W}_\mu) + \hat g' X(\boldsymbol{B}_\mu)} \bigg)\,,
\end{align}
expressed in terms of gauge eigenstates.
Here $X(\boldsymbol{\tilde W}_\mu)$ is defined analogously to $T(\boldsymbol{\tilde W}_\mu)$ in \cref{eq:Tprojection} as
\begin{align}
    X(\boldsymbol{\tilde W}_\mu) = \tilde W_{\mu}^i \mathrm{Tr} ( T_L^i \tilde X^I ) \tilde X^I\,, \quad X(\boldsymbol{B}_\mu) = B_{\mu} \mathrm{Tr} ( T_R^3 \tilde X^I ) \tilde X^I\,.
\end{align}
The resulting couplings in the mass eigenbasis can be compactly written as
\begin{align}
    \mathcal{L}_H &= c_{H R^+ R^-}^\mathrm{gauge} \cdot H \left(\mathcal{C} R_{\mu}^+\right)_i \left(\mathcal{C}^* R^{-\mu}\right)_j + \frac{1}{2} c_{H R^0 R^0}^\mathrm{gauge} \cdot H \left(\mathcal{N} R_{\mu}^0\right)_i \left(\mathcal{N} R^{0\mu}\right)_j \\
    &= c_{H R^+ R^-} \cdot H R_{i\mu}^+ R^{-\mu}_j + \frac{1}{2} c_{H R^0 R^0} \cdot H R_{i\mu}^0 R^{0\mu}_j \,,
\end{align}
with
\begin{align}
    \frac{1}{2} c_{H R^0 R^0}^\mathrm{gauge} &= 
    \resizebox{0.8\textwidth}{!}{ $\mqty(
        \frac{g'^2 (f_\pi^2 \gt^2 - 2 M_V^2 + 2 M_A^2 r^2) s_{2\theta}}{8 f_\pi \gt^2} & -\frac{g g' (f_\pi^2 \gt^2 - 2 M_V^2 + 2 M_A^2 r^2) s_{2\theta}}{8 f_\pi \gt^2} & - \frac{g' (M_A^2 - M_V^2) r c_\theta}{\sqrt{2} f_\pi \gt} & 0 & - \frac{g' \left(M_V^2 - r^2 M_A^2\right) s_\theta}{\sqrt{2} f_\pi \gt} \\
        -\frac{g g' (f_\pi^2 \gt^2 - 2 M_V^2 + 2 M_A^2 r^2) s_{2\theta}}{8 f_\pi \gt^2} & \frac{g^2 (f_\pi^2 \gt^2 - 2 M_V^2 + 2 M_A^2 r^2) s_{2\theta}}{8 f_\pi \gt^2} & \frac{g (M_A^2 - M_V^2) r c_\theta}{\sqrt{2} f_\pi \gt} & 0 & \frac{g \left(M_V^2 - r^2 M_A^2\right) s_\theta}{\sqrt{2} f_\pi \gt} \\
        - \frac{g' (M_A^2 - M_V^2) r c_\theta}{\sqrt{2} f_\pi \gt} & \frac{g (M_A^2 - M_V^2) r c_\theta}{\sqrt{2} f_\pi \gt} & 0 & 0 & \frac{r\left(M_V^2-M_A^2\right)}{f_\pi} \\
        0 & 0 & 0 & 0 & 0 \\
        - \frac{g' \left(M_V^2 - r^2 M_A^2\right) s_\theta}{\sqrt{2} f_\pi \gt} & \frac{g \left(M_V^2 - r^2 M_A^2\right) s_\theta}{\sqrt{2} f_\pi \gt} & \frac{r \left(M_V^2-M_A^2\right)}{f_\pi} & 0 & 0
    )$}\,,
\end{align}
and
\begin{align}
    c_{H R^+ R^-}^\mathrm{gauge} &= \mqty( 
        \frac{g^2 (f_\pi^2 \gt^2-2 M_V^2+2 M_A^2 r^2) s_{2\theta}}{4 f_\pi \gt^2} & -\frac{g \left(M_V^2-M_A^2\right) r c_\theta}{\sqrt{2} f_\pi \gt} & 0 & \frac{g \left(M_V^2 - r^2 M_A^2\right) s_\theta}{\sqrt{2} f_\pi \gt} \\
        -\frac{g \left(M_V^2-M_A^2\right) r c_\theta}{\sqrt{2} f_\pi \gt} & 0 & 0 & \frac{r\left(M_V^2-M_A^2\right)}{f_\pi} \\
        0 & 0 & 0 & 0 \\
        \frac{g \left(M_V^2 - r^2 M_A^2\right) s_\theta}{\sqrt{2} f_\pi \gt} & \frac{r\left(M_V^2-M_A^2\right)}{f_\pi} & 0 & 0 
    ) \,,
\end{align}
for the cosets $\SU(4)/\Sp(4)$ and $\SU(5)/\SO(5)$.
For $\SU(4)^2 / \SU(4)$ the corresponding couplings are given by
\begin{align}
    &\frac{1}{2} c_{H R^0 R^0}^\mathrm{gauge} =\nonumber \\
    &\resizebox{\textwidth}{!}{ $\mqty( \frac{g'^2 (f_\pi^2 \gt^2 - 2 M_V^2 + 2 M_A^2 r^2) s_{2\theta}}{8 f_\pi \gt^2} & -\frac{g g'
   \left(f_{\pi}^2 \gt^2+2 M_A^2 r^2-2 M_V^2\right)s_{2\theta}}{8 f_{\pi} \gt^2} &
   -\frac{g' \left(M_V^2 - r^2 M_A^2\right)s_{2\theta}}{8 f_{\pi}
   \gt} & \frac{g's_{2\theta} \left(M_V^2 - r^2 M_A^2\right)}{8
   f_{\pi} \gt} & -\frac{g' \left(M_V^2 - r^2 M_A^2\right)s^2_{\theta}}{2 \sqrt{2} f_{\pi} \gt} & \frac{g'
   \left(M_V^2-M_A^2\right)r s_{2\theta}}{8 f_{\pi} \gt} & -\frac{g'
    \left(M_V^2-M_A^2\right) r s_{2\theta}}{8 f_{\pi} \gt} &-
   \frac{g' \left(M_V^2-M_A^2\right) r c^2_{\theta}}{2 \sqrt{2}
   f_{\pi} \gt} \\
 -\frac{g g'  \left(f_{\pi}^2 \gt^2+2 M_A^2 r^2-2
   M_V^2\right)s_{2\theta}}{8 f_{\pi} \gt^2} & \frac{g^2 \left(f_{\pi}^2
   \gt^2+2 M_A^2 r^2-2 M_V^2\right)s_{2\theta}}{8 f_{\pi} \gt^2} & \frac{g \left(M_V^2 - r^2 M_A^2\right)s_{2\theta}}{8 f_{\pi} \gt} & -\frac{g  \left(M_V^2 - r^2 M_A^2\right)s_{2\theta}}{8 f_{\pi} \gt} & \frac{g
    \left(M_V^2 - r^2 M_A^2\right)s^2_{\theta}}{2 \sqrt{2} f_{\pi}
   \gt} & -\frac{g  \left(M_V^2-M_A^2\right) r s_{2\theta}}{8 f_{\pi}
   \gt} & \frac{g  \left(M_V^2-M_A^2\right) r s_{2\theta}}{8 f_{\pi}
   \gt} & \frac{g  \left(M_V^2-M_A^2\right) r c^2_{\theta}}{2 \sqrt{2}
   f_{\pi} \gt} \\
 -\frac{g'  \left(M_V^2 - r^2 M_A^2\right)s_{2\theta}}{8 f_{\pi}
   \gt} & \frac{g \left(M_V^2 - r^2 M_A^2\right)s_{2\theta}}{8
   f_{\pi} \gt} & 0 &0 & 0 & 0 & 0 & -\frac{ \left(M_V^2-M_A^2\right)r}{2
   \sqrt{2} f_{\pi}} \\
 \frac{g' \left(M_V^2 - r^2 M_A^2\right)s_{2\theta}}{8 f_{\pi}
   \gt} & -\frac{g  \left(M_V^2 - r^2 M_A^2\right)s_{2\theta}}{8
   f_{\pi} \gt} & 0 & 0 &  0&  0& 0 & \frac{\left(M_V^2-M_A^2\right)r}{2
   \sqrt{2} f_{\pi}} \\
 -\frac{g' \left(M_V^2 - r^2 M_A^2\right)s^2_{\theta}}{2 \sqrt{2}
   f_{\pi} \gt} & \frac{g  \left(M_V^2-r^2 M_A^2\right)s_{2\theta}}{2 \sqrt{2} f_{\pi} \gt} & 0 & 0& 0& \frac{
   \left(M_V^2-M_A^2\right)r}{2 \sqrt{2} f_{\pi}} & -\frac{\left(M_V^2-M_A^2\right)r}{2 \sqrt{2} f_{\pi}} & 0 \\
 \frac{g'  \left(M_V^2-M_A^2\right)r s_{2\theta}}{8 f_{\pi} \gt} &
  - \frac{g  \left(M_V^2-M_A^2\right)r s_{2\theta}}{8 f_{\pi} \gt} & 0
    &  0& \frac{ \left(M_V^2-M_A^2\right)r}{2 \sqrt{2} f_{\pi}} &0  & 0 &0  \\
 -\frac{g'  \left(M_V^2-M_A^2\right)r s_{2\theta}}{8 f_{\pi}
   \gt} & \frac{g \left(M_V^2-M_A^2\right)r s_{2\theta}}{8 f_{\pi}
   \gt} & 0&0  & -\frac{ \left(M_V^2-M_A^2\right)r}{2 \sqrt{2} f_{\pi}} &
   0&  0& 0\\
 -\frac{g'  \left(M_V^2-M_A^2\right) r c^2_{\theta}}{2 \sqrt{2}
   f_{\pi} \gt} & \frac{g  \left(M_V^2-M_A^2\right) r c^2_{\theta}}{2
   \sqrt{2} f_{\pi} \gt} & -\frac{\left(M_V^2-M_A^2\right)r}{2 \sqrt{2}
   f_{\pi}} & \frac{\left(M_V^2-M_A^2\right)r }{2 \sqrt{2} f_{\pi}} & 0 & 0 & 0 &0 
   \\
)$}\,,
\end{align}
and
\begin{align}
&c_{H R^+ R^-}^\mathrm{gauge}  = \nonumber \\
&\resizebox{\textwidth}{!}{ $\mqty(
\frac{g^2 (f_\pi^2 \gt^2-2 M_V^2+2 M_A^2 r^2) s_{2\theta}}{4 f_\pi \gt^2} & \frac{g \left(M_V^2 - r^2 M_A^2\right)s_{2\theta}}{4 f_{\pi} \gt} & -\frac{g \left(M_V^2 - r^2 M_A^2\right)s_{2\theta}}{4 f_{\pi} \gt} & -\frac{g  \left(M_V^2 - r^2 M_A^2\right)s^2_{\theta}}{\sqrt{2} f_{\pi} \gt} & -\frac{g  \left(M_V^2-M_A^2\right)r s_{2\theta}}{4 f_{\pi} \gt} & \frac{g 
   \left(M_V^2-M_A^2\right) r s_{2\theta}}{4 f_{\pi} \gt} & -\frac{g 
   \left(M_V^2-M_A^2\right) r c^2_{\theta}}{\sqrt{2} f_{\pi} \gt} \\
 \frac{g \left(M_V^2 - r^2 M_A^2\right)s_{2\theta}}{4 f_{\pi} \gt} &
    0& 0& 0 & 0 & 0 & \frac{r \left(M_V^2-M_A^2\right)}{\sqrt{2} f_{\pi}} \\
 -\frac{g  \left(M_V^2 - r^2 M_A^2\right)s_{2\theta}}{4 f_{\pi} \gt}
   & 0 & 0 & 0 &  0& 0 & -\frac{ \left(M_V^2-M_A^2\right)r}{\sqrt{2} f_{\pi}} \\
 -\frac{g \left(M_V^2 - r^2 M_A^2\right)s^2_{\theta}}{\sqrt{2} f_{\pi}
   \gt} & 0 &0 & 0 &- \frac{ \left(M_V^2-M_A^2\right)r}{\sqrt{2} f_{\pi}}
   & \frac{\left(M_V^2-M_A^2\right)r}{\sqrt{2} f_{\pi}} & 0 \\
- \frac{g \left(M_V^2-M_A^2\right)r s_{2\theta}}{4 f_{\pi} \gt} & 0
   & 0 & -\frac{\left(M_V^2-M_A^2\right)r }{\sqrt{2} f_{\pi}} & 0 & 0 &0  \\
 \frac{g  \left(M_V^2-M_A^2\right) r s_{2\theta}}{4 f_{\pi} \gt} & 0 & 0 &
   \frac{ \left(M_V^2-M_A^2\right)r}{\sqrt{2} f_{\pi}} & 0 & 0 & 0 \\
 -\frac{g  \left(M_V^2-M_A^2\right)r c^2_{\theta}}{\sqrt{2} f_{\pi}
   \gt} & \frac{ \left(M_V^2-M_A^2\right)r }{\sqrt{2} f_{\pi}} & -\frac{
   \left(M_V^2-M_A^2\right)r}{\sqrt{2} f_{\pi}} & 0 & 0 & 0 &0  \\
)$}\,.
\end{align}

\section{Further phenomenological aspects}
\label{app:pheno}

Here we collect the branching ratios plots for all three cosets as well as the individual exclusion plots for the  cosets SU(4)/Sp(4) and SU(4)$\times$SU(4)/SU(4).

\subsection{Branching ratios}
\begin{figure}[hp]
    \centering
    \includegraphics[width=.75\textwidth]{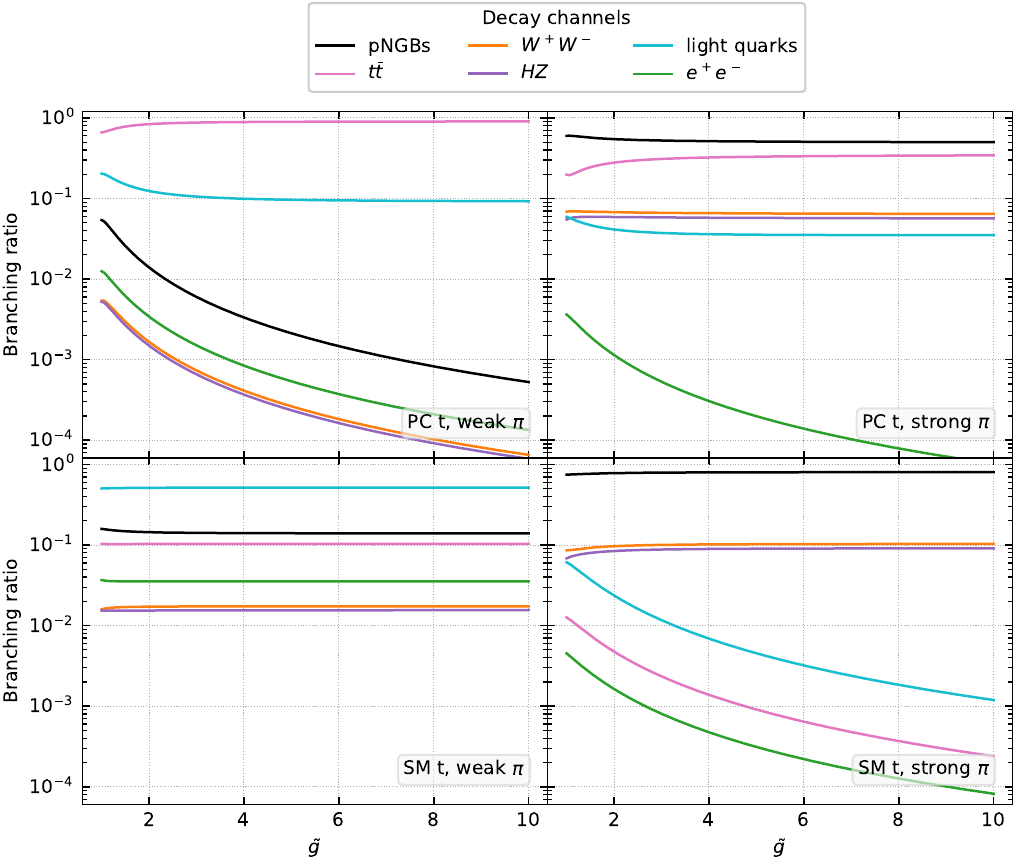}
    \caption{Branching ratios of $V_{1\mu}^0$ in the $\SU(5)/\SO(5)$ coset for the scenarios defined in \cref{sec:pheno}. We set $M_V=3\,\mathrm{TeV}$ and $M_\pi=700\,\mathrm{GeV}$.}
    \label{fig:branching-ratios-V10-SU5}
\end{figure}

\begin{figure}[hp]
    \centering
    \includegraphics[width=.75\textwidth]{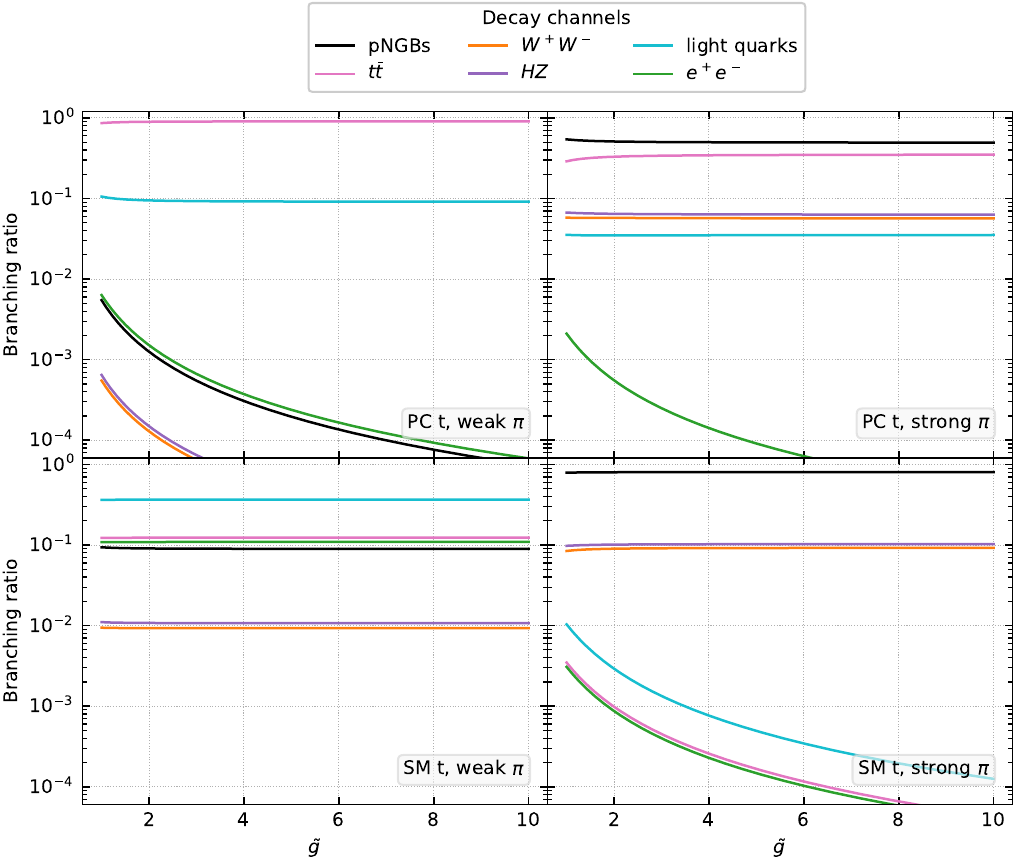}
    \caption{Branching ratios of $V_{2\mu}^0$ in the $\SU(5)/\SO(5)$ coset for the scenarios defined in \cref{sec:pheno}. We set $M_V=3\,\mathrm{TeV}$ and $M_\pi=700\,\mathrm{GeV}$.}
    \label{fig:branching-ratios-V20-SU5}
\end{figure}

\begin{figure}[hp]
    \centering
    \includegraphics[width=.75\textwidth]{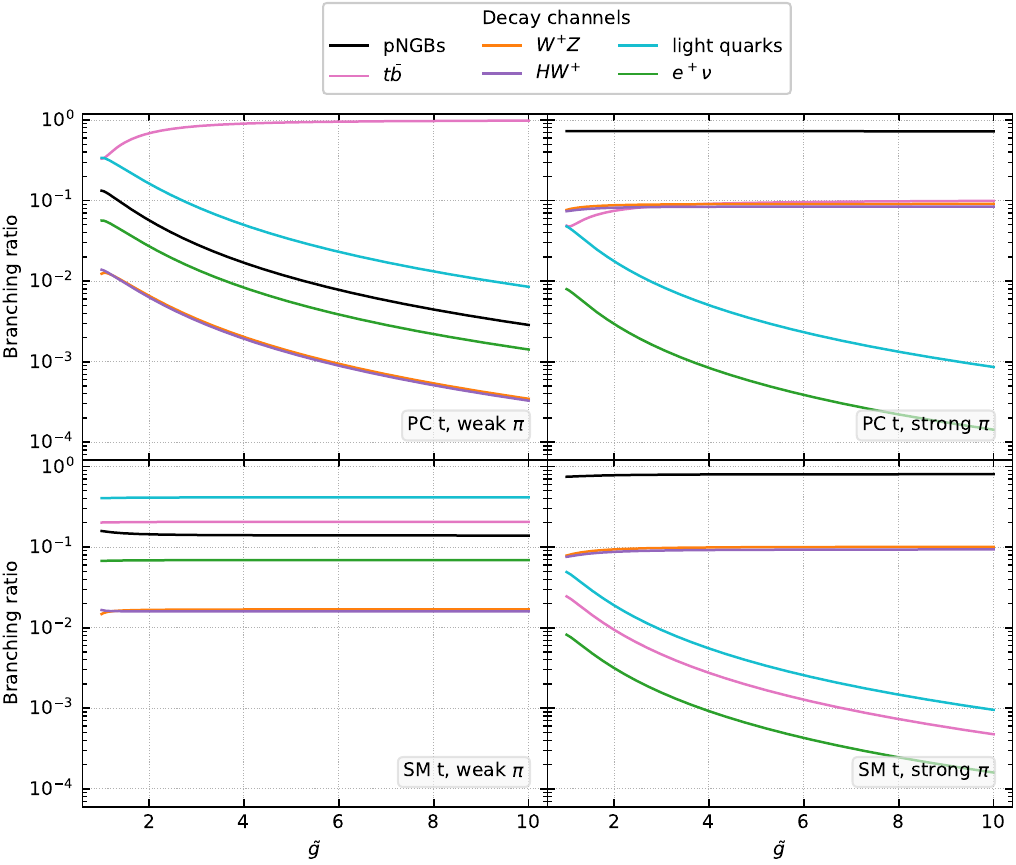}
    \caption{Branching ratios of $V_{1\mu}^+$ in the $\SU(5)/\SO(5)$ coset for the scenarios defined in \cref{sec:pheno}. We set $M_V=3\,\mathrm{TeV}$ and $M_\pi=700\,\mathrm{GeV}$.}
    \label{fig:branching-ratios-V1p-SU5}
\end{figure}

\begin{figure}[hp]
    \centering
    \includegraphics[width=.75\textwidth]{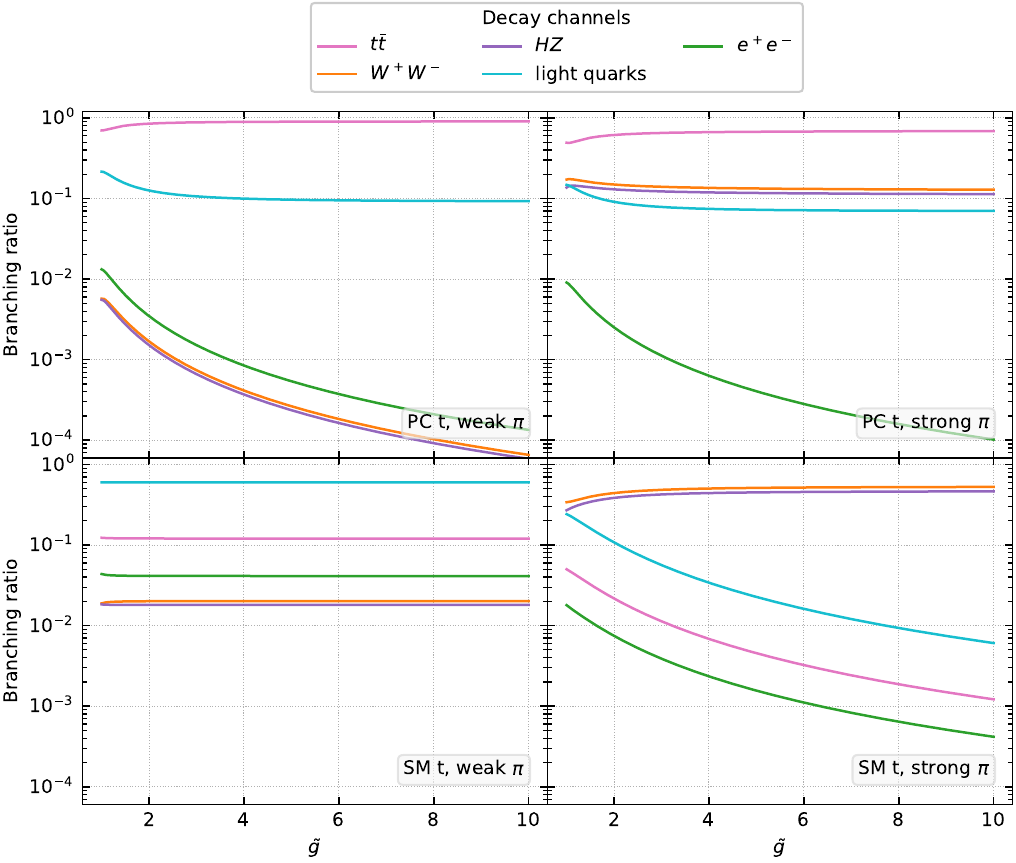}
    \caption{Branching ratios of $V_{1\mu}^0$ in the $\SU(4)/\Sp(4)$ coset for the scenarios defined in \cref{sec:pheno} and $M_V=3\,\mathrm{TeV}$.}
    \label{fig:branching-ratios-V10-SU4}
\end{figure}

\begin{figure}[hp]
    \centering
    \includegraphics[width=.75\textwidth]{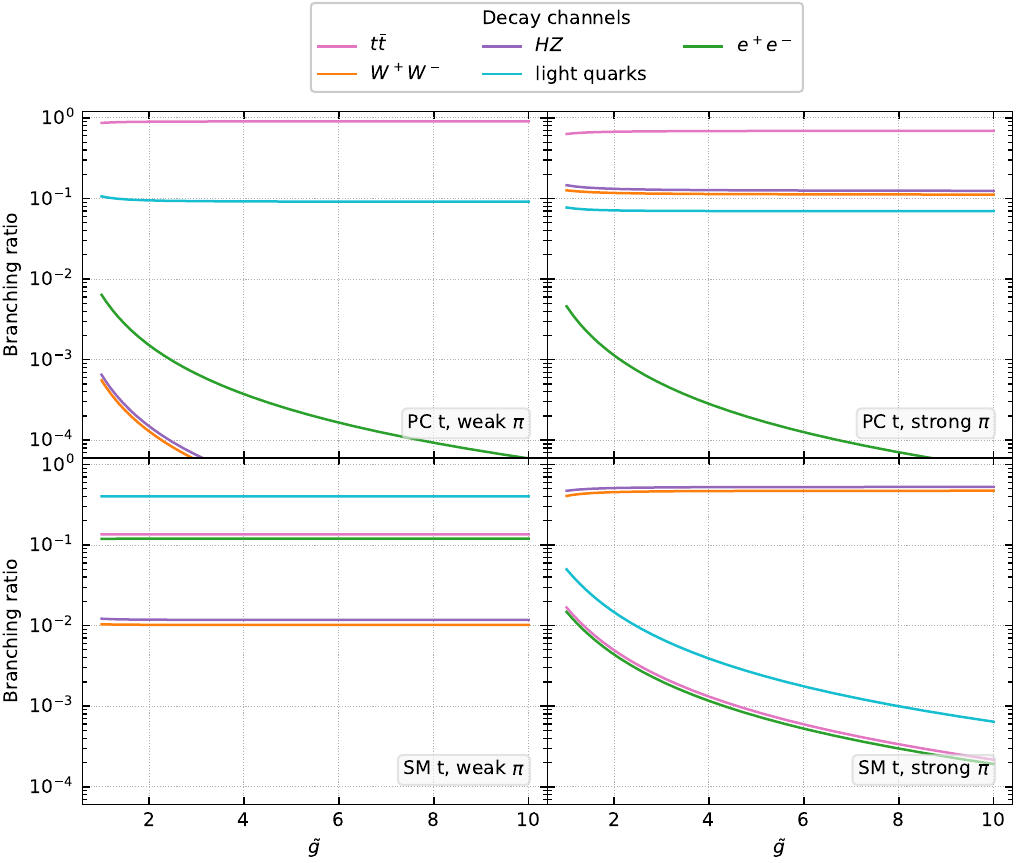}
    \caption{Branching ratios of $V_{2\mu}^0$ in the $\SU(4)/\Sp(4)$ coset for the scenarios defined in \cref{sec:pheno} and $M_V=3\,\mathrm{TeV}$.}
    \label{fig:branching-ratios-V20-SU4}
\end{figure}

\begin{figure}[hp]
    \centering
    \includegraphics[width=.75\textwidth]{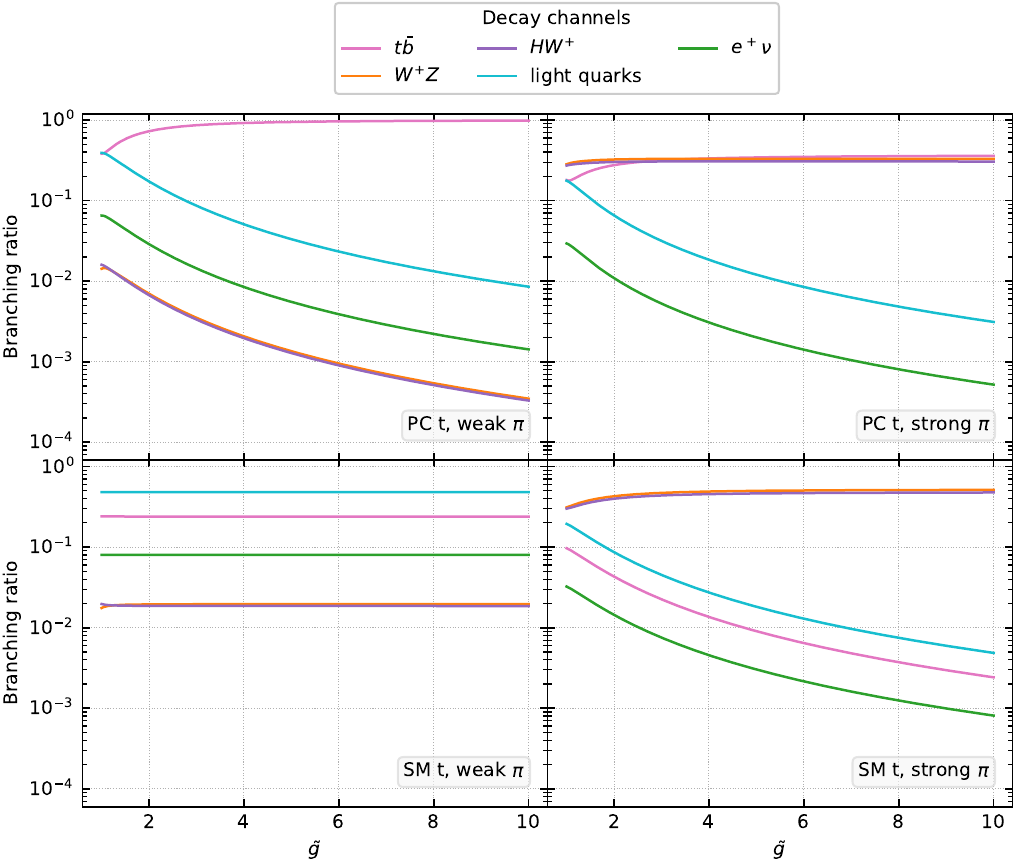}
    \caption{Branching ratios of $V_{1\mu}^+$ in the $\SU(4)/\Sp(4)$ coset for the scenarios defined in \cref{sec:pheno} and $M_V=3\,\mathrm{TeV}$.}
    \label{fig:branching-ratios-V1p-SU4}
\end{figure}

\begin{figure}[hp]
    \centering
    \includegraphics[width=.75\textwidth]{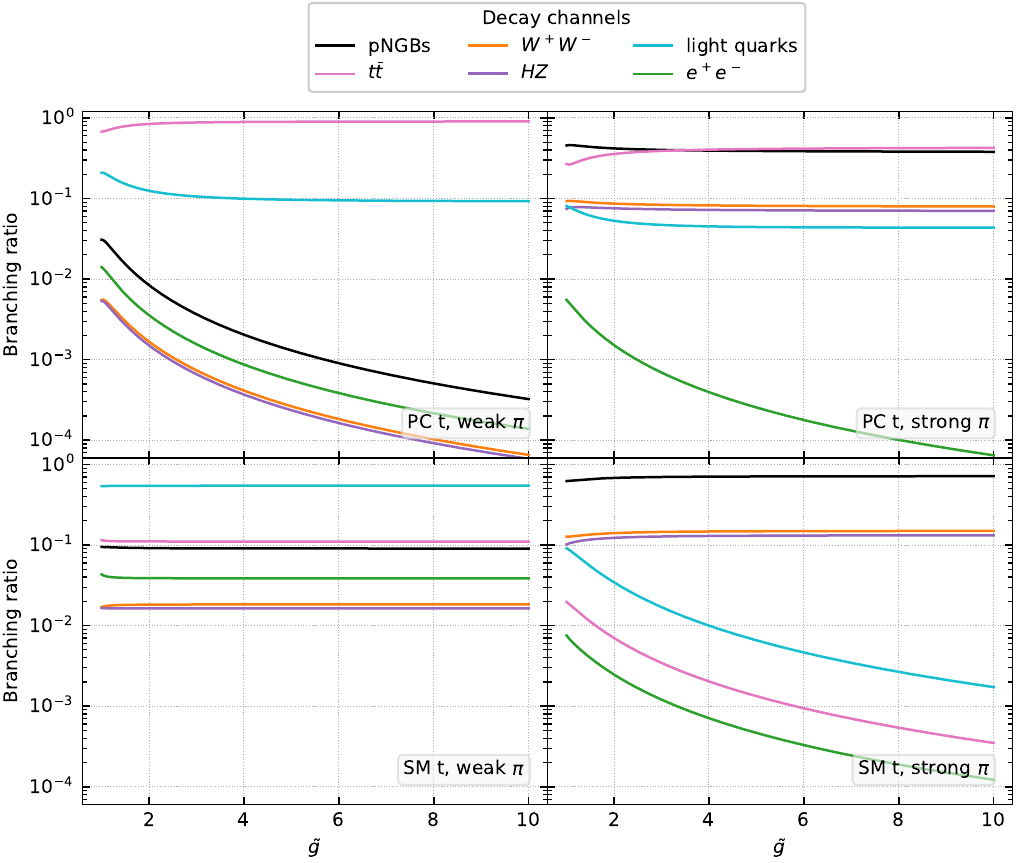}
    \caption{Branching ratios of $V_{1\mu}^0$ in the $\SU(4)\times \SU(4)/\SU(4)$ coset for the scenarios defined in \cref{sec:pheno}. We set $M_V=3\,\mathrm{TeV}$ and $M_\pi=450\,\mathrm{GeV}$.}
    \label{fig:branching-ratios-V10-SU4SU4}
\end{figure}

\begin{figure}[hp]
    \centering
    \includegraphics[width=.75\textwidth]{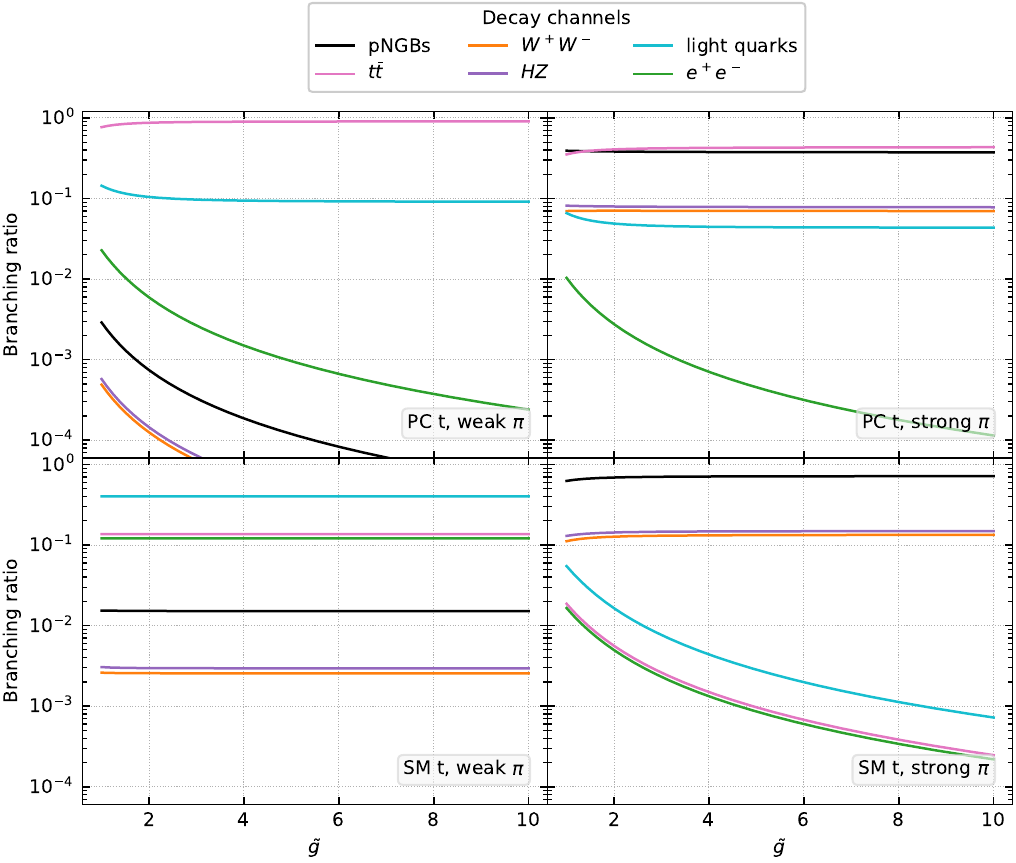}
    \caption{Branching ratios of $V_{2\mu}^0$ in the $\SU(4)\times \SU(4)/\SU(4)$ coset for the scenarios defined in \cref{sec:pheno}. We set $M_V=3\,\mathrm{TeV}$ and $M_\pi=450\,\mathrm{GeV}$.}
    \label{fig:branching-ratios-V20-SU4SU4}
\end{figure}

\begin{figure}[hp]
    \centering
    \includegraphics[width=.75\textwidth]{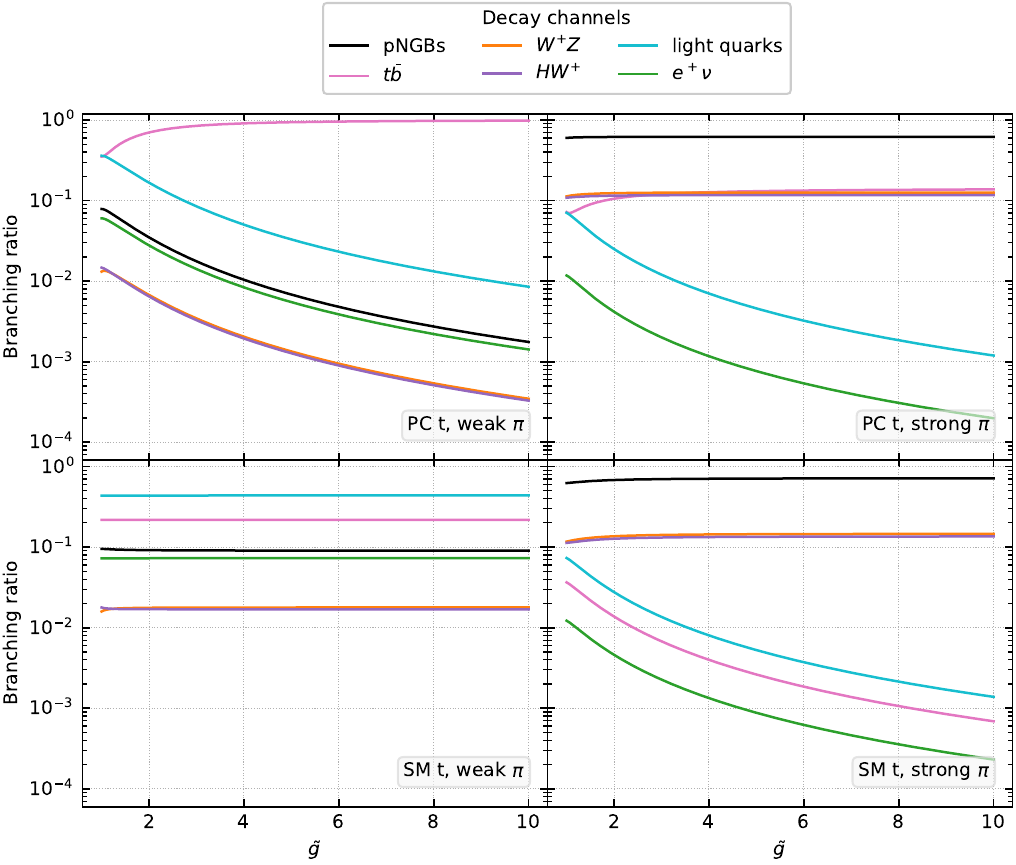}
    \caption{Branching ratios of $V_{1\mu}^+$ in the $\SU(4)\times \SU(4)/\SU(4)$ coset for the scenarios defined in \cref{sec:pheno}. We set $M_V=3\,\mathrm{TeV}$ and $M_\pi=450\,\mathrm{GeV}$.}
    \label{fig:branching-ratios-V1p-SU4SU4}
\end{figure}

Supplementing  the information on the partial widths in \cref{fig:partial-widths,fig:partial-widths-su4su4}, we present here the corresponding branching ratios.
Figures~\ref{fig:branching-ratios-V10-SU5} to \ref{fig:branching-ratios-V1p-SU5} show branching ratios of $V_{1\mu}^0$, $V_{2\mu}^0$ and $V_{1\mu}^+$ for the coset $\SU(5)/\SO(5)$. Each figure is split into four panels corresponding to the scenarios defined in \cref{sec:pheno}. 
$V_{2\mu}^0$ differs slightly from the others, as the fermion couplings are more suppressed due to the different mass mixing. This results in an enhanced dominance of the top channel (pNGB channel) in the scenario $\textbf{PC}\, \boldsymbol t,\, \mathbf{weak}\, \boldsymbol \pi$ ($\textbf{SM}\, \boldsymbol t,\, \mathbf{strong}\, \boldsymbol \pi$).
In \cref{fig:branching-ratios-V10-SU4,fig:branching-ratios-V20-SU4,fig:branching-ratios-V1p-SU4}, the corresponding branching ratios for the coset $\SU(4)/\Sp(4)$ are shown, which lack the pNGB decay channel. Therefore, the Higgs and SM gauge boson channels are dominant for $\textbf{SM}\, \boldsymbol t,\, \mathbf{strong}\, \boldsymbol \pi$.
The branching ratios for $\SU(4)^2 / \SU(4)$, shown in \cref{fig:branching-ratios-V10-SU4SU4,fig:branching-ratios-V20-SU4SU4,fig:branching-ratios-V1p-SU4SU4}, are similar to $\SU(5)/\SO(5)$. Differences can be traced back to the different pNGB content and differences in the mixing in the spin-1 sectors.

\clearpage
\subsection{Bounds on the individual channels for the cosets $\SU(4)/\Sp(4)$ and $\SU(4)^2/\SU(4)$}

\label{app:pheno-bounds}
\begin{figure}[hptb]
    \centering
    \begin{subfigure}{0.47\linewidth}
        \includegraphics[width=\linewidth]{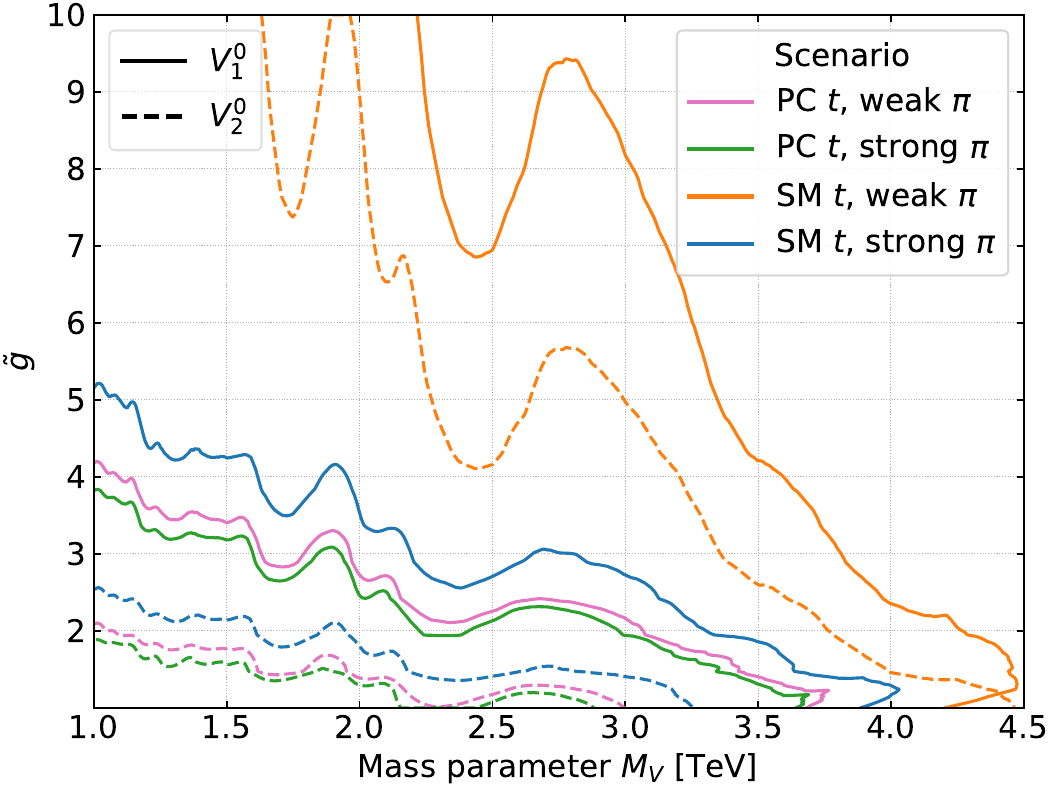}
        \caption{Bounds from $\mathcal V^0\to \ell^+ \ell^-$}
        \label{fig:bounds_su4_ll}
    \end{subfigure}\quad
    \begin{subfigure}{0.47\linewidth}
        \includegraphics[width=\linewidth]{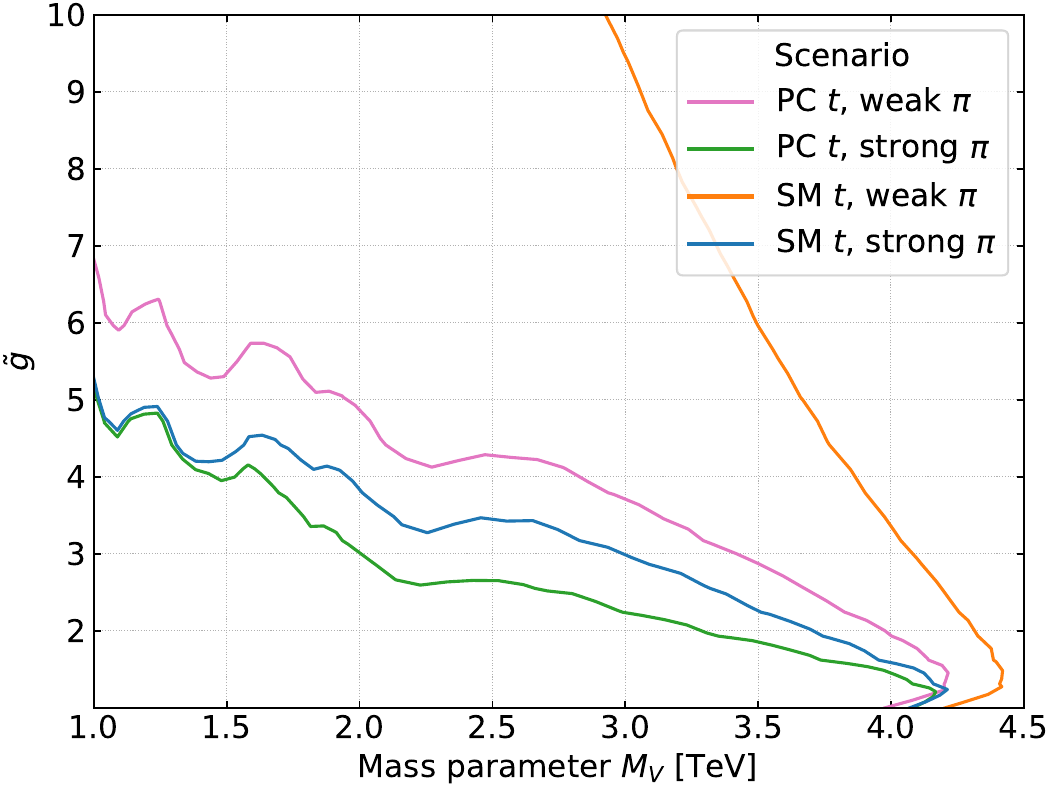}
        \caption{Bounds from $\mathcal V^\pm \to \ell^\pm \nu$}
        \label{fig:bounds_su4_lv}
    \end{subfigure}\vspace{1ex}

    \begin{subfigure}{0.47\linewidth}
        \includegraphics[width=\linewidth]{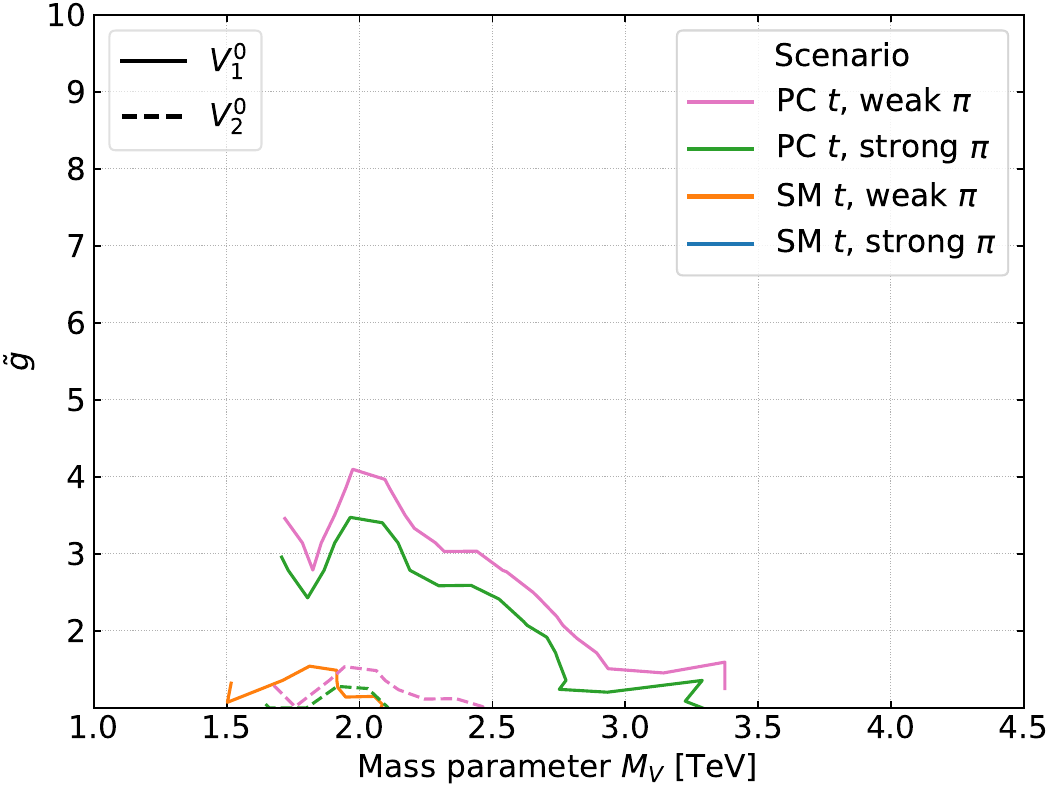}
        \caption{Bounds from $\mathcal V^0 \to t\bar t$}
        \label{fig:bounds_su4_tt}
    \end{subfigure}\quad
    \begin{subfigure}{0.47\linewidth}
        \includegraphics[width=\linewidth]{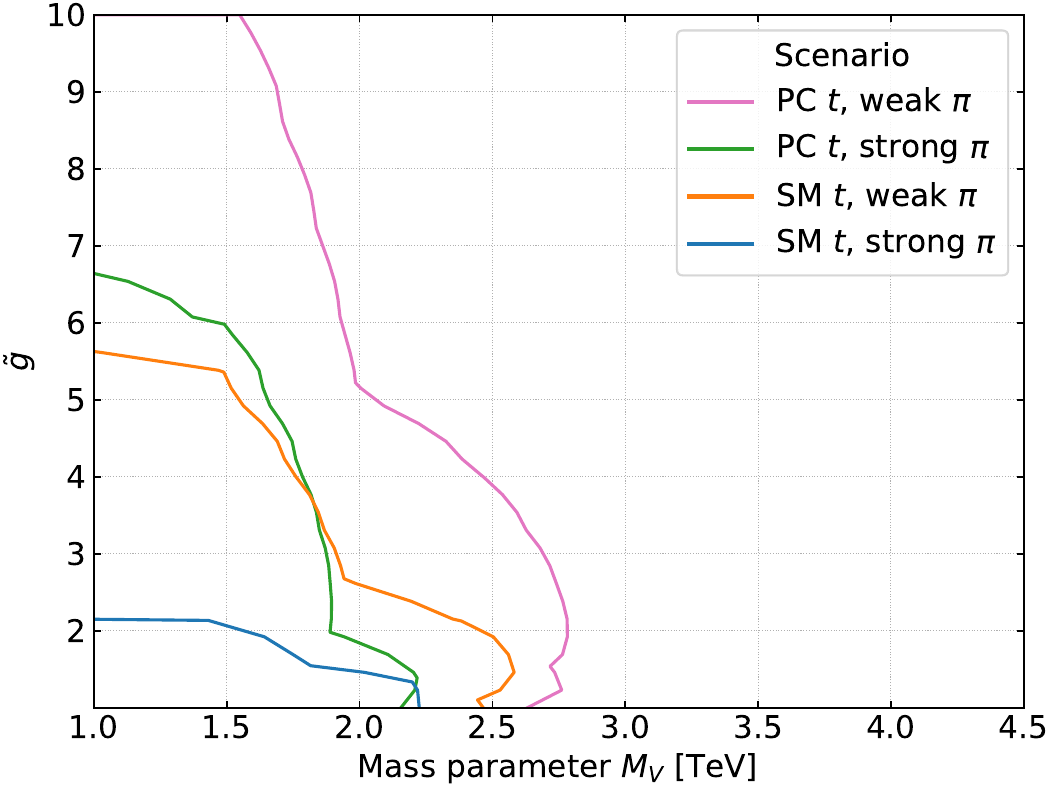}
        \caption{Bounds from $\mathcal V^\pm \to tb$}
        \label{fig:bounds_su4_tb}
    \end{subfigure}\vspace{1ex}
    
    \begin{subfigure}{0.47\linewidth}
        \includegraphics[width=\linewidth]{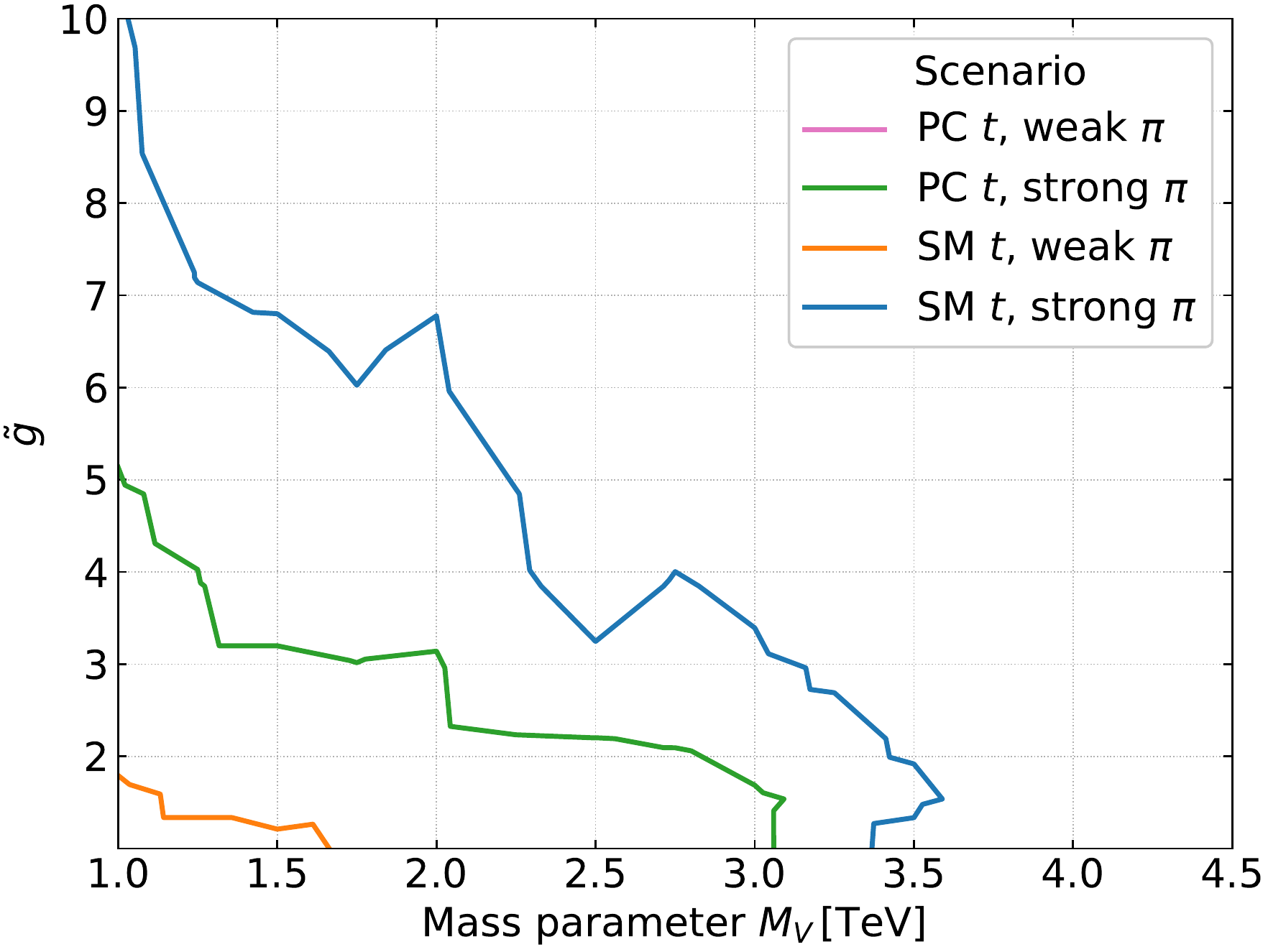}
        \caption{Bounds from $\mathcal V \to H Z, H W^\pm, W^+ W^-, W^\pm Z$}
        \label{fig:bounds_su4_higgs}
    \end{subfigure}

    \caption{Bounds on the single production of heavy vectors in the $\SU(4)/\Sp(4)$ coset.
    In the scenarios, ``SM $t$'' means the couplings of the $\mathcal V^0/ \mathcal V^\pm$ to $tt/tb$ are equal to the quark couplings to $Z/W^\pm$, whereas for ``PC $t$'' these couplings are set to 1. 
    For the \acp{pNGB}, ``weak'' and ``strong $\pi$'' refers to couplings $g_{V\pi\pi}=0$ and $g_{V\pi\pi}=4$, respectively.
    In (a)-(d) the upper limits on the cross sections are taken from direct searches \cite{ATLAS:2019erb,ATLAS:2020lks,ATLAS:2019lsy,ATLAS:2023ibb}.
    The bounds in (e) are derived from recasts of \cite{CMS:2017moi,CMS:2019xjf,Mrowietz:2020ztq,CMS:2019xud,Conte:2021xtt,CMS:2019ius,ATLAS:2020wzf,ATLAS:2022nrp,ATLAS:2021jgw,CMS:2022ubq,ATLAS:2019zci,ATLAS:2019rob}.
    }
    \label{fig:allbounds_su4sp4}
\end{figure}

We present here bounds in the $M_V-\tilde g$ plane for the SU(4)/Sp(4) and SU(4)$\times$SU(4)/SU(4) cosets. 
They have been derived using the same tools and searches as for the coset $\SU(5)/\SO(5)$, see \cref{fig:allbounds_su5so5}.
Figure \ref{fig:allbounds_su4sp4} shows the results for the coset $\SU(4)/\Sp(4)$. Panels (a) to (d) cover the decay channels into SM fermions, which are slightly stronger constrained at low masses in the \textbf{strong} $\boldsymbol{\pi}$ scenarios compared to $\SU(5)/\SO(5)$, due to the lack of the pNGB channels.
Here we recall that the corresponding coupling $g_{V \pi \pi}$
also impacts the couplings to $HV$ ($V=W,Z)$.
In \cref{fig:bounds_su4_higgs} we show the decays into two SM gauge bosons or gauge with Higgs boson. With the same reasoning as before, these are considerably stronger constrained, ranging up to $3.5~$TeV for small $\gt$.

\begin{figure}[htpb]
    \centering
    \begin{subfigure}{0.47\linewidth}
        \includegraphics[width=\linewidth]{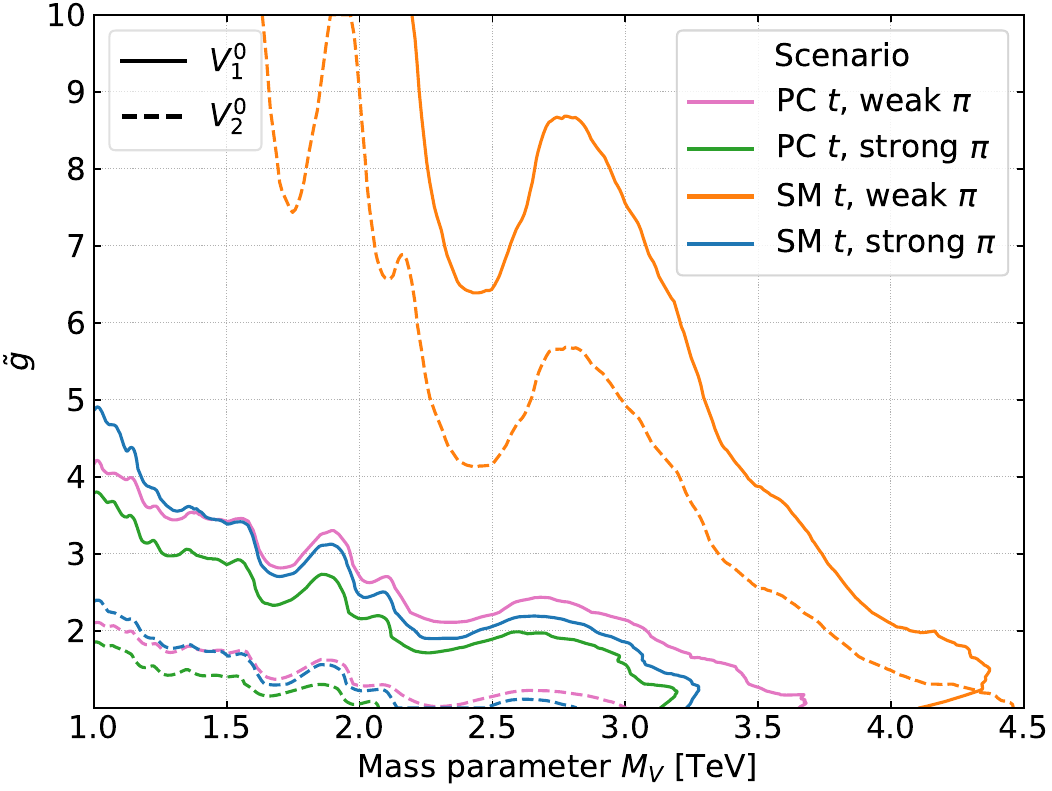}
        \caption{Bounds from $\mathcal V^0\to \ell^+ \ell^-$}
    \end{subfigure}\quad
    \begin{subfigure}{0.47\linewidth}
        \includegraphics[width=\linewidth]{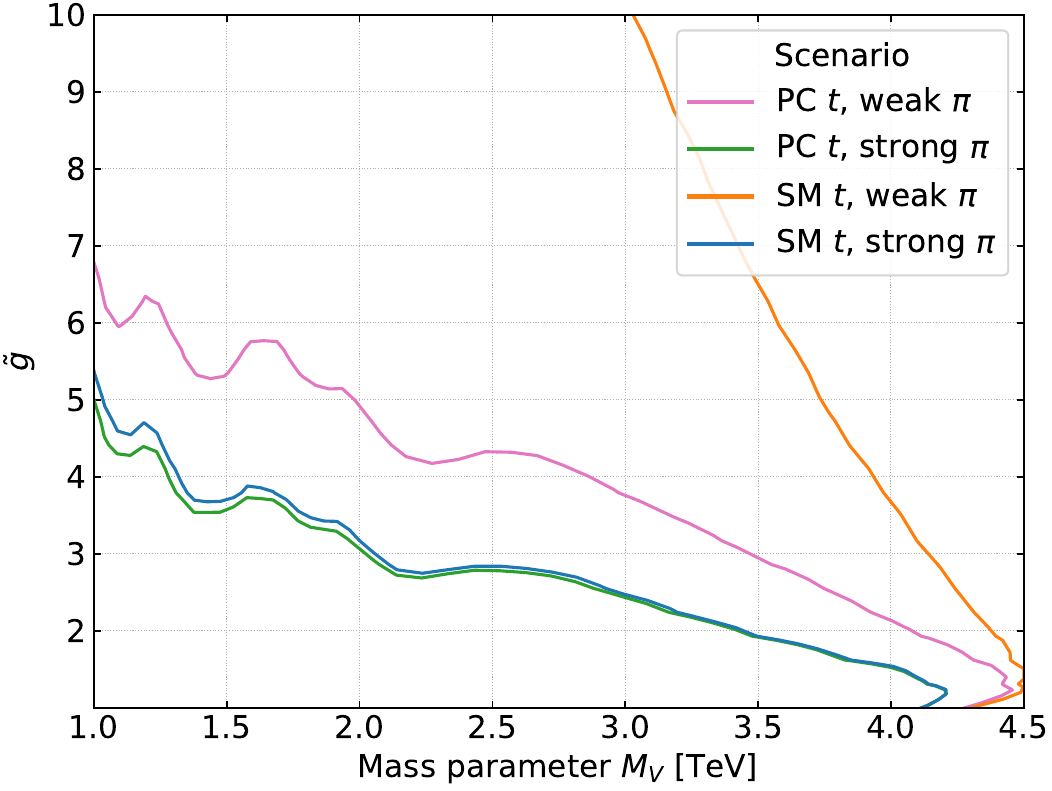}
        \caption{Bounds from $\mathcal V^\pm \to \ell^\pm \nu$ }
    \end{subfigure}\vspace{1ex}

    \begin{subfigure}{0.47\linewidth}
        \includegraphics[width=\linewidth]{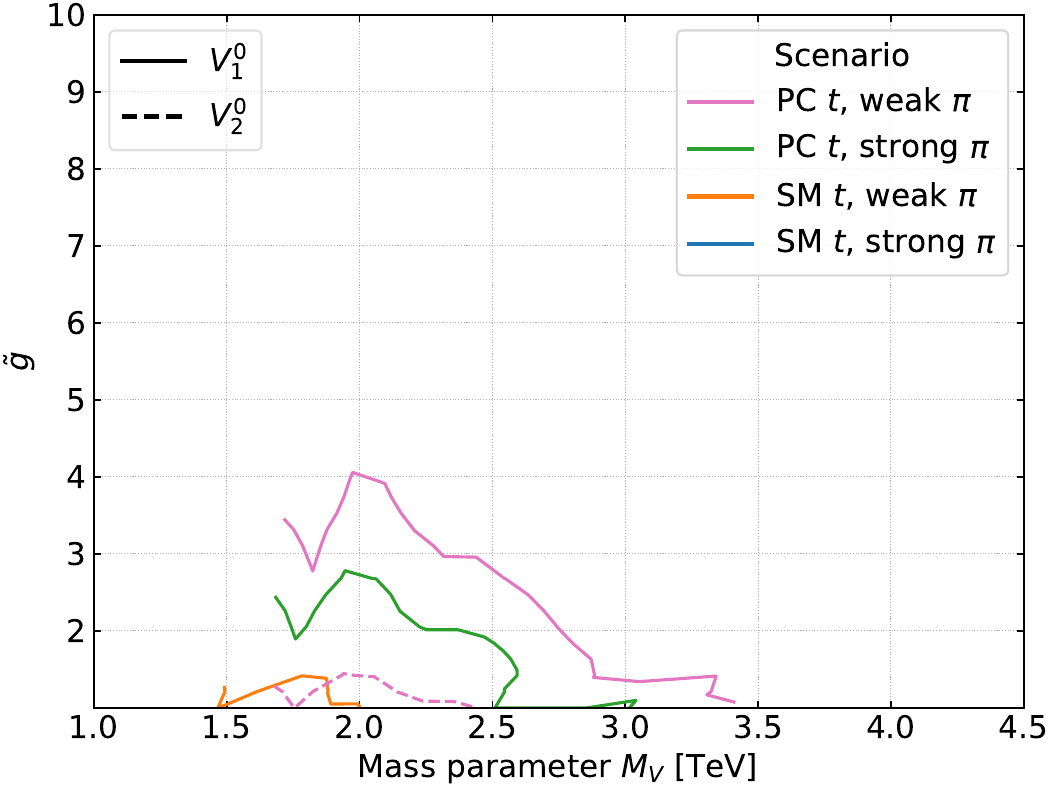}
        \caption{ Bounds from $\mathcal V^0 \to t\bar t$}
    \end{subfigure}\quad
    \begin{subfigure}{0.47\linewidth}
        \includegraphics[width=\linewidth]{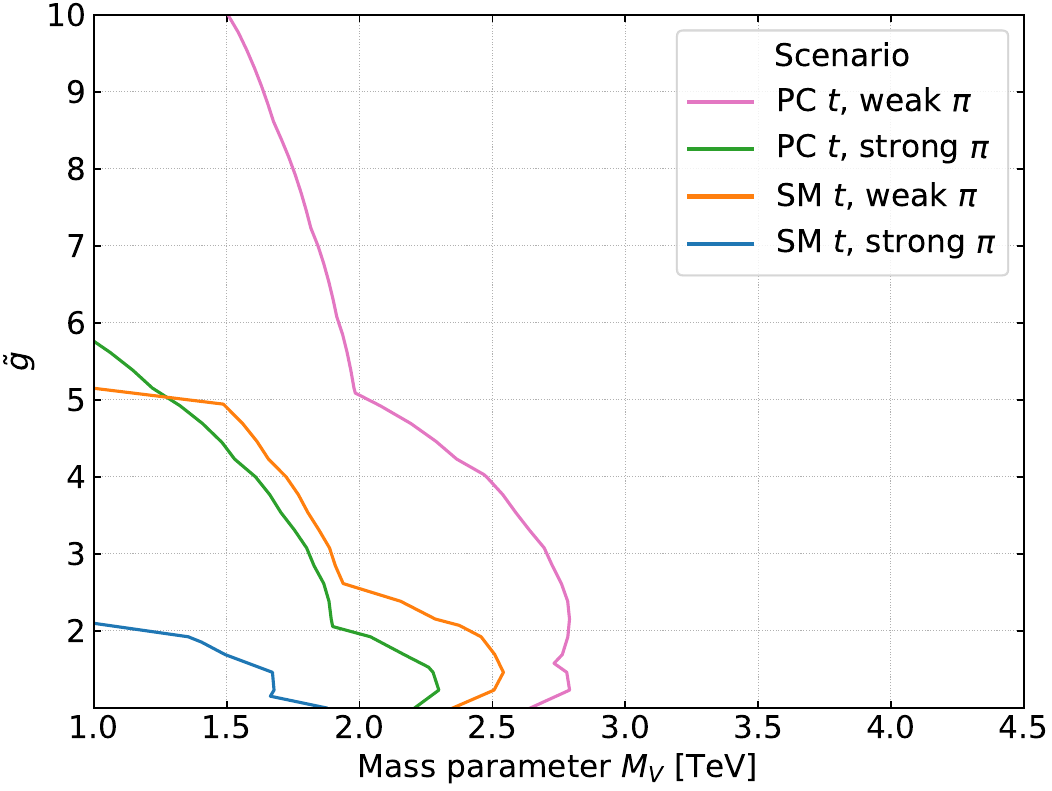}
        \caption{Bounds from $\mathcal V^\pm \to tb$}
    \end{subfigure}\vspace{1ex}
    
    \begin{subfigure}{0.47\linewidth}
        \includegraphics[width=\linewidth]{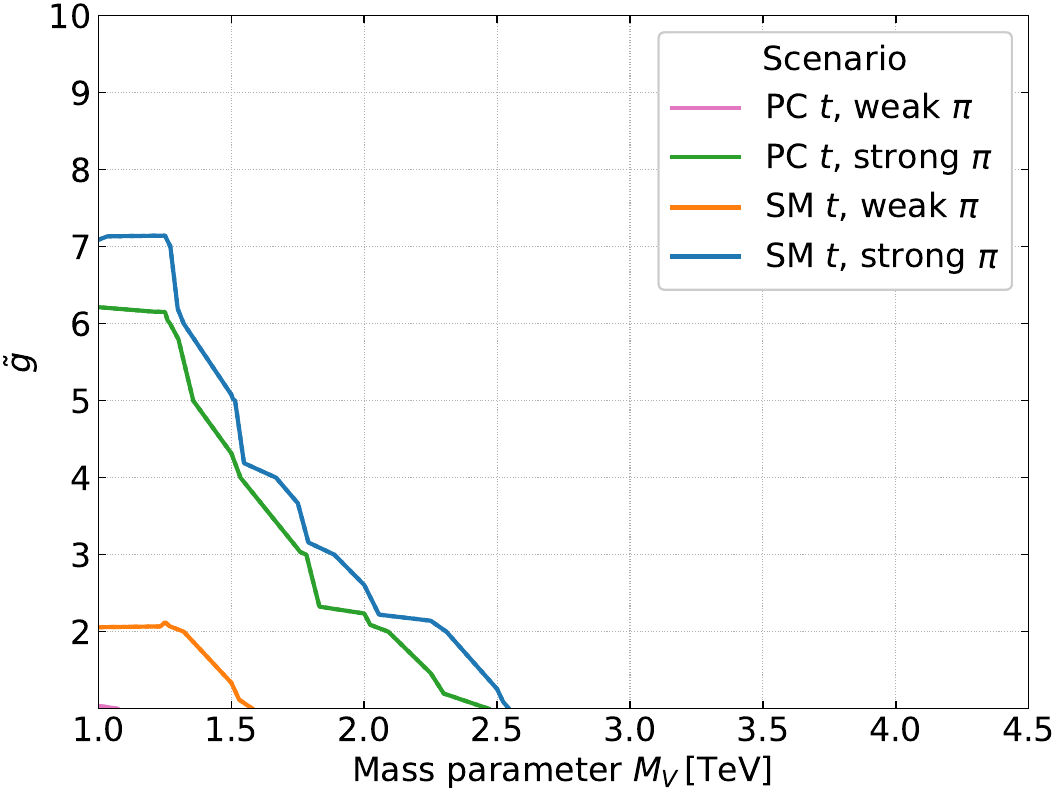}
        \caption{Bounds from $\mathcal V\to \pi\pi \to $ quarks}
        \label{fig:su4su4_pNGBs_bounds}
    \end{subfigure}\quad
    \begin{subfigure}{0.47\linewidth}
        \includegraphics[width=\linewidth]{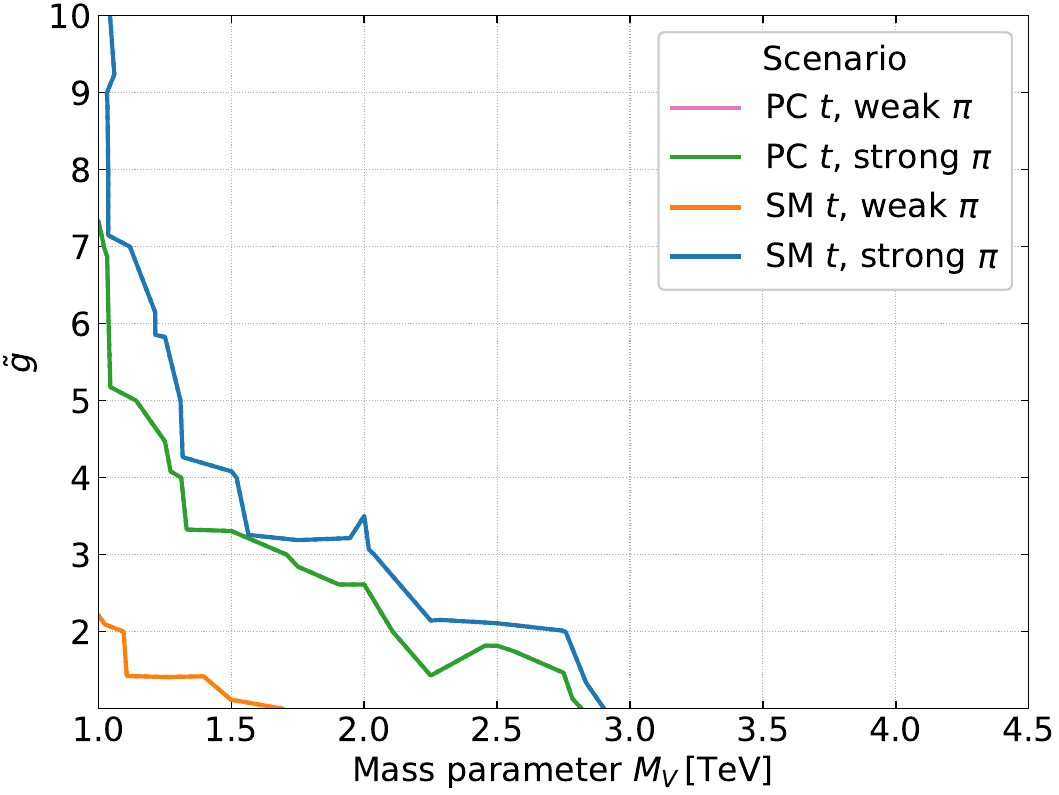}
        \caption{Bounds from $\mathcal V \to H Z, H W^\pm, W^+ W^-, W^\pm Z$}
        \label{fig:su4su4_higgs_bounds}
    \end{subfigure}
    
    \caption{Bounds on the single production of heavy vectors in the $\SU(4)^2/\SU(4)$ coset for a \ac{pNGB} mass of 450~GeV. 
    In the scenarios, ``SM $t$'' means the couplings of the $\mathcal V^0/ \mathcal V^\pm$ to $tt/tb$ are equal to the quark couplings to $Z/W^\pm$, whereas for ``PC $t$'' these couplings are set to 1. 
    For the \acp{pNGB}, ``weak'' and ``strong $\pi$'' refers to couplings $g_{V\pi\pi}=0$ and $g_{V\pi\pi}=4$, respectively.
    In (a)-(d) the upper limits on the cross sections are taken from direct searches \cite{ATLAS:2019erb,ATLAS:2020lks,ATLAS:2019lsy,ATLAS:2023ibb}.
    In (e) the bounds are derived from recasts of \cite{ATLAS:2021twp,ATLAS:2021fbt,CMS:2019zmd,CMS:2017abv}, respectively.
    The bounds in (f) are derived from recasts of \cite{CMS:2017moi,CMS:2019xjf,Mrowietz:2020ztq,CMS:2019xud,Conte:2021xtt,CMS:2019ius,ATLAS:2020wzf,ATLAS:2022nrp,ATLAS:2021jgw,CMS:2022ubq,ATLAS:2019zci,ATLAS:2019rob}.
    The regions with small $\tilde g \lesssim 2$ are not entirely reliable for scenarios with strong $\pi$ since the resonances are no longer narrow.}
    \label{fig:allbounds_su42su4}
\end{figure}

In the coset $\SU(4) \times \SU(4) / \SU(4)$, pNGB channels are present.  All scans were done with a pNGB mass of $450~$GeV.
This implies that the fermion channels are suppressed at smaller $M_V$ in the scenarios with \textbf{strong} $\boldsymbol{\pi}$.
In \cref{fig:su4su4_pNGBs_bounds}, we show the exclusion bounds derived from the decay into pNGBs.
As mentioned in \cref{sec:pheno}, these do not possess anomaly couplings to SM gauge bosons, except for the singlet $\eta$.
In the fermiophobic scenario this implies
a sizable dependence of the results on the
mass hierarchy of the additional \acp{pNGB}
similar to the case of the SU(5)/SO(5) coset \cite{Cacciapaglia:2022bax}.
This would imply a dependence on unknown parameters of the scalar potential which is beyond the scope of this paper.
Therefore, we consider here only the fermiophilic scenario. 
Compared to the corresponding bounds in $\SU(5)/\SO(5)$, the bounds get continuously stronger for smaller vector masses due to the smaller pNGB mass, avoiding the kinematic cutoff within the scanned parameter space.
The Higgs and gauge bosons channels contribute as is shown in \cref{fig:su4su4_higgs_bounds}, which is relatively similar to the $\SU(5)/\SO(5)$ case.

\clearpage
\section{Three-body decay of $\eta_3^0$}\label{sec:eta30decay}
\begin{figure}[htb]
    \centering
    \begin{subfigure}{0.25\linewidth}
        \includegraphics[width=\linewidth]{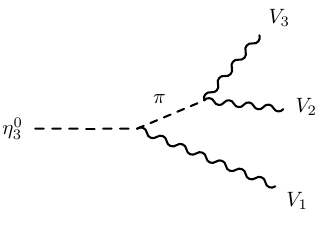}
        \caption{t-channel}
    \end{subfigure}\quad
    \begin{subfigure}{0.25\linewidth}
        \includegraphics[width=\linewidth]{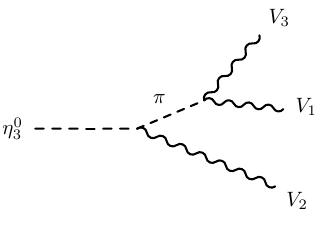}
        \caption{u-channel}
    \end{subfigure}\quad
    \begin{subfigure}{0.25\linewidth}
        \includegraphics[width=\linewidth]{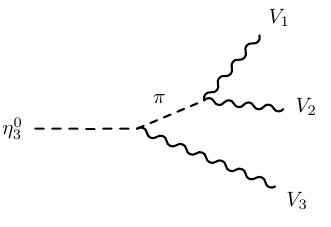}
        \caption{s-channel}
    \end{subfigure}
    \caption{Definition of t, u, s-channel diagrams for the three-body decay of $\eta_3^0$ via an off-shell pNGB $\pi$}
    \label{fig:mandelstam}
\end{figure}

The partial widths can be expressed in terms of two independent functions which we call $f$ for squared contributions and $g$ for interference terms. For further usage, we define the Mandelstam variables analog to \cref{fig:mandelstam} as
\begin{align}
    \bar t = (p_{\eta_3^0} - p_{V_1})^2\,,\qquad \bar u = (p_{\eta_3^0} - p_{V_2})^2 \,, \qquad \bar s = (p_{\eta_3^0} - p_{V_3})^2\,.
\end{align}
For simplicity of notation we use $m = m_{\eta_3^0}$ in the following and absorb the Feynman rules of both vertices into one generic coupling constant
\begin{align}
    \kappa = K^{\pi}_{V_2 V_3}  K^{\eta^3_0 \pi}_{V_1} (K^{\pi'}_{V_2 V_3})^* (K^{\eta_3^0 \pi'}_{V_1})^*\,.
\end{align}
The function $f$ is given as
\begin{align}
    f(V_1, V_2, V_3, \pi, \pi') &= \frac{\kappa}{512 \pi^3 m^3 } \int_{(m_{V_1}+m_{V_2})^2}^{(m - m_{V_3})^2} \mathrm{d}\bar{s} \int_{t_{-}(\bar{s})}^{t_{+}(\bar{s})} \mathrm{d}\bar{t} \,\frac{1}{\bar t - M_\pi^2 + i\Gamma_\pi M_\pi} \frac{1}{\bar t - M_{\pi'}^2 - i\Gamma_{\pi'} M_{\pi'}} \cdot \nonumber\\
    &\quad \left(\frac{1}{m_{V_1}^2} \left(m^2 + m_{V_1}^2 - \bar{t}\right)^2 - 4m^2\right) \cdot \left(\left(\bar{t}-m_{V_2}^4-m_{V_3}^2\right)^2 - 4m_{V_2}^2m_{V_3}^2 \right) \,,
\end{align}
whereas $g$ reads
\begin{align}
    &g(V_1, V_2, V_3, \pi, \pi') = \frac{\kappa}{512 \pi^3 m^3 } \int_{(m_{V_1}+m_{V_2})^2}^{(m - m_{V_3})^2} \mathrm{d}\bar{s} \int_{t_{-}(\bar{s})}^{t_{+}(\bar{s})} \mathrm{d}\bar{t} \,\frac{1}{\bar{t}-M_\pi^2+i\Gamma_\pi M_\pi} \frac{1}{\bar{u}-M_{\pi'}^2-i\Gamma_{\pi'} M_{\pi'}} \cdot  \nonumber \\
    &\quad \Big[
        (\bar{u} - m^2 - m_{V_2}^2) \cdot \Big(
            (\bar{t} - m^2 - m_{V_1}^2)(\bar{t} - m_{V_2}^2 - m_{V_3}^2) +
            (\bar{s} - m^2 - m_{V_3}^2)(\bar{s} - m_{V_1}^2 - m_{V_2}^2) \nonumber \\
            &\qquad -(\bar{u} - m^2 - m_{V_2}^2)(\bar{u} - m_{V_1}^2 - m_{V_3}^2) \Big) \nonumber \\
        &\quad - 2 m^2 (\bar{s} - m_{V_1}^2 - m_{V_2}^2) (\bar{t} - m_{V_2}^2 - m_{V_3}^2)
        - 2 m_{V_2}^2 (\bar{s} - m^2 - m_{V_3}^2) (\bar{t} - m^2 - m_{V_1}^2) \nonumber\\
        &\quad + 4m^2 m_{V_2}^2 (\bar{u} - m_{V_1}^2 - m_{V_3}^2)
    \Big] \,.
\end{align}
Here we integrate first over $\bar{t}$ using the $\bar{s}$ dependent integral bounds
\begin{align}
    t_{\pm} &= \frac{1}{2} \left( m^2 + m_2^2 + m_3^2 + m_4^2 - \bar{s} \right) - \frac{1}{2\bar{s}}\left( m^2 - m_4^2 \right) \left( m_2^2 - m_3^2 \right) \pm \frac{1}{2\bar{s}} \sqrt{\lambda(\bar{s}, m_2^2, m_3^2)} \sqrt{\lambda(\bar{s}, m^2, m_4^2)} \,,
\end{align}
with
\begin{align}
    \lambda(\bar{s},y,z) &= \bar{s}^2+y^2+z^2 - 2\bar{s}y - 2\bar{s}z - 2yz \,.
\end{align}

Combing all contributions the partial width $\Gamma$ of $\eta_3^0 \to W^+ + W^- + Z$ can be written as
\begin{align}
    \Gamma(W^+,W^-,Z) &= 2 f(W, W, Z,\eta_3^+,\eta_3^-) + 4 \Re[f(W, W, Z,\eta_3^+,\eta_5^-)] + 2 f(W, W, Z,\eta_5^+,\eta_5^-) \nonumber\\
    &+ 2 \Re\big[ g(W, W, Z,\eta_3^+,\eta_3^-) + g(W, W, Z,\eta_3^+,\eta_5^-)\big] \nonumber\\
    &+ 2\Re\big[g(W, W, Z,\eta_5^+,\eta_3^-) + g(W, W, Z,\eta_5^+,\eta_5^-) \big] \,,
\end{align}
where we neglected diagrams involving $\eta_{1,5}^0 \to W^+ W^-$ as those couplings are heavily suppressed by $\sin^2 \theta$.
Analogously, the expression for $\eta_3^0 \to W^+ + W^- + \gamma$ is obtained by replacing $Z$ with $\gamma$.

The neutral channels can be expressed as
\begin{align}
    \Gamma(Z,\gamma,\gamma) &= 2 f(Z, \gamma, \gamma,\eta_1^0,\eta_1^0) + 4 \Re[f(Z, \gamma, \gamma,\eta_1^0,\eta_5^0)] + 2 f(Z, \gamma, \gamma,\eta_5^0,\eta_5^0)\,, \\
    \Gamma(Z,Z,\gamma) &= f(Z, Z, \gamma,\eta_1^0,\eta_1^0) + 2 \Re[f(Z, Z, \gamma,\eta_1^0,\eta_5^0)] + f(Z, Z, \gamma,\eta_5^0,\eta_5^0) \nonumber\\
    &+ \Re\big[ g(Z, Z, \gamma, \eta_1^0,\eta_1^0) + g(Z, Z, \gamma, \eta_5^0,\eta_1^0) + g(Z, Z, \gamma, \eta_1^0,\eta_5^0) + g(Z, Z, \gamma, \eta_5^0,\eta_5^0) \big]\,, \\
    \Gamma(Z,Z,Z) &= 2 f(Z, Z, Z,\eta_1^0,\eta_1^0) + 4 \Re[f(Z, Z, Z,\eta_1^0,\eta_5^0)] + 2 f(Z, Z, Z,\eta_5^0,\eta_5^0) \nonumber\\
    &+ 4\Re\big[ g(Z, Z, Z, \eta_1^0,\eta_1^0) + g(Z, Z, Z, \eta_5^0,\eta_1^0) + g(Z, Z, Z, \eta_1^0,\eta_5^0) + g(Z, Z, Z, \eta_5^0,\eta_5^0) \big] \,.
\end{align}
For $M_\pi=700\,\mathrm{GeV}$, we calculate a total width of $7.2\cdot 10^{-9}\,\mathrm{GeV}$ with branching ratios as in \cref{tab:eta30decay}.
\begin{table}[tbh]
    \centering
    \begin{tabular}{c|ccccc}
        channel & $W^+ W^- \gamma$ & $W^+ W^- Z$ & $Z \gamma \gamma$ & $Z Z \gamma$ & $Z Z Z$ \\
        \hline
        BR [\%] & 4.3 & 70.7 & 1.5 & 1.0 & 22.5
    \end{tabular}
    \caption{The branching ratios (BR) of three-body decays via WZW terms of $\eta_3^0$ using $M_\pi =700\,\mathrm{GeV}$ and a mass splitting of $\Delta=2\,\mathrm{GeV}$.}
    \label{tab:eta30decay}
\end{table}

\clearpage

\bibliographystyle{utphys}
\bibliography{main}

\providecommand{\href}[2]{#2}\begingroup\raggedright\begin{thebibliography}{100}

\bibitem{Englert:1964et}
F.~Englert and R.~Brout, ``{Broken Symmetry and the Mass of Gauge Vector
  Mesons},'' \href{https://dx.doi.org/10.1103/PhysRevLett.13.321}{{\em Phys.
  Rev. Lett.} {\bfseries 13} (1964) 321--323}.

\bibitem{Higgs:1964pj}
P.~W. Higgs, ``{Broken Symmetries and the Masses of Gauge Bosons},''
  \href{https://dx.doi.org/10.1103/PhysRevLett.13.508}{{\em Phys. Rev. Lett.}
  {\bfseries 13} (1964) 508--509}.

\bibitem{Guralnik:1964eu}
G.~S. Guralnik, C.~R. Hagen, and T.~W.~B. Kibble, ``{Global Conservation Laws
  and Massless Particles},''
  \href{https://dx.doi.org/10.1103/PhysRevLett.13.585}{{\em Phys. Rev. Lett.}
  {\bfseries 13} (1964) 585--587}.

\bibitem{Weinberg:1975gm}
S.~Weinberg, ``{Implications of Dynamical Symmetry Breaking},''
  \href{https://dx.doi.org/10.1103/PhysRevD.19.1277}{{\em Phys. Rev. D}
  {\bfseries 13} (1976) 974--996}. [Addendum: Phys.Rev.D 19, 1277--1280
  (1979)].

\bibitem{Dimopoulos:1979es}
S.~Dimopoulos and L.~Susskind, ``{Mass Without Scalars},''
  \href{https://dx.doi.org/10.1016/0550-3213(79)90364-X}{{\em Nucl. Phys. B}
  {\bfseries 155} (1979) 237--252}.

\bibitem{Dimopoulos:1979za}
S.~Dimopoulos, L.~Susskind, and S.~Raby, ``{Technicolour},''
  \href{https://dx.doi.org/10.1063/1.32174}{{\em AIP Conf. Proc.} {\bfseries
  59} (1980) 407--452}.

\bibitem{Kaplan:1983fs}
D.~B. Kaplan and H.~Georgi, ``{SU(2) x U(1) Breaking by Vacuum Misalignment},''
  \href{https://dx.doi.org/10.1016/0370-2693(84)91177-8}{{\em Phys. Lett. B}
  {\bfseries 136} (1984) 183--186}.

\bibitem{Kaplan:1983sm}
D.~B. Kaplan, H.~Georgi, and S.~Dimopoulos, ``{Composite Higgs Scalars},''
  \href{https://dx.doi.org/10.1016/0370-2693(84)91178-X}{{\em Phys. Lett. B}
  {\bfseries 136} (1984) 187--190}.

\bibitem{Kaplan:1991dc}
D.~B. Kaplan, ``{Flavor at SSC energies: A New mechanism for dynamically
  generated fermion masses},''
  \href{https://dx.doi.org/10.1016/S0550-3213(05)80021-5}{{\em Nucl. Phys. B}
  {\bfseries 365} (1991) 259--278}.

\bibitem{Contino:2003ve}
R.~Contino, Y.~Nomura, and A.~Pomarol, ``{Higgs as a holographic
  pseudoGoldstone boson},''
  \href{https://dx.doi.org/10.1016/j.nuclphysb.2003.08.027}{{\em Nucl. Phys. B}
  {\bfseries 671} (2003) 148--174},
  \href{https://arxiv.org/abs/hep-ph/0306259}{{\ttfamily
  arXiv:hep-ph/0306259}}.

\bibitem{Agashe:2004rs}
K.~Agashe, R.~Contino, and A.~Pomarol, ``{The Minimal composite Higgs model},''
  \href{https://dx.doi.org/10.1016/j.nuclphysb.2005.04.035}{{\em Nucl. Phys. B}
  {\bfseries 719} (2005) 165--187},
  \href{https://arxiv.org/abs/hep-ph/0412089}{{\ttfamily
  arXiv:hep-ph/0412089}}.

\bibitem{Hosotani:2005nz}
Y.~Hosotani and M.~Mabe, ``{Higgs boson mass and electroweak-gravity hierarchy
  from dynamical gauge-Higgs unification in the warped spacetime},''
  \href{https://dx.doi.org/10.1016/j.physletb.2005.04.039}{{\em Phys. Lett. B}
  {\bfseries 615} (2005) 257--265},
  \href{https://arxiv.org/abs/hep-ph/0503020}{{\ttfamily
  arXiv:hep-ph/0503020}}.

\bibitem{Maldacena:1997re}
J.~M. Maldacena, ``{The Large N limit of superconformal field theories and
  supergravity},'' \href{https://dx.doi.org/10.4310/ATMP.1998.v2.n2.a1}{{\em
  Adv. Theor. Math. Phys.} {\bfseries 2} (1998) 231--252},
  \href{https://arxiv.org/abs/hep-th/9711200}{{\ttfamily
  arXiv:hep-th/9711200}}.

\bibitem{Contino:2010rs}
R.~Contino, \href{https://dx.doi.org/10.1142/9789814327183_0005}{``{The Higgs
  as a Composite Nambu-Goldstone Boson},''} in {\em {Theoretical Advanced Study
  Institute in Elementary Particle Physics}: {Physics of the Large and the
  Small}}, pp.~235--306.
\newblock 2011.
\newblock \href{https://arxiv.org/abs/1005.4269}{{\ttfamily arXiv:1005.4269
  [hep-ph]}}.

\bibitem{Cacciapaglia:2020kgq}
G.~Cacciapaglia, C.~Pica, and F.~Sannino, ``{Fundamental Composite Dynamics: A
  Review},'' \href{https://dx.doi.org/10.1016/j.physrep.2020.07.002}{{\em Phys.
  Rept.} {\bfseries 877} (2020) 1--70},
  \href{https://arxiv.org/abs/2002.04914}{{\ttfamily arXiv:2002.04914
  [hep-ph]}}.

\bibitem{Vecchi:2015fma}
L.~Vecchi, ``{A dangerous irrelevant UV-completion of the composite Higgs},''
  \href{https://dx.doi.org/10.1007/JHEP02(2017)094}{{\em JHEP} {\bfseries 02}
  (2017) 094}, \href{https://arxiv.org/abs/1506.00623}{{\ttfamily
  arXiv:1506.00623 [hep-ph]}}.

\bibitem{Ferretti:2013kya}
G.~Ferretti and D.~Karateev, ``{Fermionic UV completions of Composite Higgs
  models},'' \href{https://dx.doi.org/10.1007/JHEP03(2014)077}{{\em JHEP}
  {\bfseries 03} (2014) 077}, \href{https://arxiv.org/abs/1312.5330}{{\ttfamily
  arXiv:1312.5330 [hep-ph]}}.

\bibitem{Barnard:2013zea}
J.~Barnard, T.~Gherghetta, and T.~S. Ray, ``{UV descriptions of composite Higgs
  models without elementary scalars},''
  \href{https://dx.doi.org/10.1007/JHEP02(2014)002}{{\em JHEP} {\bfseries 02}
  (2014) 002}, \href{https://arxiv.org/abs/1311.6562}{{\ttfamily
  arXiv:1311.6562 [hep-ph]}}.

\bibitem{Cacciapaglia:2021uqh}
G.~Cacciapaglia, T.~Flacke, M.~Kunkel, and W.~Porod, ``{Phenomenology of
  unusual top partners in composite Higgs models},''
  \href{https://dx.doi.org/10.1007/JHEP02(2022)208}{{\em JHEP} {\bfseries 02}
  (2022) 208}, \href{https://arxiv.org/abs/2112.00019}{{\ttfamily
  arXiv:2112.00019 [hep-ph]}}.

\bibitem{Ferretti:2016upr}
G.~Ferretti, ``{Gauge theories of Partial Compositeness: Scenarios for Run-II
  of the LHC},'' \href{https://dx.doi.org/10.1007/JHEP06(2016)107}{{\em JHEP}
  {\bfseries 06} (2016) 107},
  \href{https://arxiv.org/abs/1604.06467}{{\ttfamily arXiv:1604.06467
  [hep-ph]}}.

\bibitem{Belyaev:2016ftv}
A.~Belyaev, G.~Cacciapaglia, H.~Cai, G.~Ferretti, T.~Flacke, A.~Parolini, and
  H.~Serodio, ``{Di-boson signatures as Standard Candles for Partial
  Compositeness},'' \href{https://dx.doi.org/10.1007/JHEP01(2017)094}{{\em
  JHEP} {\bfseries 01} (2017) 094},
  \href{https://arxiv.org/abs/1610.06591}{{\ttfamily arXiv:1610.06591
  [hep-ph]}}. [Erratum: JHEP 12, 088 (2017)].

\bibitem{Ayyar:2017uqh}
V.~Ayyar, T.~DeGrand, D.~C. Hackett, W.~I. Jay, E.~T. Neil, Y.~Shamir, and
  B.~Svetitsky, ``{Chiral Transition of SU(4) Gauge Theory with Fermions in
  Multiple Representations},''
  \href{https://dx.doi.org/10.1051/epjconf/201817508026}{{\em EPJ Web Conf.}
  {\bfseries 175} (2018) 08026},
  \href{https://arxiv.org/abs/1709.06190}{{\ttfamily arXiv:1709.06190
  [hep-lat]}}.

\bibitem{Bennett:2023wjw}
E.~Bennett, J.~Holligan, D.~K. Hong, H.~Hsiao, J.-W. Lee, C.~J.~D. Lin,
  B.~Lucini, M.~Mesiti, M.~Piai, and D.~Vadacchino, ``{Sp(2N) Lattice Gauge
  Theories and Extensions of the Standard Model of Particle Physics},''
  \href{https://dx.doi.org/10.3390/universe9050236}{{\em Universe} {\bfseries
  9} no.~5, (2023) 236}, \href{https://arxiv.org/abs/2304.01070}{{\ttfamily
  arXiv:2304.01070 [hep-lat]}}.

\bibitem{Agugliaro:2018vsu}
A.~Agugliaro, G.~Cacciapaglia, A.~Deandrea, and S.~De~Curtis, ``{Vacuum
  misalignment and pattern of scalar masses in the SU(5)/SO(5) composite Higgs
  model},'' \href{https://dx.doi.org/10.1007/JHEP02(2019)089}{{\em JHEP}
  {\bfseries 02} (2019) 089},
  \href{https://arxiv.org/abs/1808.10175}{{\ttfamily arXiv:1808.10175
  [hep-ph]}}.

\bibitem{Cacciapaglia:2022bax}
G.~Cacciapaglia, T.~Flacke, M.~Kunkel, W.~Porod, and L.~Schwarze, ``{Exploring
  extended Higgs sectors via pair production at the LHC},''
  \href{https://dx.doi.org/10.1007/JHEP12(2022)087}{{\em JHEP} {\bfseries 12}
  (2022) 087}, \href{https://arxiv.org/abs/2210.01826}{{\ttfamily
  arXiv:2210.01826 [hep-ph]}}.

\bibitem{Flacke:2023eil}
T.~Flacke, J.~H. Kim, M.~Kunkel, P.~Ko, J.~S. Pi, W.~Porod, and L.~Schwarze,
  ``{Uncovering doubly charged scalars with dominant three-body decays using
  machine learning},'' \href{https://dx.doi.org/10.1007/JHEP11(2023)009}{{\em
  JHEP} {\bfseries 11} (2023) 009},
  \href{https://arxiv.org/abs/2304.09195}{{\ttfamily arXiv:2304.09195
  [hep-ph]}}.

\bibitem{Cacciapaglia:2019bqz}
G.~Cacciapaglia, G.~Ferretti, T.~Flacke, and H.~Ser\^odio, ``{Light scalars in
  composite Higgs models},''
  \href{https://dx.doi.org/10.3389/fphy.2019.00022}{{\em Front. in Phys.}
  {\bfseries 7} (2019) 22}, \href{https://arxiv.org/abs/1902.06890}{{\ttfamily
  arXiv:1902.06890 [hep-ph]}}.

\bibitem{BuarqueFranzosi:2021kky}
D.~Buarque~Franzosi, G.~Cacciapaglia, X.~Cid~Vidal, G.~Ferretti, T.~Flacke, and
  C.~V\'azquez~Sierra, ``{Exploring new possibilities to discover a light
  pseudo-scalar at LHCb},''
  \href{https://dx.doi.org/10.1140/epjc/s10052-021-09930-y}{{\em Eur. Phys. J.
  C} {\bfseries 82} no.~1, (2022) 3},
  \href{https://arxiv.org/abs/2106.12615}{{\ttfamily arXiv:2106.12615
  [hep-ph]}}.

\bibitem{Cacciapaglia:2015eqa}
G.~Cacciapaglia, H.~Cai, A.~Deandrea, T.~Flacke, S.~J. Lee, and A.~Parolini,
  ``{Composite scalars at the LHC: the Higgs, the Sextet and the Octet},''
  \href{https://dx.doi.org/10.1007/JHEP11(2015)201}{{\em JHEP} {\bfseries 11}
  (2015) 201}, \href{https://arxiv.org/abs/1507.02283}{{\ttfamily
  arXiv:1507.02283 [hep-ph]}}.

\bibitem{Cacciapaglia:2020vyf}
G.~Cacciapaglia, A.~Deandrea, T.~Flacke, and A.~M. Iyer, ``{Gluon-Photon
  Signatures for color octet at the LHC (and beyond)},''
  \href{https://dx.doi.org/10.1007/JHEP05(2020)027}{{\em JHEP} {\bfseries 05}
  (2020) 027}, \href{https://arxiv.org/abs/2002.01474}{{\ttfamily
  arXiv:2002.01474 [hep-ph]}}.

\bibitem{Bizot:2018tds}
N.~Bizot, G.~Cacciapaglia, and T.~Flacke, ``{Common exotic decays of top
  partners},'' \href{https://dx.doi.org/10.1007/JHEP06(2018)065}{{\em JHEP}
  {\bfseries 06} (2018) 065},
  \href{https://arxiv.org/abs/1803.00021}{{\ttfamily arXiv:1803.00021
  [hep-ph]}}.

\bibitem{Xie:2019gya}
K.-P. Xie, G.~Cacciapaglia, and T.~Flacke, ``{Exotic decays of top partners
  with charge 5/3: bounds and opportunities},''
  \href{https://dx.doi.org/10.1007/JHEP10(2019)134}{{\em JHEP} {\bfseries 10}
  (2019) 134}, \href{https://arxiv.org/abs/1907.05894}{{\ttfamily
  arXiv:1907.05894 [hep-ph]}}.

\bibitem{Cacciapaglia:2019zmj}
G.~Cacciapaglia, T.~Flacke, M.~Park, and M.~Zhang, ``{Exotic decays of top
  partners: mind the search gap},''
  \href{https://dx.doi.org/10.1016/j.physletb.2019.135015}{{\em Phys. Lett. B}
  {\bfseries 798} (2019) 135015},
  \href{https://arxiv.org/abs/1908.07524}{{\ttfamily arXiv:1908.07524
  [hep-ph]}}.

\bibitem{BuarqueFranzosi:2016ooy}
D.~Buarque~Franzosi, G.~Cacciapaglia, H.~Cai, A.~Deandrea, and M.~Frandsen,
  ``{Vector and Axial-vector resonances in composite models of the Higgs
  boson},'' \href{https://dx.doi.org/10.1007/JHEP11(2016)076}{{\em JHEP}
  {\bfseries 11} (2016) 076},
  \href{https://arxiv.org/abs/1605.01363}{{\ttfamily arXiv:1605.01363
  [hep-ph]}}.

\bibitem{Cacciapaglia:2024wdn}
G.~Cacciapaglia, A.~Deandrea, M.~Kunkel, and W.~Porod, ``{Coloured spin-1
  states in composite Higgs models},''
  \href{https://dx.doi.org/10.1007/JHEP06(2024)092}{{\em JHEP} {\bfseries 06}
  (2024) 092}, \href{https://arxiv.org/abs/2404.02198}{{\ttfamily
  arXiv:2404.02198 [hep-ph]}}.

\bibitem{Bennett:2017kga}
E.~Bennett, D.~K. Hong, J.-W. Lee, C.~J.~D. Lin, B.~Lucini, M.~Piai, and
  D.~Vadacchino, ``{Sp(4) gauge theory on the lattice: towards SU(4)/Sp(4)
  composite Higgs (and beyond)},''
  \href{https://dx.doi.org/10.1007/JHEP03(2018)185}{{\em JHEP} {\bfseries 03}
  (2018) 185}, \href{https://arxiv.org/abs/1712.04220}{{\ttfamily
  arXiv:1712.04220 [hep-lat]}}.

\bibitem{Bennett:2019cxd}
E.~Bennett, D.~K. Hong, J.-W. Lee, C.-J.~D. Lin, B.~Lucini, M.~Mesiti, M.~Piai,
  J.~Rantaharju, and D.~Vadacchino, ``{$Sp(4)$ gauge theories on the lattice:
  quenched fundamental and antisymmetric fermions},''
  \href{https://dx.doi.org/10.1103/PhysRevD.101.074516}{{\em Phys. Rev. D}
  {\bfseries 101} no.~7, (2020) 074516},
  \href{https://arxiv.org/abs/1912.06505}{{\ttfamily arXiv:1912.06505
  [hep-lat]}}.

\bibitem{Bennett:2019jzz}
E.~Bennett, D.~K. Hong, J.-W. Lee, C.~J.~D. Lin, B.~Lucini, M.~Piai, and
  D.~Vadacchino, ``{Sp(4) gauge theories on the lattice: $N_f=2$ dynamical
  fundamental fermions},''
  \href{https://dx.doi.org/10.1007/JHEP12(2019)053}{{\em JHEP} {\bfseries 12}
  (2019) 053}, \href{https://arxiv.org/abs/1909.12662}{{\ttfamily
  arXiv:1909.12662 [hep-lat]}}.

\bibitem{Bennett:2020hqd}
E.~Bennett, J.~Holligan, D.~K. Hong, J.-W. Lee, C.~J.~D. Lin, B.~Lucini,
  M.~Piai, and D.~Vadacchino, ``{Color dependence of tensor and scalar glueball
  masses in Yang-Mills theories},''
  \href{https://dx.doi.org/10.1103/PhysRevD.102.011501}{{\em Phys. Rev. D}
  {\bfseries 102} no.~1, (2020) 011501},
  \href{https://arxiv.org/abs/2004.11063}{{\ttfamily arXiv:2004.11063
  [hep-lat]}}.

\bibitem{Bennett:2020qtj}
E.~Bennett, J.~Holligan, D.~K. Hong, J.-W. Lee, C.~J.~D. Lin, B.~Lucini,
  M.~Piai, and D.~Vadacchino, ``{Glueballs and strings in $Sp(2N)$ Yang-Mills
  theories},'' \href{https://dx.doi.org/10.1103/PhysRevD.103.054509}{{\em Phys.
  Rev. D} {\bfseries 103} no.~5, (2021) 054509},
  \href{https://arxiv.org/abs/2010.15781}{{\ttfamily arXiv:2010.15781
  [hep-lat]}}.

\bibitem{Bennett:2022yfa}
E.~Bennett, D.~K. Hong, H.~Hsiao, J.-W. Lee, C.~J.~D. Lin, B.~Lucini,
  M.~Mesiti, M.~Piai, and D.~Vadacchino, ``{Lattice studies of the Sp(4) gauge
  theory with two fundamental and three antisymmetric Dirac fermions},''
  \href{https://dx.doi.org/10.1103/PhysRevD.106.014501}{{\em Phys. Rev. D}
  {\bfseries 106} no.~1, (2022) 014501},
  \href{https://arxiv.org/abs/2202.05516}{{\ttfamily arXiv:2202.05516
  [hep-lat]}}.

\bibitem{Kulkarni:2022bvh}
S.~Kulkarni, A.~Maas, S.~Mee, M.~Nikolic, J.~Pradler, and F.~Zierler,
  ``{Low-energy effective description of dark $Sp(4)$ theories},''
  \href{https://dx.doi.org/10.21468/SciPostPhys.14.3.044}{{\em SciPost Phys.}
  {\bfseries 14} no.~3, (2023) 044},
  \href{https://arxiv.org/abs/2202.05191}{{\ttfamily arXiv:2202.05191
  [hep-ph]}}.

\bibitem{Bennett:2023gbe}
E.~Bennett {\em et~al.}, ``{Symplectic lattice gauge theories in the grid
  framework: Approaching the conformal window},''
  \href{https://dx.doi.org/10.1103/PhysRevD.108.094508}{{\em Phys. Rev. D}
  {\bfseries 108} no.~9, (2023) 094508},
  \href{https://arxiv.org/abs/2306.11649}{{\ttfamily arXiv:2306.11649
  [hep-lat]}}.

\bibitem{Bennett:2023mhh}
E.~Bennett, D.~K. Hong, H.~Hsiao, J.-W. Lee, C.~J.~D. Lin, B.~Lucini, M.~Piai,
  and D.~Vadacchino, ``{Lattice investigations of the chimera baryon spectrum
  in the Sp(4) gauge theory},''
  \href{https://dx.doi.org/10.1103/PhysRevD.109.094512}{{\em Phys. Rev. D}
  {\bfseries 109} no.~9, (2024) 094512},
  \href{https://arxiv.org/abs/2311.14663}{{\ttfamily arXiv:2311.14663
  [hep-lat]}}.

\bibitem{Bennett:2023qwx}
E.~Bennett, J.~Holligan, D.~K. Hong, J.-W. Lee, C.~J.~D. Lin, B.~Lucini,
  M.~Piai, and D.~Vadacchino, ``{Spectrum of mesons in quenched Sp(2N) gauge
  theories},'' \href{https://dx.doi.org/10.1103/PhysRevD.109.094517}{{\em Phys.
  Rev. D} {\bfseries 109} no.~9, (2024) 094517},
  \href{https://arxiv.org/abs/2312.08465}{{\ttfamily arXiv:2312.08465
  [hep-lat]}}.

\bibitem{Bennett:2024wda}
E.~Bennett, N.~Forzano, D.~K. Hong, H.~Hsiao, J.-W. Lee, C.~J.~D. Lin,
  B.~Lucini, M.~Piai, D.~Vadacchino, and F.~Zierler, ``{Mixing between flavor
  singlets in lattice gauge theories coupled to matter fields in multiple
  representations},''
  \href{https://dx.doi.org/10.1103/PhysRevD.110.074504}{{\em Phys. Rev. D}
  {\bfseries 110} no.~7, (2024) 074504},
  \href{https://arxiv.org/abs/2405.05765}{{\ttfamily arXiv:2405.05765
  [hep-lat]}}.

\bibitem{Ayyar:2017qdf}
V.~Ayyar, T.~DeGrand, M.~Golterman, D.~C. Hackett, W.~I. Jay, E.~T. Neil,
  Y.~Shamir, and B.~Svetitsky, ``{Spectroscopy of SU(4) composite Higgs theory
  with two distinct fermion representations},''
  \href{https://dx.doi.org/10.1103/PhysRevD.97.074505}{{\em Phys. Rev. D}
  {\bfseries 97} no.~7, (2018) 074505},
  \href{https://arxiv.org/abs/1710.00806}{{\ttfamily arXiv:1710.00806
  [hep-lat]}}.

\bibitem{Ayyar:2018glg}
V.~Ayyar, T.~DeGrand, D.~C. Hackett, W.~I. Jay, E.~T. Neil, Y.~Shamir, and
  B.~Svetitsky, ``{Partial compositeness and baryon matrix elements on the
  lattice},'' \href{https://dx.doi.org/10.1103/PhysRevD.99.094502}{{\em Phys.
  Rev. D} {\bfseries 99} no.~9, (2019) 094502},
  \href{https://arxiv.org/abs/1812.02727}{{\ttfamily arXiv:1812.02727
  [hep-ph]}}.

\bibitem{Ayyar:2018ppa}
V.~Ayyar, T.~DeGrand, D.~C. Hackett, W.~I. Jay, E.~T. Neil, Y.~Shamir, and
  B.~Svetitsky, ``{Finite-temperature phase structure of SU(4) gauge theory
  with multiple fermion representations},''
  \href{https://dx.doi.org/10.1103/PhysRevD.97.114502}{{\em Phys. Rev. D}
  {\bfseries 97} no.~11, (2018) 114502},
  \href{https://arxiv.org/abs/1802.09644}{{\ttfamily arXiv:1802.09644
  [hep-lat]}}.

\bibitem{Ayyar:2018zuk}
V.~Ayyar, T.~Degrand, D.~C. Hackett, W.~I. Jay, E.~T. Neil, Y.~Shamir, and
  B.~Svetitsky, ``{Baryon spectrum of SU(4) composite Higgs theory with two
  distinct fermion representations},''
  \href{https://dx.doi.org/10.1103/PhysRevD.97.114505}{{\em Phys. Rev. D}
  {\bfseries 97} no.~11, (2018) 114505},
  \href{https://arxiv.org/abs/1801.05809}{{\ttfamily arXiv:1801.05809
  [hep-ph]}}.

\bibitem{Ayyar:2019exp}
V.~Ayyar, M.~F. Golterman, D.~C. Hackett, W.~Jay, E.~T. Neil, Y.~Shamir, and
  B.~Svetitsky, ``{Radiative Contribution to the Composite-Higgs Potential in a
  Two-Representation Lattice Model},''
  \href{https://dx.doi.org/10.1103/PhysRevD.99.094504}{{\em Phys. Rev. D}
  {\bfseries 99} no.~9, (2019) 094504},
  \href{https://arxiv.org/abs/1903.02535}{{\ttfamily arXiv:1903.02535
  [hep-lat]}}.

\bibitem{Golterman:2020pyx}
M.~Golterman, W.~I. Jay, E.~T. Neil, Y.~Shamir, and B.~Svetitsky, ``{Low-energy
  constant $L_{10}$ in a two-representation lattice theory},''
  \href{https://dx.doi.org/10.1103/PhysRevD.103.074509}{{\em Phys. Rev. D}
  {\bfseries 103} no.~7, (2021) 074509},
  \href{https://arxiv.org/abs/2010.01920}{{\ttfamily arXiv:2010.01920
  [hep-lat]}}.

\bibitem{Hasenfratz:2023sqa}
A.~Hasenfratz, E.~T. Neil, Y.~Shamir, B.~Svetitsky, and O.~Witzel, ``{Infrared
  fixed point and anomalous dimensions in a composite Higgs model},''
  \href{https://dx.doi.org/10.1103/PhysRevD.107.114504}{{\em Phys. Rev. D}
  {\bfseries 107} no.~11, (2023) 114504},
  \href{https://arxiv.org/abs/2304.11729}{{\ttfamily arXiv:2304.11729
  [hep-lat]}}.

\bibitem{Erdmenger:2020lvq}
J.~Erdmenger, N.~Evans, W.~Porod, and K.~S. Rigatos, ``{Gauge/gravity dynamics
  for composite Higgs models and the top mass},''
  \href{https://dx.doi.org/10.1103/PhysRevLett.126.071602}{{\em Phys. Rev.
  Lett.} {\bfseries 126} no.~7, (2021) 071602},
  \href{https://arxiv.org/abs/2009.10737}{{\ttfamily arXiv:2009.10737
  [hep-ph]}}.

\bibitem{Erdmenger:2020flu}
J.~Erdmenger, N.~Evans, W.~Porod, and K.~S. Rigatos, ``{Gauge/gravity dual
  dynamics for the strongly coupled sector of composite Higgs models},''
  \href{https://dx.doi.org/10.1007/JHEP02(2021)058}{{\em JHEP} {\bfseries 02}
  (2021) 058}, \href{https://arxiv.org/abs/2010.10279}{{\ttfamily
  arXiv:2010.10279 [hep-ph]}}.

\bibitem{Elander:2020nyd}
D.~Elander, M.~Frigerio, M.~Knecht, and J.-L. Kneur, ``{Holographic models of
  composite Higgs in the Veneziano limit. Part I. Bosonic sector},''
  \href{https://dx.doi.org/10.1007/JHEP03(2021)182}{{\em JHEP} {\bfseries 03}
  (2021) 182}, \href{https://arxiv.org/abs/2011.03003}{{\ttfamily
  arXiv:2011.03003 [hep-ph]}}.

\bibitem{Elander:2021bmt}
D.~Elander, M.~Frigerio, M.~Knecht, and J.-L. Kneur, ``{Holographic models of
  composite Higgs in the Veneziano limit. Part II. Fermionic sector},''
  \href{https://dx.doi.org/10.1007/JHEP05(2022)066}{{\em JHEP} {\bfseries 05}
  (2022) 066}, \href{https://arxiv.org/abs/2112.14740}{{\ttfamily
  arXiv:2112.14740 [hep-ph]}}.

\bibitem{Elander:2023aow}
D.~Elander, A.~Fatemiabhari, and M.~Piai, ``{Toward minimal composite Higgs
  models from regular geometries in bottom-up holography},''
  \href{https://dx.doi.org/10.1103/PhysRevD.107.115021}{{\em Phys. Rev. D}
  {\bfseries 107} no.~11, (2023) 115021},
  \href{https://arxiv.org/abs/2303.00541}{{\ttfamily arXiv:2303.00541
  [hep-th]}}.

\bibitem{Erdmenger:2023hkl}
J.~Erdmenger, N.~Evans, Y.~Liu, and W.~Porod, ``{Holographic Non-Abelian
  Flavour Symmetry Breaking},''
  \href{https://dx.doi.org/10.3390/universe9060289}{{\em Universe} {\bfseries
  9} no.~6, (2023) 289}, \href{https://arxiv.org/abs/2304.09190}{{\ttfamily
  arXiv:2304.09190 [hep-th]}}.

\bibitem{Erdmenger:2024dxf}
J.~Erdmenger, N.~Evans, Y.~Liu, and W.~Porod, ``{Holography for Sp(2N$_{c}$)
  gauge dynamics: from composite Higgs to technicolour},''
  \href{https://dx.doi.org/10.1007/JHEP07(2024)169}{{\em JHEP} {\bfseries 07}
  (2024) 169}, \href{https://arxiv.org/abs/2404.14480}{{\ttfamily
  arXiv:2404.14480 [hep-ph]}}.

\bibitem{Belyaev:2008yj}
A.~Belyaev, R.~Foadi, M.~T. Frandsen, M.~Jarvinen, F.~Sannino, and A.~Pukhov,
  ``{Technicolor Walks at the LHC},''
  \href{https://dx.doi.org/10.1103/PhysRevD.79.035006}{{\em Phys. Rev. D}
  {\bfseries 79} (2009) 035006},
  \href{https://arxiv.org/abs/0809.0793}{{\ttfamily arXiv:0809.0793 [hep-ph]}}.

\bibitem{Becciolini:2014eba}
D.~Becciolini, D.~Buarque~Franzosi, R.~Foadi, M.~T. Frandsen, T.~Hapola, and
  F.~Sannino, ``{Custodial Vector Model},''
  \href{https://dx.doi.org/10.1103/PhysRevD.92.015013}{{\em Phys. Rev. D}
  {\bfseries 92} no.~1, (2015) 015013},
  \href{https://arxiv.org/abs/1410.6492}{{\ttfamily arXiv:1410.6492 [hep-ph]}}.
  [Addendum: Phys.Rev.D 92, 079904 (2015)].

\bibitem{BuarqueFranzosi:2023xux}
D.~Buarque~Franzosi, ``{Towards the precise description of Composite Higgs
  models at colliders},'' \href{https://arxiv.org/abs/2302.02422}{{\ttfamily
  arXiv:2302.02422 [hep-ph]}}.

\bibitem{Bando:1987br}
M.~Bando, T.~Kugo, and K.~Yamawaki, ``{Nonlinear Realization and Hidden Local
  Symmetries},'' \href{https://dx.doi.org/10.1016/0370-1573(88)90019-1}{{\em
  Phys. Rept.} {\bfseries 164} (1988) 217--314}.

\bibitem{Coleman:1969sm}
S.~R. Coleman, J.~Wess, and B.~Zumino, ``{Structure of phenomenological
  Lagrangians. 1.},'' \href{https://dx.doi.org/10.1103/PhysRev.177.2239}{{\em
  Phys. Rev.} {\bfseries 177} (1969) 2239--2247}.

\bibitem{Callan:1969sn}
C.~G. Callan, Jr., S.~R. Coleman, J.~Wess, and B.~Zumino, ``{Structure of
  phenomenological Lagrangians. 2.},''
  \href{https://dx.doi.org/10.1103/PhysRev.177.2247}{{\em Phys. Rev.}
  {\bfseries 177} (1969) 2247--2250}.

\bibitem{2nd_paper}
R.~Caliri, J.~Hadlik, M.~Kunkel, and W.~Porod.
\newblock In preparation.

\bibitem{Banerjee:2022izw}
A.~Banerjee, D.~B. Franzosi, and G.~Ferretti, ``{Modelling vector-like quarks
  in partial compositeness framework},''
  \href{https://dx.doi.org/10.1007/JHEP03(2022)200}{{\em JHEP} {\bfseries 03}
  (2022) 200}, \href{https://arxiv.org/abs/2202.00037}{{\ttfamily
  arXiv:2202.00037 [hep-ph]}}.

\bibitem{ATLAS:2019erb}
{\bfseries ATLAS} Collaboration, G.~Aad {\em et~al.}, ``{Search for high-mass
  dilepton resonances using 139 fb$^{-1}$ of $pp$ collision data collected at
  $\sqrt{s}=$13 TeV with the ATLAS detector},''
  \href{https://dx.doi.org/10.1016/j.physletb.2019.07.016}{{\em Phys. Lett. B}
  {\bfseries 796} (2019) 68--87},
  \href{https://arxiv.org/abs/1903.06248}{{\ttfamily arXiv:1903.06248
  [hep-ex]}}.

\bibitem{ATLAS:2020lks}
{\bfseries ATLAS} Collaboration, G.~Aad {\em et~al.}, ``{Search for $
  t\overline{t} $ resonances in fully hadronic final states in $pp$ collisions
  at $ \sqrt{s} $ = 13 TeV with the ATLAS detector},''
  \href{https://dx.doi.org/10.1007/JHEP10(2020)061}{{\em JHEP} {\bfseries 10}
  (2020) 061}, \href{https://arxiv.org/abs/2005.05138}{{\ttfamily
  arXiv:2005.05138 [hep-ex]}}.

\bibitem{ATLAS:2019lsy}
{\bfseries ATLAS} Collaboration, G.~Aad {\em et~al.}, ``{Search for a heavy
  charged boson in events with a charged lepton and missing transverse momentum
  from $pp$ collisions at $\sqrt{s} = 13$ TeV with the ATLAS detector},''
  \href{https://dx.doi.org/10.1103/PhysRevD.100.052013}{{\em Phys. Rev. D}
  {\bfseries 100} no.~5, (2019) 052013},
  \href{https://arxiv.org/abs/1906.05609}{{\ttfamily arXiv:1906.05609
  [hep-ex]}}.

\bibitem{ATLAS:2023ibb}
{\bfseries ATLAS} Collaboration, G.~Aad {\em et~al.}, ``{Search for
  vector-boson resonances decaying into a top quark and a bottom quark using pp
  collisions at $ \sqrt{s} $ = 13 TeV with the ATLAS detector},''
  \href{https://dx.doi.org/10.1007/JHEP12(2023)073}{{\em JHEP} {\bfseries 12}
  (2023) 073}, \href{https://arxiv.org/abs/2308.08521}{{\ttfamily
  arXiv:2308.08521 [hep-ex]}}.

\bibitem{Alloul:2013bka}
A.~Alloul, N.~D. Christensen, C.~Degrande, C.~Duhr, and B.~Fuks, ``{FeynRules
  2.0 - A complete toolbox for tree-level phenomenology},''
  \href{https://dx.doi.org/10.1016/j.cpc.2014.04.012}{{\em Comput. Phys.
  Commun.} {\bfseries 185} (2014) 2250--2300},
  \href{https://arxiv.org/abs/1310.1921}{{\ttfamily arXiv:1310.1921 [hep-ph]}}.

\bibitem{Degrande:2011ua}
C.~Degrande, C.~Duhr, B.~Fuks, D.~Grellscheid, O.~Mattelaer, and T.~Reiter,
  ``{UFO - The Universal FeynRules Output},''
  \href{https://dx.doi.org/10.1016/j.cpc.2012.01.022}{{\em Comput. Phys.
  Commun.} {\bfseries 183} (2012) 1201--1214},
  \href{https://arxiv.org/abs/1108.2040}{{\ttfamily arXiv:1108.2040 [hep-ph]}}.

\bibitem{Alwall:2014hca}
J.~Alwall, R.~Frederix, S.~Frixione, V.~Hirschi, F.~Maltoni, O.~Mattelaer,
  H.~S. Shao, T.~Stelzer, P.~Torrielli, and M.~Zaro, ``{The automated
  computation of tree-level and next-to-leading order differential cross
  sections, and their matching to parton shower simulations},''
  \href{https://dx.doi.org/10.1007/JHEP07(2014)079}{{\em JHEP} {\bfseries 07}
  (2014) 079}, \href{https://arxiv.org/abs/1405.0301}{{\ttfamily
  arXiv:1405.0301 [hep-ph]}}.

\bibitem{Ball:2012cx}
R.~D. Ball {\em et~al.}, ``{Parton distributions with LHC data},''
  \href{https://dx.doi.org/10.1016/j.nuclphysb.2012.10.003}{{\em Nucl. Phys. B}
  {\bfseries 867} (2013) 244--289},
  \href{https://arxiv.org/abs/1207.1303}{{\ttfamily arXiv:1207.1303 [hep-ph]}}.

\bibitem{Buckley:2014ana}
A.~Buckley, J.~Ferrando, S.~Lloyd, K.~Nordstr\"om, B.~Page, M.~R\"ufenacht,
  M.~Sch\"onherr, and G.~Watt, ``{LHAPDF6: parton density access in the LHC
  precision era},''
  \href{https://dx.doi.org/10.1140/epjc/s10052-015-3318-8}{{\em Eur. Phys. J.
  C} {\bfseries 75} (2015) 132},
  \href{https://arxiv.org/abs/1412.7420}{{\ttfamily arXiv:1412.7420 [hep-ph]}}.

\bibitem{Sjostrand:2014zea}
T.~Sj\"ostrand, S.~Ask, J.~R. Christiansen, R.~Corke, N.~Desai, P.~Ilten,
  S.~Mrenna, S.~Prestel, C.~O. Rasmussen, and P.~Z. Skands, ``{An introduction
  to PYTHIA 8.2}'' \href{https://dx.doi.org/10.1016/j.cpc.2015.01.024}{{\em
  Comput. Phys. Commun.} {\bfseries 191} (2015) 159--177},
  \href{https://arxiv.org/abs/1410.3012}{{\ttfamily arXiv:1410.3012 [hep-ph]}}.

\bibitem{Dobbs:2001ck}
M.~Dobbs and J.~B. Hansen, ``{The HepMC C++ Monte Carlo event record for High
  Energy Physics},''
  \href{https://dx.doi.org/10.1016/S0010-4655(00)00189-2}{{\em Comput. Phys.
  Commun.} {\bfseries 134} (2001) 41--46}.

\bibitem{Conte:2012fm}
E.~Conte, B.~Fuks, and G.~Serret, ``{MadAnalysis 5, A User-Friendly Framework
  for Collider Phenomenology},''
  \href{https://dx.doi.org/10.1016/j.cpc.2012.09.009}{{\em Comput. Phys.
  Commun.} {\bfseries 184} (2013) 222--256},
  \href{https://arxiv.org/abs/1206.1599}{{\ttfamily arXiv:1206.1599 [hep-ph]}}.

\bibitem{Conte:2014zja}
E.~Conte, B.~Dumont, B.~Fuks, and C.~Wymant, ``{Designing and recasting LHC
  analyses with MadAnalysis 5},''
  \href{https://dx.doi.org/10.1140/epjc/s10052-014-3103-0}{{\em Eur. Phys. J.
  C} {\bfseries 74} no.~10, (2014) 3103},
  \href{https://arxiv.org/abs/1405.3982}{{\ttfamily arXiv:1405.3982 [hep-ph]}}.

\bibitem{Dumont:2014tja}
B.~Dumont, B.~Fuks, S.~Kraml, S.~Bein, G.~Chalons, E.~Conte, S.~Kulkarni,
  D.~Sengupta, and C.~Wymant, ``{Toward a public analysis database for LHC new
  physics searches using MADANALYSIS 5},''
  \href{https://dx.doi.org/10.1140/epjc/s10052-014-3242-3}{{\em Eur. Phys. J.
  C} {\bfseries 75} no.~2, (2015) 56},
  \href{https://arxiv.org/abs/1407.3278}{{\ttfamily arXiv:1407.3278 [hep-ph]}}.

\bibitem{Conte:2018vmg}
E.~Conte and B.~Fuks, ``{Confronting new physics theories to LHC data with
  MADANALYSIS 5},'' \href{https://dx.doi.org/10.1142/S0217751X18300272}{{\em
  Int. J. Mod. Phys. A} {\bfseries 33} no.~28, (2018) 1830027},
  \href{https://arxiv.org/abs/1808.00480}{{\ttfamily arXiv:1808.00480
  [hep-ph]}}.

\bibitem{Drees:2013wra}
M.~Drees, H.~Dreiner, D.~Schmeier, J.~Tattersall, and J.~S. Kim, ``{CheckMATE:
  Confronting your Favourite New Physics Model with LHC Data},''
  \href{https://dx.doi.org/10.1016/j.cpc.2014.10.018}{{\em Comput. Phys.
  Commun.} {\bfseries 187} (2015) 227--265},
  \href{https://arxiv.org/abs/1312.2591}{{\ttfamily arXiv:1312.2591 [hep-ph]}}.

\bibitem{Dercks:2016npn}
D.~Dercks, N.~Desai, J.~S. Kim, K.~Rolbiecki, J.~Tattersall, and T.~Weber,
  ``{CheckMATE 2: From the model to the limit},''
  \href{https://dx.doi.org/10.1016/j.cpc.2017.08.021}{{\em Comput. Phys.
  Commun.} {\bfseries 221} (2017) 383--418},
  \href{https://arxiv.org/abs/1611.09856}{{\ttfamily arXiv:1611.09856
  [hep-ph]}}.

\bibitem{Cacciari:2008gp}
M.~Cacciari, G.~P. Salam, and G.~Soyez, ``{The anti-$k_t$ jet clustering
  algorithm},'' \href{https://dx.doi.org/10.1088/1126-6708/2008/04/063}{{\em
  JHEP} {\bfseries 04} (2008) 063},
  \href{https://arxiv.org/abs/0802.1189}{{\ttfamily arXiv:0802.1189 [hep-ph]}}.

\bibitem{Cacciari:2011ma}
M.~Cacciari, G.~P. Salam, and G.~Soyez, ``{FastJet User Manual},''
  \href{https://dx.doi.org/10.1140/epjc/s10052-012-1896-2}{{\em Eur. Phys. J.
  C} {\bfseries 72} (2012) 1896},
  \href{https://arxiv.org/abs/1111.6097}{{\ttfamily arXiv:1111.6097 [hep-ph]}}.

\bibitem{deFavereau:2013fsa}
{\bfseries DELPHES 3} Collaboration, J.~de~Favereau, C.~Delaere, P.~Demin,
  A.~Giammanco, V.~Lema\^\i{}tre, A.~Mertens, and M.~Selvaggi, ``{DELPHES 3, A
  modular framework for fast simulation of a generic collider experiment},''
  \href{https://dx.doi.org/10.1007/JHEP02(2014)057}{{\em JHEP} {\bfseries 02}
  (2014) 057}, \href{https://arxiv.org/abs/1307.6346}{{\ttfamily
  arXiv:1307.6346 [hep-ex]}}.

\bibitem{Read:2002hq}
A.~L. Read, ``{Presentation of search results: The $CL_s$ technique},''
  \href{https://dx.doi.org/10.1088/0954-3899/28/10/313}{{\em J. Phys. G}
  {\bfseries 28} (2002) 2693--2704}.

\bibitem{Bierlich:2019rhm}
C.~Bierlich {\em et~al.}, ``{Robust Independent Validation of Experiment and
  Theory: Rivet version 3},''
  \href{https://dx.doi.org/10.21468/SciPostPhys.8.2.026}{{\em SciPost Phys.}
  {\bfseries 8} (2020) 026}, \href{https://arxiv.org/abs/1912.05451}{{\ttfamily
  arXiv:1912.05451 [hep-ph]}}.

\bibitem{Butterworth:2019wnt}
J.~Butterworth, ``{BSM constraints from model-independent measurements: A
  Contur Update},''
  \href{https://dx.doi.org/10.1088/1742-6596/1271/1/012013}{{\em J. Phys. Conf.
  Ser.} {\bfseries 1271} no.~1, (2019) 012013},
  \href{https://arxiv.org/abs/1902.03067}{{\ttfamily arXiv:1902.03067
  [hep-ph]}}.

\bibitem{Buckley:2021neu}
A.~Buckley {\em et~al.}, ``{Testing new physics models with global comparisons
  to collider measurements: the Contur toolkit},''
  \href{https://dx.doi.org/10.21468/SciPostPhysCore.4.2.013}{{\em SciPost Phys.
  Core} {\bfseries 4} (2021) 013},
  \href{https://arxiv.org/abs/2102.04377}{{\ttfamily arXiv:2102.04377
  [hep-ph]}}.

\bibitem{ATLAS:2018nud}
{\bfseries ATLAS} Collaboration, M.~Aaboud {\em et~al.}, ``{Search for photonic
  signatures of gauge-mediated supersymmetry in 13 TeV $pp$ collisions with the
  ATLAS detector},'' \href{https://dx.doi.org/10.1103/PhysRevD.97.092006}{{\em
  Phys. Rev. D} {\bfseries 97} no.~9, (2018) 092006},
  \href{https://arxiv.org/abs/1802.03158}{{\ttfamily arXiv:1802.03158
  [hep-ex]}}.

\bibitem{ATLAS:2021twp}
{\bfseries ATLAS} Collaboration, G.~Aad {\em et~al.}, ``{Search for squarks and
  gluinos in final states with one isolated lepton, jets, and missing
  transverse momentum at $\sqrt{s}=13$~ with the ATLAS detector},''
  \href{https://dx.doi.org/10.1140/epjc/s10052-021-09748-8}{{\em Eur. Phys. J.
  C} {\bfseries 81} no.~7, (2021) 600},
  \href{https://arxiv.org/abs/2101.01629}{{\ttfamily arXiv:2101.01629
  [hep-ex]}}. [Erratum: Eur.Phys.J.C 81, 956 (2021)].

\bibitem{ATLAS:2021fbt}
{\bfseries ATLAS} Collaboration, G.~Aad {\em et~al.}, ``{Search for
  R-parity-violating supersymmetry in a final state containing leptons and many
  jets with the ATLAS experiment using $\sqrt{s} = 13 \text{ TeV}$
  proton\textendash{}proton collision data},''
  \href{https://dx.doi.org/10.1140/epjc/s10052-021-09761-x}{{\em Eur. Phys. J.
  C} {\bfseries 81} no.~11, (2021) 1023},
  \href{https://arxiv.org/abs/2106.09609}{{\ttfamily arXiv:2106.09609
  [hep-ex]}}.

\bibitem{CMS:2019zmd}
{\bfseries CMS} Collaboration, T.~C. Collaboration {\em et~al.}, ``{Search for
  supersymmetry in proton-proton collisions at 13 TeV in final states with jets
  and missing transverse momentum},''
  \href{https://dx.doi.org/10.1007/JHEP10(2019)244}{{\em JHEP} {\bfseries 10}
  (2019) 244}, \href{https://arxiv.org/abs/1908.04722}{{\ttfamily
  arXiv:1908.04722 [hep-ex]}}.

\bibitem{CMS:2017abv}
{\bfseries CMS} Collaboration, A.~M. Sirunyan {\em et~al.}, ``{Search for
  supersymmetry in multijet events with missing transverse momentum in
  proton-proton collisions at 13 TeV},''
  \href{https://dx.doi.org/10.1103/PhysRevD.96.032003}{{\em Phys. Rev. D}
  {\bfseries 96} no.~3, (2017) 032003},
  \href{https://arxiv.org/abs/1704.07781}{{\ttfamily arXiv:1704.07781
  [hep-ex]}}.

\bibitem{CMS:2017moi}
{\bfseries CMS} Collaboration, A.~M. Sirunyan {\em et~al.}, ``{Search for
  electroweak production of charginos and neutralinos in multilepton final
  states in proton-proton collisions at $\sqrt{s}=$ 13 TeV},''
  \href{https://dx.doi.org/10.1007/JHEP03(2018)166}{{\em JHEP} {\bfseries 03}
  (2018) 166}, \href{https://arxiv.org/abs/1709.05406}{{\ttfamily
  arXiv:1709.05406 [hep-ex]}}.

\bibitem{CMS:2019xjf}
{\bfseries CMS} Collaboration, ``{Search for supersymmetry in proton-proton
  collisions at 13 TeV in final states with jets and missing transverse
  momentum},''. \href{https://cds.cern.ch/record/2682103}{CMS PAS SUS-19-006}.

\bibitem{Mrowietz:2020ztq}
M.~Mrowietz, S.~Bein, and J.~Sonneveld, ``{Implementation of the CMS-SUS-19-006
  analysis in the MadAnalysis 5 framework (supersymmetry with large hadronic
  activity and missing transverse energy; 137 fb${}^{-1}$)},''
  \href{https://dx.doi.org/10.1142/S0217732321410078}{{\em Mod. Phys. Lett. A}
  {\bfseries 36} no.~01, (2021) 2141007}.

\bibitem{CMS:2019xud}
{\bfseries CMS} Collaboration, ``{Search for new physics in multilepton final
  states in pp collisions at $\sqrt{s}=13~\mathrm{TeV}$},''.
  \href{https://cds.cern.ch/record/2668721}{CMS PAS EXO-19-002}.

\bibitem{Conte:2021xtt}
E.~Conte and R.~Ducrocq, ``{Implementation of the CMS-EXO-19-002 search in the
  MadAnalysis 5 framework (physics beyond the Standard Model with multilepton
  final states; 137 fb$^{-1}$)},''
  \href{https://dx.doi.org/10.1142/S0217732321410121}{{\em Mod. Phys. Lett. A}
  {\bfseries 36} no.~01, (2021) 2141012}.

\bibitem{CMS:2019ius}
{\bfseries CMS} Collaboration, ``{Search for supersymmetry with a compressed
  mass spectrum in events with a soft $\tau$ lepton, a highly energetic jet,
  and large missing transverse momentum in proton-proton collisions at
  $\sqrt{s}=13~\mathrm{TeV}$},''. \href{https://cds.cern.ch/record/2684821}{CMS
  PAS SUS-19-002}.

\bibitem{ATLAS:2020wzf}
{\bfseries ATLAS} Collaboration, ``{Search for new phenomena in events with
  jets and missing transverse momentum in p p collisions at $\sqrt{s}$ = 13 TeV
  with the ATLAS detector},''.
  \href{http://cds.cern.ch/record/2728058}{ATLAS-CONF-2020-040}.

\bibitem{ATLAS:2022nrp}
{\bfseries ATLAS} Collaboration, G.~Aad {\em et~al.}, ``{Cross-section
  measurements for the production of a Z boson in association with
  high-transverse-momentum jets in pp collisions at $ \sqrt{s} $ = 13 TeV with
  the ATLAS detector},'' \href{https://dx.doi.org/10.1007/JHEP06(2023)080}{{\em
  JHEP} {\bfseries 06} (2023) 080},
  \href{https://arxiv.org/abs/2205.02597}{{\ttfamily arXiv:2205.02597
  [hep-ex]}}.

\bibitem{ATLAS:2021jgw}
{\bfseries ATLAS} Collaboration, G.~Aad {\em et~al.}, ``{Measurements of
  $W^+W^-+\ge 1~$jet production cross-sections in $pp$ collisions at
  $\sqrt{s}=13~$TeV with the ATLAS detector},''
  \href{https://dx.doi.org/10.1007/JHEP06(2021)003}{{\em JHEP} {\bfseries 06}
  (2021) 003}, \href{https://arxiv.org/abs/2103.10319}{{\ttfamily
  arXiv:2103.10319 [hep-ex]}}.

\bibitem{CMS:2022ubq}
{\bfseries CMS} Collaboration, A.~Tumasyan {\em et~al.}, ``{Measurement of the
  mass dependence of the transverse momentum of lepton pairs in Drell-Yan
  production in proton-proton collisions at $\sqrt{s}$ = 13 TeV},''
  \href{https://dx.doi.org/10.1140/epjc/s10052-023-11631-7}{{\em Eur. Phys. J.
  C} {\bfseries 83} no.~7, (2023) 628},
  \href{https://arxiv.org/abs/2205.04897}{{\ttfamily arXiv:2205.04897
  [hep-ex]}}.

\bibitem{ATLAS:2019zci}
{\bfseries ATLAS} Collaboration, G.~Aad {\em et~al.}, ``{Measurement of the
  transverse momentum distribution of Drell\textendash{}Yan lepton pairs in
  proton\textendash{}proton collisions at $\sqrt{s}=13$ TeV with the ATLAS
  detector},'' \href{https://dx.doi.org/10.1140/epjc/s10052-020-8001-z}{{\em
  Eur. Phys. J. C} {\bfseries 80} no.~7, (2020) 616},
  \href{https://arxiv.org/abs/1912.02844}{{\ttfamily arXiv:1912.02844
  [hep-ex]}}.

\bibitem{ATLAS:2019rob}
{\bfseries ATLAS} Collaboration, M.~Aaboud {\em et~al.}, ``{Measurement of
  fiducial and differential $W^+W^-$ production cross-sections at $\sqrt{s}=13$
  TeV with the ATLAS detector},''
  \href{https://dx.doi.org/10.1140/epjc/s10052-019-7371-6}{{\em Eur. Phys. J.
  C} {\bfseries 79} no.~10, (2019) 884},
  \href{https://arxiv.org/abs/1905.04242}{{\ttfamily arXiv:1905.04242
  [hep-ex]}}.

\bibitem{cmssus16033}
F.~Ambrogi and J.~Sonneveld,
``{MadAnalysis5 recast of CMS-SUS-16-033},''.
%%CITATION = INSPIRE-1685439;%%.

\bibitem{Cacciapaglia:2014uja}
G.~Cacciapaglia and F.~Sannino, ``{Fundamental Composite (Goldstone) Higgs
  Dynamics},'' \href{https://dx.doi.org/10.1007/JHEP04(2014)111}{{\em JHEP}
  {\bfseries 04} (2014) 111}, \href{https://arxiv.org/abs/1402.0233}{{\ttfamily
  arXiv:1402.0233 [hep-ph]}}.

\bibitem{Preskill:1980mz}
J.~Preskill, ``{Subgroup Alignment in Hypercolor Theories},''
  \href{https://dx.doi.org/10.1016/0550-3213(81)90265-0}{{\em Nucl. Phys. B}
  {\bfseries 177} (1981) 21--59}.

\end{thebibliography}\endgroup

\end{document}